\documentclass[amsmath,amssymb,aps,pre,superscriptaddress,onecolumn,floatfix,longbibliography]{revtex4-2}
\usepackage{graphicx}
\usepackage{amsmath}
\usepackage{amssymb}
\usepackage{amsfonts}
\usepackage{tabularx}
\usepackage{float}
\usepackage[T1]{fontenc}
\usepackage{bbold}
\usepackage{bm}
\usepackage{hyperref}
\usepackage{subcaption}

\newcommand{\be}{\begin{equation}}
\newcommand{\ee}{\end{equation}}
\newcommand{\bea}{\begin{eqnarray}}
\newcommand{\eea}{\end{eqnarray}}
\DeclareMathOperator*{\SumInt}{%
\mathchoice%
  {\ooalign{$\displaystyle\sum$\cr\hidewidth$\displaystyle\int$\hidewidth\cr}}
  {\ooalign{\raisebox{.14\height}{\scalebox{.7}{$\textstyle\sum$}}\cr\hidewidth$\textstyle\int$\hidewidth\cr}}
  {\ooalign{\raisebox{.2\height}{\scalebox{.6}{$\scriptstyle\sum$}}\cr$\scriptstyle\int$\cr}}
  {\ooalign{\raisebox{.2\height}{\scalebox{.6}{$\scriptstyle\sum$}}\cr$\scriptstyle\int$\cr}}
}

\begin{document}

\title{Superdiffusion, normal diffusion and chaos in semiclassical Bose-Hubbard chains}

\author{Dragan Markovi\'c}
\email{vokramnagard@gmail.com}
\affiliation{Friedrich-Alexander-Universit\"at Erlangen-N\"urnberg, Department of Mathematics, Cauerstr. 11, D-91058 Erlangen, Germany}
\affiliation{Center for the Study of Complex Systems, Institute of Physics Belgrade, University of Belgrade, Pregrevica 118, 11080 Belgrade, Serbia}

\author{Mihailo \v{C}ubrovi\'c}
\email{cubrovic@ipb.ac.rs}
\affiliation{Center for the Study of Complex Systems, Institute of Physics Belgrade, University of Belgrade, Pregrevica 118, 11080 Belgrade, Serbia}

\date{\today}

\begin{abstract}
We study the evolution of two-point correlation functions of one-dimensional Bose-Hubbard model in the semiclassical regime in the framework of Truncated Wigner Approximation (TWA) with quantum jumps as first-order corrections. At early times, the correlation functions show strong superdiffusion with universal integer exponents determined solely by the initial conditions and completely insensitive to system parameters and chaos. Only after a long time this regime crosses over to normal diffusion regime which is most robust when nonintegrability is strong. For strong nonintegrability, the system ends up in a homogeneous state while for weak nonintegrability the oscillations and inhomogeneities persist, despite the fact that chaos is nearly always strong and only weakly depends on the nonintegrability parameter. We conclude that the superdiffusive regime is neither prethermalized nor a precursor to thermalization but a novel early-time phenomenon related to a special scaling symmetry of the Bose-Hubbard Hamiltonian.
\end{abstract}

\maketitle

\section{\label{secint}Introduction}

The interplay of chaos, transport and ergodicity in many-body systems is a very old puzzle, yet despite some deep knowledge we have gained so far there are still many unknowns. Some universal theorems guarantee the existence of diffusion, hydrodynamics and ergodicity in sufficiently chaotic systems, but as it often goes with rigorous results, their actual domain of applicability is not clear, and we often do not know how much the specifics of the system at hand influence the results. Systems with mixed phase space, with both regular and chaotic structures, are notoriously difficult to understand, but even in strongly chaotic systems one may find surprises, e.g. the absence of thermalization \cite{Lin:2017} or recurrences from quantum scars \cite{Lin:2019}.

This theoretical framework has ample applications in experimentally relevant systems. The typical playground are the cold atom systems, where one can construct simple model Hamiltonians and fine-tune the interactions. In cold-atom systems, quantities of relevance for chaos and transport such as conductivity or out-of-time ordered correlator can be measured and compared to theory. Atom-laser interactions are another field where one can experimentally assess both transport and chaos. Further examples are provided by cavity quantum electrodynamics, quantum billiards and other mesoscopic systems. The goal is always to measure some transport-related quantity such as conductivity, thermal conductivity or density gradient and to relate it to some microscopic chaos quantity such as quantum fidelity or time-disordered correlators.

In this work we attempt to relate microscopic chaos to transport in one-dimensional Bose-Hubbard chains, and then to draw some universal lessons from it. The universality, experimental and phenomenological relevance of the Bose-Hubbard model has inspired many works on chaos, and still more works on transport. In a nutshell, the Bose-Hubbard the system is in general chaotic and in parallel with expected hydrodynamic normal diffusion it can also exhibit some unusual transport properties such as superdiffusion \cite{Kolovsky:2004,Kolovsky:2007,Graefe:2007,Trimborn:2008,Kolovsky:2016,Fischer:PhysRevA.93.043620,Sorg:PhysRevA.90.033606,Kolovsky:2020,McCormack:2021Photo...8..554M,Altman:PhysRevLett.98.180601,Pausch2body,PauschSpectrum,Dag:2022vqb,Dahan:2022classical,PauschOptimalRoute,Nakerst:2022prc,Markovic:2023gtx,Minganti2023dissipativephase,Ferrari:2023zfe,Ferrari:2024ogi,Lu:2025bqn,Pausch:2025zgd}. We look at chains with large total occupation number $N$, which are amenable to a semiclassical description in terms of a controlled expansion organized around the Truncated Wigner Approximation (TWA). Although we clearly cannot see some deep quantum effects in this regime, it has the advantage that it allows us to study long chains and the dependence of equilibration and thermalization on chain length, which will prove quite important.

Deep in the quantum regime, there are two main aspects of chaos: energy level statistics \cite{Haake:book} and the time-disordered correlation functions \cite{Swingle:2018ekw}. The connection between these is a subject of active research and is far form clear yet. However, in the semiclassical approximation, we do not have access to the discrete structure of the energy spectrum, and the chaos that we find is not much of a quantum chaos: we study it through Lyapunov exponents, power spectrum and other usual classical indicators. Nevertheless, the fact that we consider an ensemble of orbits distributed according to the Wigner pseudodistribution, and also the quantum jump corrections, provides for genuinely new physics compared to purely classical analysis. It is truly semiclassical chaos, neither completely classical nor quantum.

There is already a substantial body of work done on the semiclassical Bose-Hubbard systems, relying largely on the Wigner pseudodistribution (which makes the essence of the TWA approach) or its close relative, the Husimi function. This line of work was started already in \cite{PolkovnikSachdev2002,Polkovnik2003,Polkovnik2003Main} and subsequently used in many of the papers cited above, though mostly in the leading-order approximation, without quantum jumps. Quantum jumps are often assumed to lead only to small corrections, in sharp contrast with the discontinuous change of superdiffusion exponents that we find. Recently, however, the importance of quantum jump corrections was explored in detail in \cite{Ferrari:2023zfe,Ferrari:2024ogi} (although these works deal with a dissipative case, very different from an isolated chain, but still point out to the importance of quantum corrections). These studies show that the Wigner pseudodistribution itself gives insight into the structure of the chaotic states. In this paper we take the complementary position, studying mainly the expectation values and their evolution.

In our previous work \cite{Markovic:2023gtx} we have specifically studied the semiclassical chaos in parallel with the anomalous transport, i.e. the evolution of the distribution function of occupation numbers along the chain. In that work we have considered purely classical trajectories for an ensemble of initial conditions, thus corresponding to the leading-order TWA ("pure TWA") \cite{Polkovnik2003Main,Polkovnik2003,PolkovnikAnnals,PolkovnikSciPost,Polkovnik2018}. The surprising findings on transport in \cite{Markovic:2023gtx} can be summarized as follows:
\begin{enumerate}
    \item At short timescales transport is either highly anomalous (superdiffusive), with particle number distribution evolving according to power laws $t^\alpha$ where the integer exponent $\alpha$ is determined solely by initial conditions, or totally absent (the number distribution oscillates without spreading). Which of the two scenarios is realized is also determined by the initial conditions.
    \item At very long timescales, superdiffusion will sometimes become normal but not always.
    \item Chaos is nearly always strong but there is some (very much expected) dependence on the chain length (long chains have a vanishing measure of stability islands) and nonintegrability (the more nonintegrable, the more chaotic). Strangely, transport is completely insensitive to these same factors: anomalous diffusion is universal and blind to the strength of chaos.
\end{enumerate}
Thus, even though strong chaos is normally meant to imply normal diffusion, in \cite{Markovic:2023gtx} we have always seen superdiffusion which is usually associated either with nonlocal interactions or with integrable structures. At first the physics beyond this phenomenology was not clear: one could hypothesize either that the system is in prethermalized state until normal diffusion kicks in, or that long-range quantum correlations survive even in the semiclassical limit, but we did not have clear arguments for either.

In this work, we explain the nonstandard transport properties as a special early-time phenomenon which is not related either to chaos or to later normal diffusion: it is a consequence of the scaling properties of the Bose-Hubbard Hamiltonian. On the other hand, we find that normal diffusion and thermalization are indeed correlated with the strength of chaos. These realizations have been possible also due to the inclusion of leading quantum corrections modeled by quantum jumps within the framework of TWA. While higher-order TWA is still a semiclassical approach in the sense that it is an expansion in $\hbar$, it goes beyond the leading-order calculations, containing in an explicit way the nonlocal quantum correlations. Somewhat unexpectedly, it will turn out that these have a substantial influence on quantitative aspects of superdiffusion, without changing it qualitatively. More importantly, the normal diffusion appears nearly universally once the quantum jumps are included. Together with some analytical arguments, it is this comparison to the normal diffusion regime that leads to the understanding of the superdiffusion regime as a novel, universal, early-time phenomenon which is insensitive both to chaos and to late-time normal diffusion and equilibration. 

The plan of the paper is the following. The next section introduces the general formalism of TWA for the Bose-Hubbard chain. Section \ref{secres} considers the transport of particle density and contains universal and general scaling laws for superdiffusive transport as well as the description of the transition to long-time normal diffusion and thermalization. Section \ref{secother} explains how other observables may have very different transport properties. In Section \ref{secchaos} we relate the transport to chaos and mixing. Section \ref{secconc} sums up the conclusions.

\section{\label{secsetup}Setup and formalism}

\subsection{\label{secsetupbh}Bose-Hubbard chain and classical variables}

The Hamiltonian of the Bose-Hubbard model in one dimension with $L$ sites reads:
\be
H=\sum_{j=1}^L\left[-J\left(b_j^\dagger b_{j+1}+b_jb_{j+1}^\dagger\right)+\frac{U}{2}n_j\left(n_j-1\right)-\mu n_j\right],~~n_j\equiv b_j^\dagger b_j\label{bh0}
\ee
The hopping parameter, the on-site Coulomb repulsion and the chemical potential are $J$, $U$ and $\mu$ respectively. When considering the semiclassical limit and also the quantum corrections to it (which we do in the next section), it is convenient to introduce two different sets of variables. The first step is two rescale the creation and annihilation operators $b_j^\dagger,b_j$ by the total number $N=\sum_{j=1}^L n_j$ and take the limit $N\to\infty$, when the classical limit becomes exact. This yields
\begin{equation}
\label{bh}H=\sum_{j=1}^L\left[-J\left(\psi_j^\dagger \psi_{j+1}+\psi_j\psi^\dagger_{j+1}\right)+\frac{U}{2}\psi_j^\dagger\psi_j\left(\psi_j^\dagger\psi_j-1\right)-\mu \psi_j^\dagger\psi_j\right],
\end{equation}
where we have now rescaled $U\mapsto UN$. The commutator of the new variables is $\left[\psi^*_j,\psi_j\right]=1/N\to 0$, thus $\psi^*_j,\psi_j$ become classical quantities. Two more useful sets of variables are the canonical variables $(P_j,Q_j)$ and the number-phase variables $(I_j,\phi_j)$. The canonical variables and the corresponding Hamiltonian are
\bea
&&P_j=\frac{1}{\sqrt{2}}\left(\psi_j+\psi_j^*\right),~~Q_j=\frac{i}{\sqrt{2}}\left(\psi_j-\psi_j^*\right)\label{bhpq}\\
&&H=\sum_{j=1}^L\left[-J\left(P_jP_{j+1}+Q_jQ_{j+1}\right)+\frac{U}{8}\left(P_j^2+Q_j^2\right)^2-\frac{\mu}{2}\left(P_j^2+Q_j^2\right)\right],\label{bhpqham}
\eea
yielding the equations of motion
\bea
&&\dot{P}_j=-J(Q_{j-1}+Q_{j+1})+Q_j\left(\frac{U}{2}\left(P_j^2+Q_j^2\right)-\mu\right)\label{bhpqeoms1}\\
&&\dot{Q}_j=J(P_{j-1}+P_{j+1})-P_j\left(\frac{U}{2}\left(P_j^2+Q_j^2\right)-\mu\right).\label{bhpqeoms2}
\eea
The number-phase variables are particularly useful when considering dynamics and chaos: they have a clear physical interpretation and in the integrable limit $J=0$ they correspond to action-angle variables \footnote{With slight abuse of terminology, the name action-angle variables is also used in the literature (including our previous work \cite{Markovic:2023gtx}) in nonintegrable cases, the logic being that if the system is sufficiently close to integrable the actions are still adiabatic invariants.}. The variables (in terms of $P_j,Q_j$ defined above) and the Hamiltonian read:
\bea
\label{bhiphi}&&P_j=\sqrt{2I_j}\sin\phi_j,~~Q_j=\sqrt{2I_j}\cos\phi_j\\
\label{bhiphiham}&&H=\sum_{j=1}^L\left(\frac{U}{2}I_j^2-\mu I_j\right)-2J\sum_{j=1}^L\sqrt{I_jI_{j+1}}\cos\left(\phi_j-\phi_{j+1}\right).
\eea
This results in the equations of motion:
\bea
&&\dot{I}_j=2J\left(\sqrt{I_jI_{j-1}}\sin\left(\phi_{j-1}-\phi_j\right)+\sqrt{I_jI_{j+1}}\sin\left(\phi_{j+1}-\phi_j\right)\right)\label{eomphi}\label{bhiphieoms1}\\
&&\dot{\phi}_j=-\mu+UI_j-J\left(\sqrt{\frac{I_{j-1}}{I_j}}\cos\left(\phi_j-\phi_{j-1}\right)+\sqrt{\frac{I_{j+1}}{I_j}}\cos\left(\phi_j-\phi_{j+1}\right)\right)\label{eomi}\label{bhiphieoms2}
\eea
As it has to be, the right-hand side of (\ref{bhiphieoms1}) vanishes when $J=0$.

\subsection{\label{secsetuptwa}Truncated Wigner Approximation and beyond: general formalism}

Now we formulate the framework of TWA for the Bose-Hubbard Hamiltonian. We will use both the canonical variables (Eqs.~\ref{bhpq}-\ref{bhpqham}) and the number-phase variables (Eqs.~\ref{bhiphi}-\ref{bhiphiham}). The outcome will be an expansion in powers of $\hbar$ of correlation functions (we focus on one-point functions, i.e. expectation values, and two-point functions relevant for transport). Nice reviews of TWA and beyond are found in \cite{PolkovnikAnnals,PolkovnikSciPost}. Early implementations and applications to the Bose-Hubbard model are \cite{PolkovnikSachdev2002,Polkovnik2003,Polkovnik2003Main,Sinatra:2002JPhB...35.3599S}. Later applications to the Bose-Hubbard model are \cite{Pelster:2022ScPP...12...51M} and several references cited in the Introduction.

We will first consider the coordinate-momentum basis Wigner pseudodistribution $W(P,Q)$. The idea is that one would like to construct a "probability distribution function" in phase space -- but because of the uncertainty relations for conjugate pairs one cannot really construct such a distribution function and the best we can do is a pseudodistribution, normalized to unity but not positive definite:
\be 
\int\frac{d^DQ}{(2\pi)^D}\int\frac{d^DP}{(2\pi)^D}W(P,Q)=1.\label{wnorm}
\ee
It has the form of a Fourier transform of the density matrix $\rho$ over one of the variables (being an operator, it depends on two variables):
\bea
W(P,Q)=\int\frac{d^D\xi}{(2\pi)^D}\langle Q-\xi/2\vert\rho\vert Q+\xi/2\rangle e^{\imath P\xi}=\int\frac{d^D\kappa}{(2\pi)^D}\langle P-\kappa/2\vert\rho\vert P+\kappa/2\rangle e^{-\imath Q\kappa}\label{wpq}
\eea
The equation of motion for $W$ is obtained from the Moyal bracket:
\be 
\frac{\partial W}{\partial t}=\lbrace H_W,W\rbrace_M\equiv 2H_W\sin\frac{\Lambda}{2}W,\label{weom0}
\ee
where $H_W$ is the Wigner Hamiltonian (correctly ordered $H$), $\lbrace\rbrace_M$ is the Moyal bracket and
\bea 
\Lambda\equiv\frac{\overleftarrow{\partial}}{\partial Q}\frac{\overrightarrow{\partial}}{\partial P}-\frac{\overleftarrow{\partial}}{\partial P}\frac{\overrightarrow{\partial}}{\partial Q}\Rightarrow A\Lambda B=\frac{\partial A}{\partial Q}\frac{\partial B}{\partial P}-\frac{\partial A}{\partial P}\frac{\partial B}{\partial Q}=\lbrace A,B\rbrace.\label{moyal}
\eea
The leading approximation for the Moyal bracket is just the Poisson bracket for the classical EOM. Then $W$ becomes the phase space distribution, which is positive definite, satisfies the Liouville equation and is conserved on (on-shell) trajectories $P(t)$ and $Q(t)$, the solutions of classical EOM. Still, it goes beyond the classical limit precisely because it is a distribution -- it yields the evolution of expectation values over this distribution, it spreads and evolves just like (a first approximation of) a quantum probability amplitude. This, lowest-order approximation already works much better than the integration of a \emph{single} classical trajectory and has proven useful in many works \cite{PolkovnikSachdev2002,Polkovnik2003,Polkovnik2003Main,Dag:2022vqb,Markovic:2023gtx}. This is the TWA in the strict sense, whereas higher-order corrections are typically called "beyond-TWA".

At higher order $W$ is not positive definite and although it satisfies an analytic equation of motion and changes continuously in time, there are no orbits (characteristics) $P(t)$ and $Q(t)$ on which it is conserved hence if we simulate the evolution by equations for $P,Q$ we need to introduce "quantum jumps". In doing this we mainly follow \cite{PolkovnikAnnals} and the recent work \cite{Ferrari:2024ogi}.


The final formal ingredient we need is the trick to write $H_W$ and equation of motion for $W$ for any given $H$ making use of the Bopp operators
\be
Q\mapsto\hat{Q}\equiv Q+\frac{\imath}{2}\frac{\partial}{\partial P}=Q-\frac{\imath}{2}\frac{\overleftarrow{\partial}}{\partial P},~~P\mapsto\hat{P}\equiv P-\frac{\imath}{2}\frac{\partial}{\partial Q}=P+\frac{\imath}{2}\frac{\overleftarrow{\partial}}{\partial Q}\label{bopp}
\ee
Once we have the equation of motion for $W$, the first-order term in $\hbar$ in the expansion of the Moyal bracket yields drift terms (first order in $\partial_P,\partial_Q$) and diffusion terms (second order in $\partial_P,\partial_Q$). Second-order term in $\hbar$, which is third- and higher-order in $P,Q$, is only present if the Hamiltonian contains quartic (e.g. Coulomb repulsion in the Hubbard model) and higher interactions. Strict TWA corresponds to cutting off these terms so we only have a Fokker-Planck equation (diffusion in phase space). Its equivalent Langevin equation has a deterministic term from first-order and a stochastic term from second-order contributions in $\partial_P,\partial_Q$. If no second-order term is present the equivalent Langevin equation is in fact noiseless.


\subsection{\label{secsetupbhtwa}TWA equations for the Bose-Hubbard chain}

Applying the Bopp operators (\ref{bopp}), the Wigner Hamiltonian becomes:
\be
\label{wbh} H_W=\sum_{j=1}^L\left[-J\left(\psi_j^* \psi_{j+1}+\psi_j\psi^*_{j+1}\right)+\frac{U}{2}|\psi_j|^4-\Tilde{\mu}|\psi_j|^2\right],~~\Tilde{\mu}\equiv\mu+U
\ee
The equation of motion for the Wigner distribution function (\ref{weom0}) can be packaged into the form ($\lbrace,\rbrace_c$ is the classical Poisson bracket):
\begin{equation}
    \imath\hbar\frac{\partial W}{\partial t}=\{H_W,W\}_c+\frac{1}{8}\sum_{k,l,m}\frac{\partial^3 H_W}{\partial \psi_k\partial\psi^*_l\partial\psi^*_m}\frac{\partial^3 W}{\partial \psi^*_k\partial\psi_l\partial\psi_m}-\mathrm{c.c.}\label{weom}
\end{equation}
This is all: while in general one expects an infinite series from the Moyal bracket, since the highest-order term in our Hamiltonian is of fourth order there are no terms with derivatives of order five or higher. We will now transform the above equation from $\psi$ variables to the canonical pair $(P_j,Q_j)$, according to Eqs.~(\ref{bhpqham}). In these coordinates we get:
\begin{equation}
    \label{EOMpq} \hbar\frac{\partial W}{\partial t}=\sum_j \frac{\partial }{\partial P_j}(f_{Q_j}W)-\frac{\partial}{\partial Q_j}(f_{P_j}W)\\ + \frac{U}{8}\frac{\partial^3}{\partial Q_j^3}(P_j W)-\frac{U}{8}\frac{\partial^3}{\partial P_j^3}(Q_j W)+\frac{U}{8}\frac{\partial^3}{\partial Q_j \partial P_j^2}(P_j W)-\frac{U}{8}\frac{\partial^3}{\partial P_j \partial Q_j^2}(Q_j W),
\end{equation}
where we use the notation
\be
f_{x_j}\equiv -J\left(x_{j-1}+x_{j+1}\right)+\frac{U}{2}\left(Q_j^2+P_j^2\right)x_k-\Tilde{\mu}x_k.
\ee
The drift terms in this equation are $\frac{f_{Q_k}}{\hbar}$ and $-\frac{f_{P_k}}{\hbar}$ and diffusion terms are absent; however, we have third-derivative terms.

Taking the expectation value of an operator over the Wigner pseudodistribution and making use of Eq.~(\ref{EOMpq}) we can evaluate the mean values of an observable $\hat{O}$ by the following recipe (for a detailed derivation one can consult e.g. the review \cite{PolkovnikAnnals}, Eq.~(239)):
\begin{multline}
    \langle \hat{O}(\hat{\mathbf{P}},\hat{\mathbf{Q}},t)\rangle=\int d\mathbf{P_0}\int d\mathbf{Q_0}W_0(\mathbf{P_0},\mathbf{Q_0})\times\\
    \times\left[1+\frac{U}{8\hbar}\sum_j\sum_A\int dS_j\int dR_j\left(Q_j\left(\tau_A\right)R_j+S_jP_j\left(\tau_A\right)\right)\left(R_j^2+S_j^2-4\right)\exp\bigg(-\frac{S_j^2+R_j^2}{2}\bigg)\right]O(\mathbf{P},\mathbf{Q},t),\label{opexpval}
\end{multline}
where the hats emphasize that we take the expectation values of quantum operators, as opposed to the non-hatted variables on the right-hand side which denote just classical quantities as functions of classical coordinates and momenta $\mathbf{Q}$ and $\mathbf{P}$. At times $\tau_A$ quantum jumps occur, which evolve the canonical variables as
\be
Q_j(\tau_A) \longrightarrow Q_j(\tau_A)+S_j\sqrt[3]{\Delta \tau}\equiv Q_j(\tau_A)+\delta Q_j(\tau_A),~~P_j(\tau_A)\longrightarrow P_j(\tau_A)+R_j\sqrt[3]{\Delta \tau}\equiv P_j(\tau_A)+\delta P_j(\tau_A),\label{twajumps}
\ee
where $\Delta\tau\equiv\tau_{A+1}-\tau_A$ and $\sum\Delta\tau=t$. Notice that the last two terms in Eq.~(\ref{opexpval}) in fact do not contribute to jumps at this order because they are proportional to the first moment of the stochastic variables ($R_k$ or $S_k$), which vanishes.

At this point, one may decide to use Eq.~(\ref{opexpval}) directly to simulate evolution through a stochastic differential equation as in \cite{Ferrari:2024ogi}, or alternatively we can integrate over the jumps, taking into account the moments of the jump distribution, given in \cite{PolkovnikAnnals}. The choice involves a tradeoff: the former requires adding stochastically generated terms to the right-hand side of the equations and averaging over them, the latter requires us to solve an integro-differential equation, integrating the right-hand side over the function $\hat{O}(\hat{P},\hat{Q},t)$.

From now on we will not explicitly write the dependence of the operators on the initial values $\mathbf{P}_0$ and $\mathbf{Q}_0$ so as not to clutter the notation (we will still write it for $W_0$ because that quantity only depends on initial and boundary conditions and does not evolve). Applying the formula (\ref{opexpval}) to the action (number) operator we get:
\begin{equation}
    \langle\hat{I}_j (t)\rangle=\int d\mathbf{P_0}\int d\mathbf{Q_0}W_0(\mathbf{P_0},\mathbf{Q_0})\frac{P_j(t)^2+Q^2_j(t)}{2}.\label{i1exp}
\end{equation}
Therefore, single-point correlators for the actions do not receive any contribution from the jumps (this is because the expression $I_j\propto P_j^2+Q_j^2$ only contains first and second moments of the jumps, which vanish). For the square of the action the story is different (because the third moments of the jumps are nonzero):
\begin{eqnarray}
&&\langle\hat{I}^2_j(t)\rangle=\int d\mathbf{P_0}\int d\mathbf{Q_0}W_0(\mathbf{P_0},\mathbf{Q_0})\left(\frac{P_j(t)^2+Q_j(t)^2}{2}\right)^2+\nonumber\\ &+&\frac{U}{8\hbar}\int d\mathbf{P_0}\int d\mathbf{Q_0}W_0(\mathbf{P_0},\mathbf{Q_0})\SumInt_{A,k}\left(Q_k\left(\tau_A\right)R_k+P_k\left(\tau_A\right)S_k\right)\frac{1}{4}\left[\left(P_j+\delta P_j)^2+(Q_j+\delta Q_j\right)^2\right]^2.\label{i2exp0}
\end{eqnarray}
In the above, we have denoted
\be
\SumInt_{A,k}=\sum_{A,k}\int dS_k\int dR_k\left(R_k^2+S_k^2-4\right)\exp\left(-\frac{S_k^2+R_k^2}{2}\right).\label{sumint}
\ee
The nontrivial part of the above integral then becomes:
\begin{eqnarray}
&&\frac{U}{32\hbar}\int d\mathbf{P_0}\int d\mathbf{Q_0}W_0(\mathbf{P_0},\mathbf{Q_0})\SumInt_{A,k}\left(Q_k\left(\tau_A\right)R_k+P_k\left(\tau_A\right)S_k\right)\times 4 \left(Q_j\left(\tau_A\right)\delta Q_j^3+P_j\left(\tau_A\right)\delta P_j^3\right)=\nonumber\\
&=&\frac{U}{8\hbar}\int d\mathbf{P_0}\int d\mathbf{Q_0}W_0(\mathbf{P_0},\mathbf{Q_0})\SumInt_{A,k}\left(Q_k\left(\tau_A\right)R_k+P_k\left(\tau_A\right)S_k\right)\times \left(Q_j\left(\tau_A\right)S_j^3+P_j\left(\tau_A\right)R_j^3\right)\Delta\tau=\nonumber\\
&=&\frac{U}{8\hbar}\int d\mathbf{P_0}\int d\mathbf{Q_0}W_0\left(\mathbf{P_0},\mathbf{Q_0}\right)\SumInt_{A,k}\left(P_j\left(\tau_A\right)Q_k\left(\tau_A\right)R_k^4+P_k(\tau_A)Q_j\left(\tau_A\right)S_k^4\right)
\end{eqnarray}
The integral over the jumps in Eq.~(\ref{sumint}) becomes elementary by introducing polar coordinates and evaluates to $12\pi$ \footnote{Explicitly: $\int\int dS_jdR_j R_j^4 (R_j^2+S_j^2-4)\exp\left(-\frac{R_j^2+S_j^2}{2}\right)=\int r dr\int d\varphi r^4 \cos^4\varphi (r^4-4)e^{-\frac{r^2}{2}}=\frac{3\pi}{4}\times 16=12\pi$.}. At the end of the day, $I_j^2$ evolves as
\begin{eqnarray}
&&\langle \hat{I}^2_j(t)\rangle=\int d\mathbf{P_0}\int d\mathbf{Q_0}W_0(\mathbf{P_0},\mathbf{Q_0})\left(\frac{P^2_j(t)+Q^2_j(t)}{2}\right)^2+\nonumber\\&+&\frac{3U\Delta\tau\pi}{2\hbar}\int d\mathbf{P_0}\int d\mathbf{Q_0}W_0(\mathbf{P_0},\mathbf{Q_0})\sum_{A,k}\left(P_j\left(t\right)Q_k\left(\tau_A\right)+Q_j\left(t\right)P_k\left(\tau_A\right)\right).\label{i2exp}
\end{eqnarray}
The practical way to evaluate Eq.~(\ref{i2exp}) always starts from integrating an ensemble of classical trajectories as in pure TWA (corresponding to the first line of the equation). Then one has to make the choice that we discussed after Eq.~(\ref{opexpval}). One can either explicitly perform the sum over quantum jumps in the second line, drawing the jumps from an appropriate distribution given in \cite{PolkovnikAnnals} (the result converges to a stable value as $\Delta\tau$ is taken smaller and smaller, and subsequent $\tau_n$ in the sum come closer and closer). The other option is to take the limit $\Delta\tau\to 0$ in the sum and turn it into the integral over classical trajectories:
\begin{eqnarray}
&&\langle \hat{I}^2_j(t)\rangle=\int d\mathbf{P_0}\int d\mathbf{Q_0}W_0(\mathbf{P_0},\mathbf{Q_0})\left(\frac{P^2_j(t)+Q^2_j(t)}{2}\right)^2+\nonumber\\&+&\frac{3U\pi}{2\hbar}\int d\mathbf{P_0}\int d\mathbf{Q_0}W_0(\mathbf{P_0},\mathbf{Q_0})\sum_k\int dt'\left(P_j\left(t\right)Q_k\left(t'\right)+Q_j\left(t\right)P_k\left(t'\right)\right).\label{i2expint}
\end{eqnarray}
We find the second formulation more efficient for numerical work. In Appendix \ref{secappequiv} we demonstrate that the two formulations indeed give nearly identical results.

The above result is readily generalized to mixed two-point correlation functions relating two different sites at two different time instants. Following the same derivation as in Eqs.~(\ref{opexpval}-\ref{i2exp}) we arrive at:
\begin{eqnarray}
    &&\langle I_i(t_1)I_j(t_2)\rangle=\int d\mathbf{P_0}\int d\mathbf{Q_0}W_0(\mathbf{P_0},\mathbf{Q_0})
    \left(\frac{P_i^2(t_1)+Q^2_i(t_1)}{2}\right)\left(\frac{P^2_j(t_2)+Q^2_j(t_2)}{2}\right)+\nonumber\\
    &&\frac{3U\Delta\tau\pi }{4\hbar}\int d\mathbf{P_0}\int d\mathbf{Q_0}W_0(\mathbf{P_0},\mathbf{Q_0})\sum_{A,k}\left[\left(P_i\left(t_1\right)+P_j\left(t_2\right)\right)Q_k\left(\tau_A\right)+\left(Q_i\left(t_1\right)+Q_j\left(t_2\right)\right)P_k\left(\tau_A\right)\right]\label{i2expmixed}
\end{eqnarray}
The sum over the jumps $A$ can again be rewritten as an integral, in the same way as in Eq.~(\ref{i2expint}).

The final question is what initial conditions to choose for the Wigner function (which determines also the operator VEVs). Classically, it can be anything; quantum-mechanically, it has to be a consistent equilibrium solution to Eq.~(\ref{weom}). The physically motivated solution is the coherent state in canonical variables $P_n,Q_n$ which respects the translation and phase invariance. This means
\be
W_0(\mathbf{P},\mathbf{Q})=\mathcal{N}\times e^{-\frac{\left(\mathbf{P}-\bar{\mathbf{P}}\right)^2}{\sigma_P^2}-\frac{\left(\mathbf{Q}-\bar{\mathbf{Q}}\right)^2}{\sigma_Q^2}},\label{w0coherent}
\ee
i.e. the (in general squeezed) coherent state is a Gaussian with widths $\sigma_P$ and $\sigma_Q$ in the momenta and coordinates respectively. For $\sigma_P\to 0$ and $\sigma_Q\to\infty$ we get a fully localized wavepacket at some sites determined by $\bar{\mathbf{Q}}$ -- the classical case of initially fully localized particles, which was found in \cite{Markovic:2023gtx} to yield anomalous (superdiffusive) scaling laws. For boundary conditions we impose no flux through the edges of the chain, i.e. no sinks or sources at the sites $1$ and $L$ so that the total number is conserved.

\section{\label{secres}Transport and superdiffusion}

\subsection{\label{secresnum}Superdiffusion of number density}

Our most robust result is the superdiffusive transport of the connected two-point correlator (which is of course nothing but the dispersion of the number/action):
\be
D_{mn}\equiv\langle I_mI_n\rangle^2-\langle I_m\rangle\langle I_n\rangle\label{dmn}
\ee
for appropriately chosen (but sufficiently generic) initial conditions. In other words, the superdiffusion noticed in our earlier work \cite{Markovic:2023gtx} survives the quantum-jump correction to TWA but, as we shall see, the anomalous exponents change. To remind, in \cite{Markovic:2023gtx} we have considered the zeroth-order TWA, i.e. the distribution function of classical orbits, and found superdiffusion whenever a subset of sites is initially filled and the others are completely empty. For such configurations, we have found that if the site $n$ initially had nonzero filling $I_n$, the second moment of the number density at sites $I_{n+l}$ and $I_{n-l}$, i.e. sites at distance $l$ from the site $n$, disperses as
\be
D_{n\pm l,n\pm l}=\langle I^2_{n\pm l}\rangle-\langle I_{n\pm l}\rangle^2\sim t^{4l}.\label{oldie}
\ee
If no site is initially empty, then there is no transport at all and the moments $\langle I_m\rangle$ and $\langle I_m^2\rangle$ just oscillate for all $m$ with no long-term trend. 

\begin{figure}[H]
(a)\includegraphics[width=0.27\textwidth]{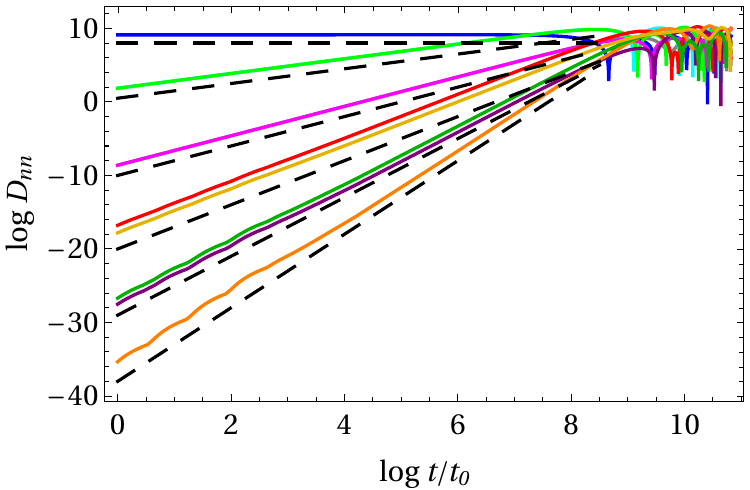}
(b)\includegraphics[width=0.27\textwidth]{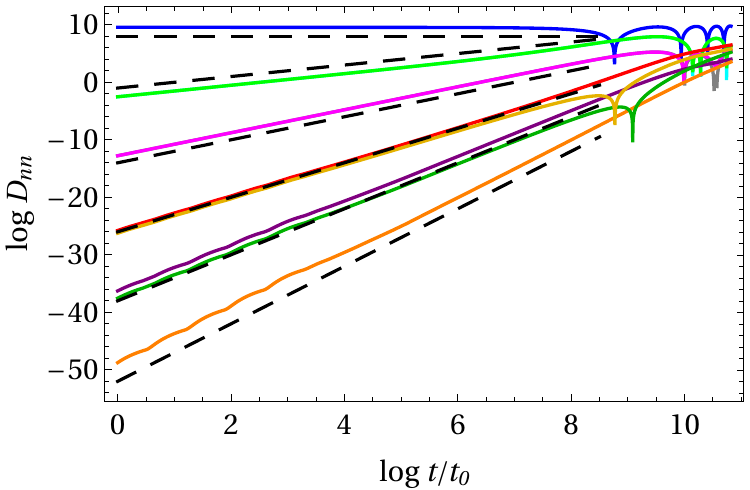}
(c)\includegraphics[width=0.37\textwidth]{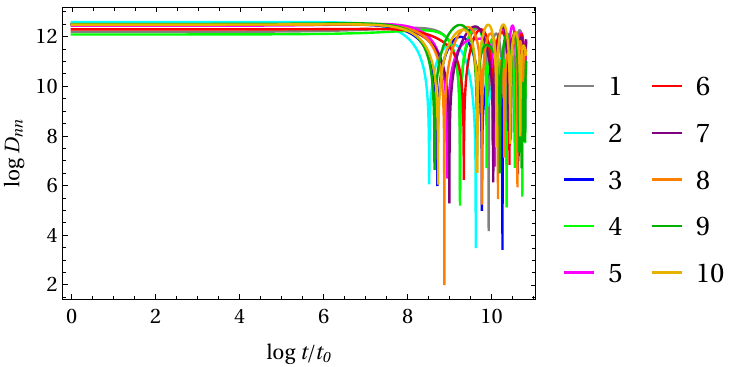}
\caption{\label{superfluidTWA} Log-log plot of the dispersion $D_{nn}=\langle I_n^2\rangle-\langle I_n\rangle^2$ in a chain of length $L=10$ in three different cases: (a) $U/J=0.1$ and initially filled site $n=3$, the rest empty (b) $U/J=10$ and initially filled site $n=3$, the rest empty (c) $U/J=0.1$ and all sites initially filled with uniformly distributed random fillings. In all three cases we have $\mu/J=0.05$. In (a) and (b), black dashed lines confirm the law $D(I_n^2(t))\sim t^l$, with $l\in\mathbb{N}$ being the distance from the initially filled site $n=3$. The color code (lighter and darker lines) denotes the site number.}
\end{figure}

Now we generalize this result to mixed two-site correlators (with $m\neq n$ in Eq.~(\ref{dmn})), and to beyond-TWA (i.e., quantum jump) calculations. We find numerically that the mixed correlators in the presence of leading-order corrections (Eqs.~\ref{i2exp}-\ref{i2expmixed}) scale as
\be
D_{mn}=\langle I_mI_n\rangle-\langle I_m\rangle\langle I_n\rangle\sim t^{\mathrm{min}(l_m,l_n)},\label{newbiemix}
\ee
where $l_m$ and $l_n$ are distances of the sites $m$ and $n$ respectively from their nearest initially filled sites. The scaling for single-site correlators agrees with the above result with $m=n$:
\be
D_{nn}=\langle I^2_{n\pm l}\rangle-\langle I_{n\pm l}\rangle^2\sim t^l,\label{newbie}
\ee
with $l$ being the distance from $n$ to the nearest filled site. Therefore, even with quantum-jump corrections the picture remains similar as before: if all occupation numbers are initially nonempty, there is no transport; if some are empty and some not, then there is a superdiffusive scaling law. But quantitatively the exponents are changed significantly, from $4l$ to $l$, i.e. the transport is "less superdiffusive": for nearest-neighbor sites with $l=1$ we in fact have normal diffusion $\langle I_{n\pm 1}^2\rangle\sim t$. The result (\ref{newbie}) is illustrated in Fig.~\ref{superfluidTWA}(a-b) for two different regimes, corresponding to the superfluid and Mott regime (small vs. large $U/J$) respectively (there is no phase transition between them as the system is one-dimensional). The initial state is of the form (\ref{w0coherent}) with $\mathbf{Q}_0$ and $\mathbf{P}_0$ centered at a single site $n=3$, i.e.
\be
\mathbf{Q}_0=(0,0,q,0,0,0,0,0,0,0),~\mathbf{P}_0=(0,0,p,0,0,0,0,0,0,0),~~p^2+q^2=N.
\ee
For such boundary condition, where $M$ states ($0<M<L$) are initially filled and the remaining $L-M>0$ states are empty, the law (\ref{newbie}) is perfectly satisfied in both regimes. Also, unlike the pure TWA calculation of \cite{Markovic:2023gtx}, the presence of a boundary does not modify the basic law (\ref{newbie}).

In Fig.~\ref{superfluidTWA}(c) we plot the evolution for a broad initial distribution, again of the form (\ref{w0coherent}) but with wavepackets at every site, so the initial occupations of all sites are nonzero and of comparable magnitude (i.e., the vectors $\mathbf{P}_0$ and $\mathbf{Q}_0$ have all nonzero components). Here, as we can see, there is no transport at all.

All these results do not depend on the exact initial filling of the sites, as long as the filling is nonzero. This is demonstrated in in Fig.~\ref{restransTWA} in Appendix \ref{secappfigs}.

\begin{figure}[H]
(a)\includegraphics[width=0.3\textwidth]{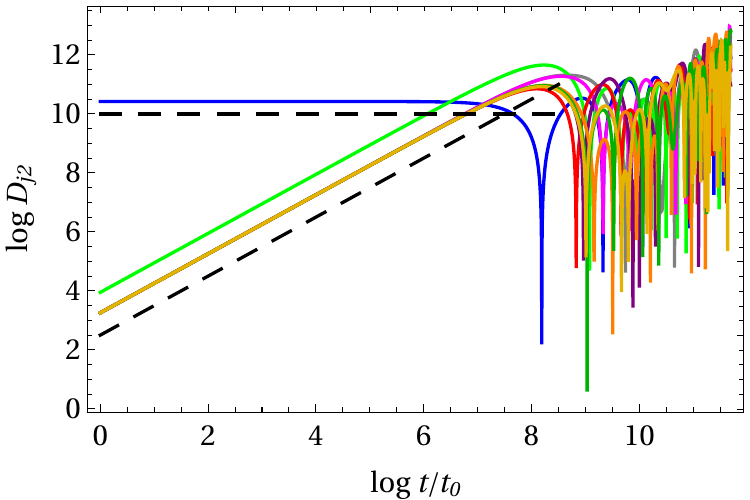}
(b)\includegraphics[width=0.3\textwidth]{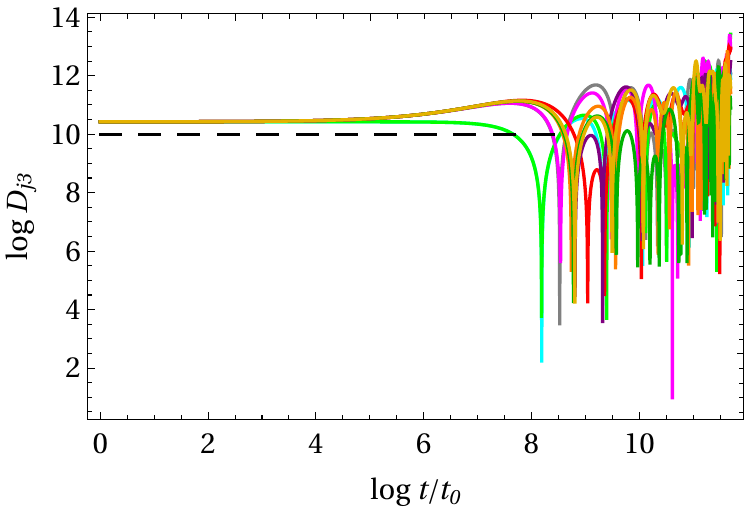}
(c)\includegraphics[width=0.3\textwidth]{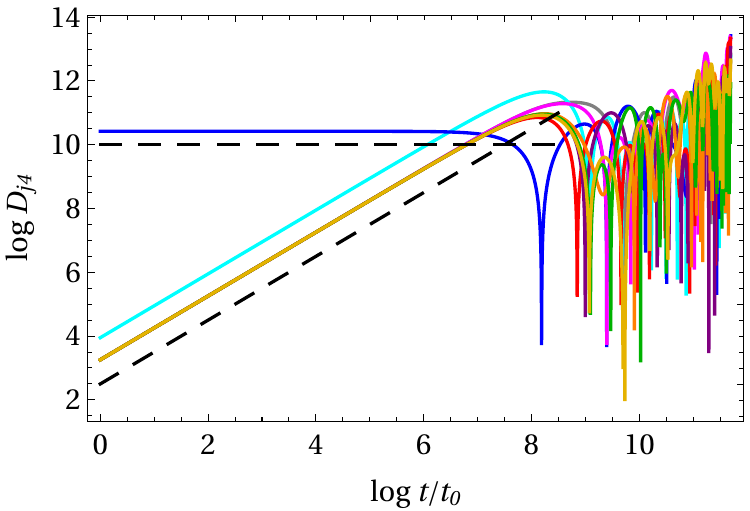}
(d)\includegraphics[width=0.3\textwidth]{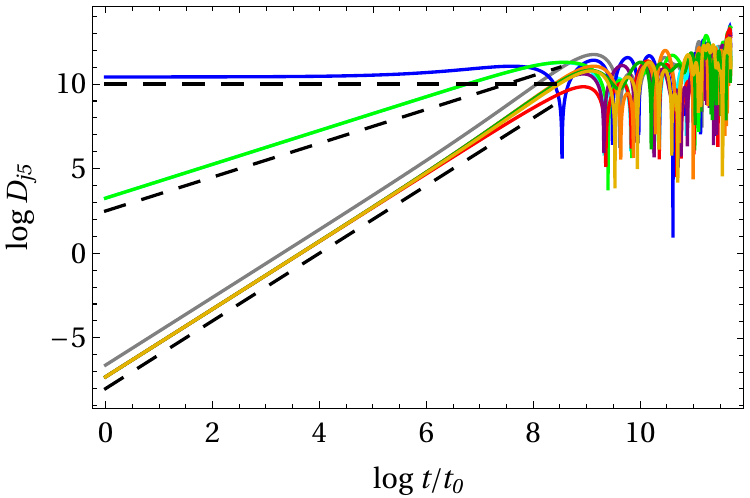}
(e)\includegraphics[width=0.3\textwidth]{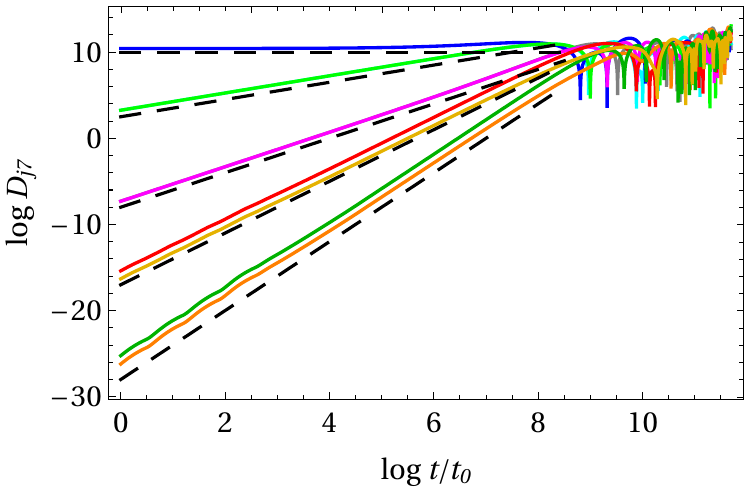}
(f)\includegraphics[width=0.3\textwidth]{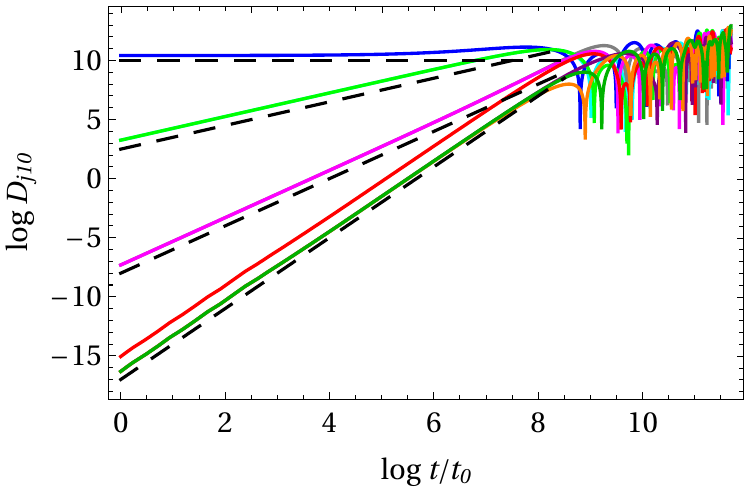}
\caption{\label{mixtransTWA1} Log-log plot of two-site correlation functions $D_{nj}\equiv\langle I_nI_j\rangle-\langle I_n\rangle\langle I_j\rangle$ with $n=2,3,4,5,7,10$ (panels a-e) and $j$ ranging over the whole chain ($1$ to $10$) in a chain of length $L=10$, with 
$U/J=0.1$, $\mu/J=0.05$. The only initially filled site is $n=3$. The color code for $j$ is the same as in the previous figures.}
\end{figure}

We also test the two-site scaling law (\ref{newbiemix}). In Fig.~\ref{mixtransTWA1} we find nearly perfect agreement with the law (\ref{newbiemix}), for two different initial configurations. For additional evidence, see also Fig.~\ref{mixtransTWA3} in Appendix \ref{secappfigs}.

Altogether, the behavior of $D_{mn}$ is robustly superdiffusive with universal integer exponents; indeed, quantum jumps make the superdiffusion even more robust than the leading-order semiclassical calculation as the exponents are now insensitive to resonances and boundary effects, and they satisfy the superdiffusion law also for $m\neq n$. Furthermore, the results in Fig.~\ref{superfluidTWA} are completely insensitive to the wavepacket widths $\sigma_P$ and $\sigma_Q$ of the initial coherent state (\ref{w0coherent}): from very broad ($\sigma_P,\sigma_Q\gg 1$) to very narrow ($\sigma_P,\sigma_Q\ll 1$) states, from isotropic ($\sigma_P=\sigma_Q$) to highly squeezed ($\sigma_P\gg\sigma_Q$ or vise versa), the log-log plots are visually indistinguishable from those given in the figures, where we have used $\sigma_P=\sigma_Q=1/\sqrt{2}$. Hence the superdiffusion does not depend on "how quantum" the initial wavepacket is, it is just important \emph{where the packet is initially centered}. The increased robustness (insensitivity to resonances and finite-size effects) comes from more restricted initial conditions as opposed to purely classical calculations (coherent states of canonical variables $P$ and $Q$ with finite width). Finally, the values of the exponents differ from the leading-order calculation: $4l$ vs. $l$ (cf. Eq.~(\ref{oldie}) vs. Eq.~(\ref{newbie})). This difference comes solely from quantum jumps as we will argue in the next subsection.

\subsection{A simple derivation of the superdiffusion law}

We are able to provide a simple analytical justification for the law (\ref{newbie}), consistent with the argument from self-similarity given in \cite{Markovic:2023gtx} but simpler as it does not require a canonical transformation to first-order locally integrable Hamiltonian. This simpler approach becomes necessary to understand how the scaling changes due to quantum jumps: it would be too complicated to rewrite Eqs.~(\ref{i2exp}-\ref{i2expmixed}) in locally integrable canonical variables \footnote{On the other hand, the simplified derivation has some disadvantages. It is less rigorous and more handwaving, and it does not capture the classically resonant case; but since the resonant case loses its special scaling in the presence of quantum corrections which are of prime importance in this work, we do not regard this as important.}.

\subsubsection{Superdiffusion from classical equations of motion}

We will first reproduce the classical (i.e., pure TWA) scaling of Eq.~(\ref{oldie}) and then use it as an input for the quantum jump corrections. This subsubsection thus repeats the conclusions from \cite{Markovic:2023gtx} in a less rigorous but more quantitative formulation (in addition to scaling exponents we will get also the solution to the equations of motion in early-time leading-order approximation).

Assume that the site $n$ is initially filled with occupation $N_0$, and that all the other sites are empty. The idea is to look for approximate solutions of the action-angle equations of motion (\ref{bhiphieoms1}-\ref{bhiphieoms2}), expanding over the small occupations of the sites with $j\neq n$. For the site $n$, since $I_n(0)=N_0$, $I_{n\pm 1}(0)=0$, at not too long times $t$ we have $\epsilon\equiv I_{n\pm 1}\ll I_n$. On the other hand, the angles live on a circle and there is no reason to suppose they take any special value at $t\to 0$; in a highly chaotic regime their motion is very fast and erratic, in a weakly chaotic regime it is closer to quasiperiodic but in any case their early values $\phi_j(t\to 0)$ are not expected to be close to 0 or show any specific scaling with time. This in turn means that the sines and cosines of their differences are also approximately constant at short times and do not scale with $t$. At the end we will justify the consistency of this assumption, but for now let us take it on faith and solve the equations.

Consider first the action equation (\ref{bhiphieoms1}) for $j=n$. At early times, we can write it as
\be
\dot{I}_n=2J\sqrt{\epsilon I_n}\left(\sin\chi_n-\sin\chi_{n+1}\right)
\ee
where $\chi_j\equiv\phi_j-\phi_{j-1}$. The formal solution is
\be
I_n(t)\approx\left(\sqrt{N_0}-2J\sqrt{\epsilon}\int_0^tdt'\left(\sin\chi_n\left(t'\right)-\sin\chi_{n+1}\left(t'\right)\right)\right)^2
\ee
Whatever the integral on the right-hand side, it multiplies $\sqrt{\epsilon}$ so at early times it is a small correction to $N_0$ of order unity. Therefore $I_n(t)\approx N_0-\ldots$ and there is no scaling behavior. For $j=n\pm 1$ we get
\be
\dot{I}_{n\pm 1}=2J\sqrt{N_0 I_{n\pm 1}}\sin\chi_{n\pm 1}+O\left(\sqrt{\epsilon}\right),
\ee
therefore up to higher-order corrections in $\sqrt{\epsilon}$ the solution is
\be
I_{n\pm 1}=J^2N_0\left(\int_0^tdt'\sin\chi_{n\pm 1}\left(t'\right)\right)^2
\ee
Now invoke our assumption that $\chi$'s behave as $t^0$ at early times. Then the integral scales as $t$ and $I_{n\pm 1}\sim t^2$. Finally, we can show by induction that for $j=n\pm l$, i.e. $l$ sites away from the initially filled site $n$, the scaling is $I_n\sim t^{2l}$. Assuming this scaling for $j=n\pm l$ as the induction hypothesis, Eq.~(\ref{bhiphieoms1}) for $j=n\pm (l+1)$ yields
\be
\dot{I}_{n\pm (l+1)}=2J\sqrt{I_{n\pm l}I_{n\pm (l+1)}}\sin\chi_{n\pm(l+1)}\sim 2J\sqrt{I_{n\pm (l+1)}}t^l\sin\chi_{n\pm(l+1)},
\ee
thus, assuming again that the sines of the angles do not scale, we get
\be
I_{n\pm (l+1)}\approx Jc_{n\pm l}\left(\int_0^tdt'\left(t'\right)^l\sin\chi_{n\pm (l+1)}\right)^2\propto\frac{Jc_{n\pm l}}{(l+1)^2}t^{2l+2},
\ee
where we have written $I_{n\pm l}=c_{n\pm l}t^{n\pm l}$. We have thus proven the expected result, $I_{n\pm (l+1)}\propto t^{2l+2}$. We could also derive the recursive relations for the coefficients $c_{n\pm l}$ but that would hardly be useful because our approximations are quite crude; we are content to derive the scaling.

The last point is to justify the assumption that the angles show no scaling behavior at small $t$. To see this, consider first the equation for $\phi_n$. The $\mu$-term and the term $UI_n\approx UN_0$ are constants at leading order and the remaining terms are linear in $t$ and thus subleading (from $I_{n\pm 1}\sim t^2$). Integrating the right-hand side, we get $\phi_n\approx\phi_n(0)-(\mu-UN_0)t+\ldots$, so for a general $\phi_n(0)$ the angle is of order unity. Now consider $\phi_{n\pm l}$ and use induction again. The $\mu$-term is a constant, the second term is of order $U\epsilon\to 0$ for early times, and the third and fourth term are proportional to $t$ and $1/t$ respectively. For $t$ small, the last term dominates. Therefore, we only keep this term and get
\be
\dot{\phi}_{n\pm l}=-\frac{J}{t}\cos(\phi_{n\pm l}-\phi_{n\pm (l-1)}).
\ee
By hypothesis, $\phi_{n\pm (l-1)}$ is roughly a constant. The above equation can then be integrated exactly to find a rational function of $t^J$, which has no simple scaling behavior. Expanding in small $t$, the leading-order solution is
\be
\phi_{n\pm l}-\phi_{n\pm (l-1)}\propto\arctan\tanh\frac{J\log t}{2},
\ee
which for $\to 0$ gives $\phi_{n\pm l}\sim \pi/2-\phi_{n\pm (l-1)}$, which in general makes the sine and cosine of the angle constant at leading order, as we expected.

\subsubsection{Superdiffusion from quantum jumps}

Now that we have reproduced $I_{n\pm l}\sim t^{2l}$ and thus $I_{n\pm l}^2\sim t^{4l}$ in the pure TWA case \footnote{Of course, since we average over $W_0$, it is in general not true that $\langle I_{n\pm l}^2\rangle=\langle I_{n\pm l}\rangle^2$. But in the simple approximation that we use the averaging does not change the naive result.}, let us evaluate the quantum-corrected Eq.~(\ref{i2expint}). The first term is just the average of the classical result over the Wigner pseudodistribution -- but in our simple approach the scaling is by construction insensitive to details of $W_0$ (which is indeed the case, according to the numerics). The second term can be estimated by noticing that, from Eqs.~(\ref{bhpq}) and the results in the previous subsubsection, $P_n\sim Q_n\sim t^l$. We then get
\bea
D_{n\pm l,n\pm l}&\approx &t^{4l}+2\sum_{l'=0}^{n-1}t^l\int_0^tdt'(t')^{l'}+2\sum_{l'=1}^{L-l'}t^l\int_0^tdt'(t')^{l'}=\nonumber\\
&=&t^{4l}-t^{L+l+2-n}\Phi(t,1,L+2-l)-t^l\left(1+t^{n+1}\left(\Phi\left(t,1,n+1\right)+2\log\left(1-t\right)\right)\right).
\eea
By $\Phi$ we denote the Lerch transcendent (generalized zeta function). At early times, the above behaves as (we ignore the coefficients in front and write just the scaling):
\be
D_{n\pm l,n\pm l}\sim t^{4l}+t^l.\label{scallawfin}
\ee
Therefore, quantum jumps spoil the exact scaling -- we have a sum of two power laws. But for small $t$ the second term will dominate, hence we get the $t^l$ scaling from the numerics.

The regime of validity of this scaling is $t<1$ in appropriate units, because we have derived it by expanding the occupation numbers for short times. The natural unit of time is $1/J$, and in our numerics we always take $J=c/t_0$. Here, $t_0$ is the computational unit of time, and $c=2\times 10^{-4}$ can be thought of as the relative (dimensionless) spacing between energy levels, or in general some microscopic scale. The simple superdiffusive scaling is thus valid for $t/t_0=J/c\times t\lesssim 5000\approx \exp(8.5)$. In our plots, the superdiffusive scaling indeed ceases between $t/t_0\approx 4000$ and $t/t_0\approx 15000$, a decent agreement. From the numerics, this crossover time seems independent of $U/J$ but it grows with the system size $L$. This size dependence is logical as it takes more time for a larger system to start thermalizing; however, we cannot reproduce this $L$-scaling with the above simple derivation, which basically assumes $t/t_0=O(1)$ and is not sensitive to the prefactor of order unity.

We can thus bring the following preliminary conclusion: \emph{Superdiffusion is an early-time phenomenon, unrelated to late-time hydrodynamic diffusion and thermalization and likely unrelated to the strength of chaos (dependent on $U/J$). It is strongly dependent on initial conditions (but it does hold for a rather broad class of initial configurations).}

\subsection{\label{secresnumdiff}Normal diffusion of number density}

We will now study the normal diffusion, typically signifying the onset of the hydrodynamical regime and thermalization. It turns out it is indeed closely related to thermalization and requires strong enough nonintegrability -- unlike superdiffusion which is a special effect determined by the specific form of the Bose-Hubbard Hamiltonian.

\begin{figure}[H]
(a)\includegraphics[width=0.27\textwidth]{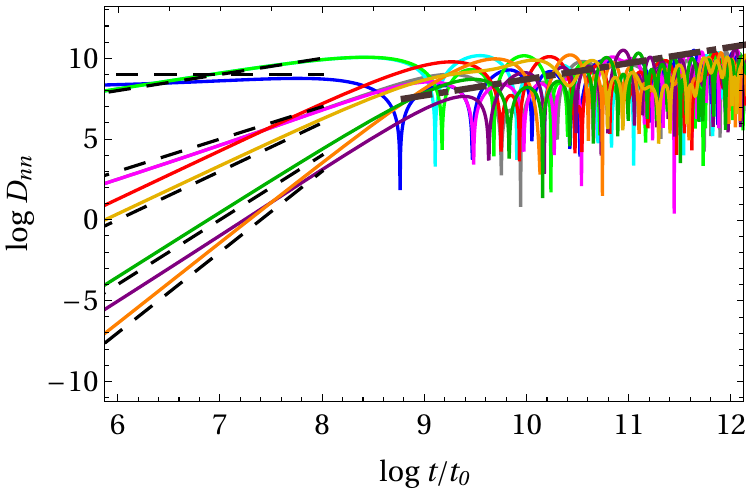}
(b)\includegraphics[width=0.27\textwidth]{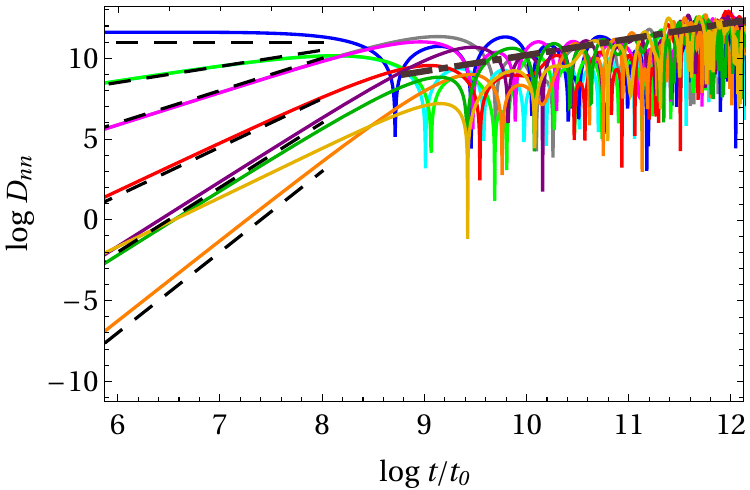}
(c)\includegraphics[width=0.27\textwidth]{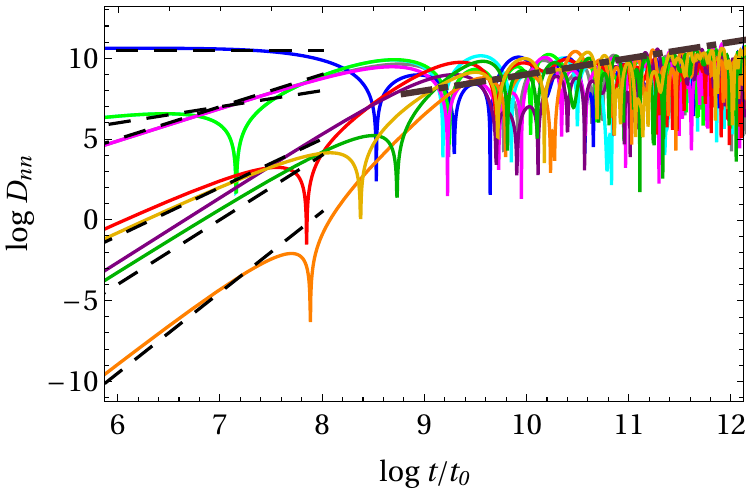}\\
(d)\includegraphics[width=0.27\textwidth]{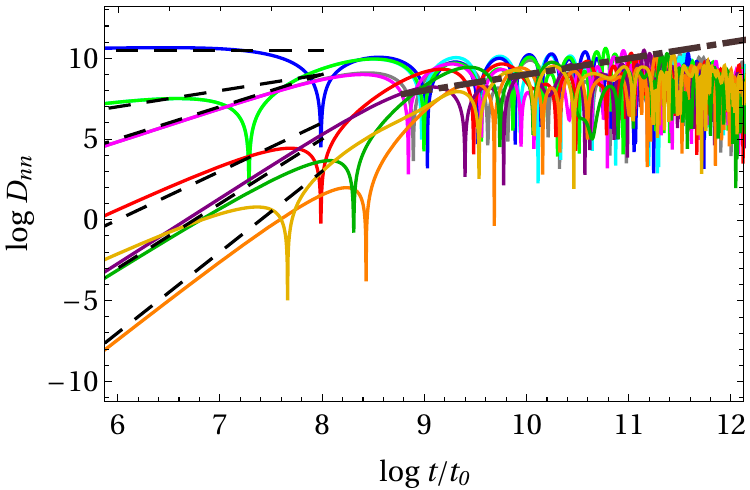}
(e)\includegraphics[width=0.27\textwidth]{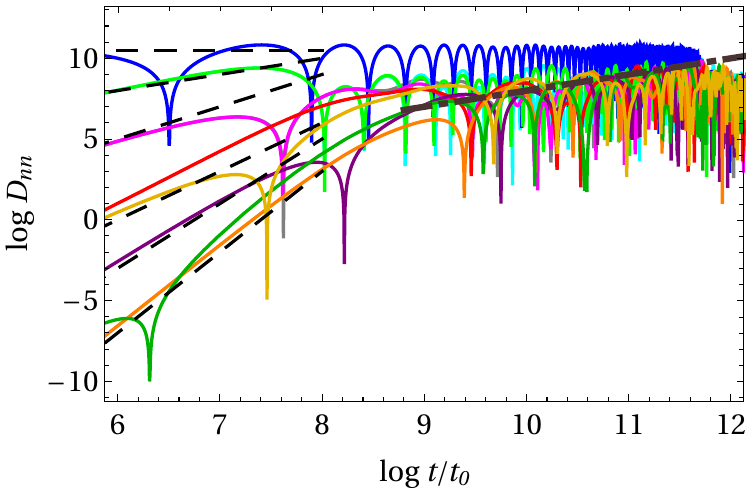}
(f)\includegraphics[width=0.37\textwidth]{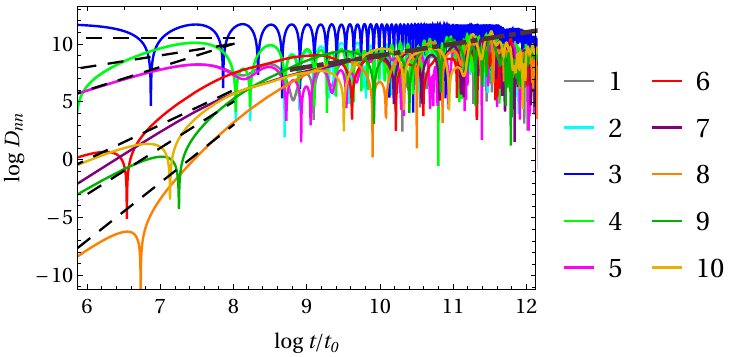}
\caption{\label{normdiff} Zoom-in on the normal diffusion regime that kicks in after long times, for $L=10$ and $U/J=0.1,0.5,1.0,2.0,4.0,10.0$ (a-f), with $n=3$ as the initially filled site. In all cases $\mu=0$. Thin black dashed lines delineate the superdiffusion laws of Eq.~(\ref{newbie}) whereas the thick gray dash-dotted line is the fit to the normal diffusion law $D_{nn}\sim t$. The trend of normal diffusion is always present but in (e) and (f) the site $n=3$ is an exception, and for larger $U/J$ values in general there is some discrepancy at long times. The color code (light to dark) is the same as in all previous figures.}
\end{figure}

Normal diffusion always sets in after superdiffusion has come to an end. In Fig.~\ref{normdiff} we zoom in onto the late, normal diffusion regime for increasing $U/J$ values. While the normal diffusion is apparently present for all six (finite) values of $U/J$, for large $U/J$ there is some deviation from the linear dispersion growth at late times, and the initially filled site ($n=3$) definitely does not diffuse for large $U/J$, i.e. deep into the localized regime. This is not so surprising but it does emphasize the very different character of the anomalous regime, which is always present. Note that the normal diffusion regime can also be seen in Fig.~\ref{superfluidTWA}(a-b)  but in that figure we did not want to draw the attention away from the anomalous regime and thus did not draw the linear fit. Finally, note that  Fig.~\ref{superfluidTWA}(c), with the near-uniform initial state, exhibits neither normal nor anomalous diffusion: if the state is already close to uniform there is no transport.

\begin{figure}[H]
(a)\includegraphics[width=0.41\textwidth]{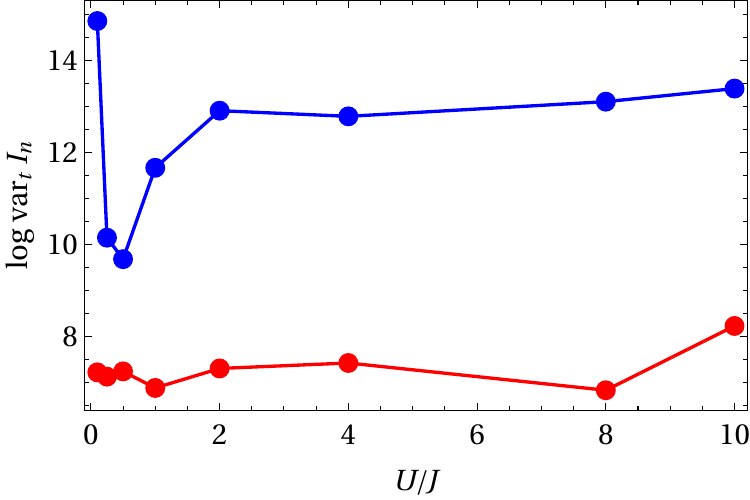}
(b)\includegraphics[width=0.52\textwidth]{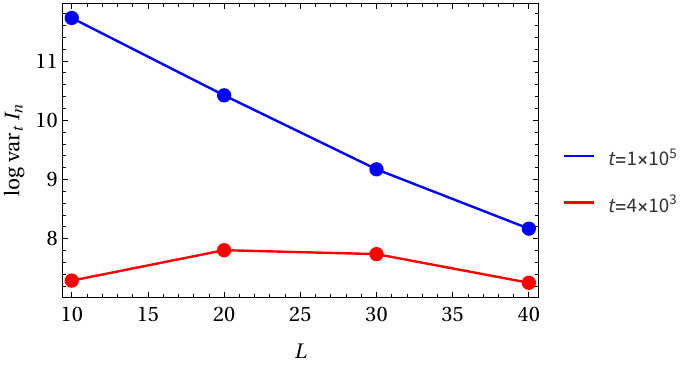}
\caption{\label{vartplot} (a) Integrated temporal fluctuation summed over the whole chain of length $L=10$ (a), as a function of $U/J$ taking values $0.1,0.2,0.5,1.0,2.0,4.0,8.0,10.0$. The blue (darker) line is for the whole integration time of the equations of motion (which encompasses also the normal diffusion regime and equilibration) while the red (brighter) line includes only the early superdiffusion. Full equilibration is only reached for intermediate $U/J$ values (the optimal value being $U/J=0.5$), while small and large $U/J$ exhibit persistent fluctuations (notice that the scale of $\mathrm{var}_t$ is logarithmic). In contrast, at early times there is no strong dependence on $U/J$. (b) Dependence of the temporal variation on the chain length, for $U/J=1$ and the initially filled site $n=3$. Longer chains have smaller variation over the whole integration time, confirming the intuition that diffusion leads to a thermal equilibrium, which most easily happens in large systems. The variation in the superdiffusion regime is nearly insensitive to system size.}
\end{figure}

The outcome of diffusion should be the equilibration of the system and a state close to ergodic. A quantitative way to characterize equilibration is proposed in \cite{Pausch:2025zgd}: it is the integrated temporal fluctuation of the number density per site. We find it most useful to consider the sum over all sites as this characterizes the equilibration of the whole chain:
\be
\mathrm{var}_t=\frac{1}{t_f}\sum_{n=1}^L\frac{\int dt'\left(D_{nn}\left(t'\right)-\frac{N}{L}\right)^2}{D_{nn}(t_f)}.\label{vart}
\ee
In Fig.~\ref{vartplot}(a) we study the dependence of the fluctuations on $U/J$ at short (superdiffusion) and long (normal diffusion) timescales. We find that the mixing from superdiffusion is roughly the same at all $U/J$ whereas normal diffusion is by a few orders of magnitude more effective at intermediate $U/J$ values. This suggests that equilibrium is only reached for sufficiently nonintegrable dynamics -- we know that the Bose-Hubbard Hamiltonian is integrable for $U/J=0$ and $U/J=\infty$, thus $U/J\sim 1$ should be the most chaotic regime. Similar conclusions can be reached qualitatively simply by plotting the evolution of occupation numbers as a function of time; this is shown in Appendix \ref{secappfigs}.

In order to understand that the superdiffusive regime is a nonthermal regime, it is helpful to look at the normal diffusion regime for chains of different length. In Fig.~\ref{vartplot}(b) we study the temporal variation $\mathrm{var}_t$ from Eq.~(\ref{vart}) for chains of length $L=10,20,30,40$ with the same initial configuration and parameters. The variation over long times drops almost exponentially with the systems size \footnote{This is somewhat unexpected, one would rather expect a power law. But in order to be sure one would also need to study more than four $L$ values.}, whereas the variation over the superdiffusive timescale is insensitive to $L$ -- thus the superdiffusive regime has nothing to do with thermalization or equilibration, it works the same way in a system of any size.

\begin{figure}[H]
(a)\includegraphics[width=0.45\textwidth]{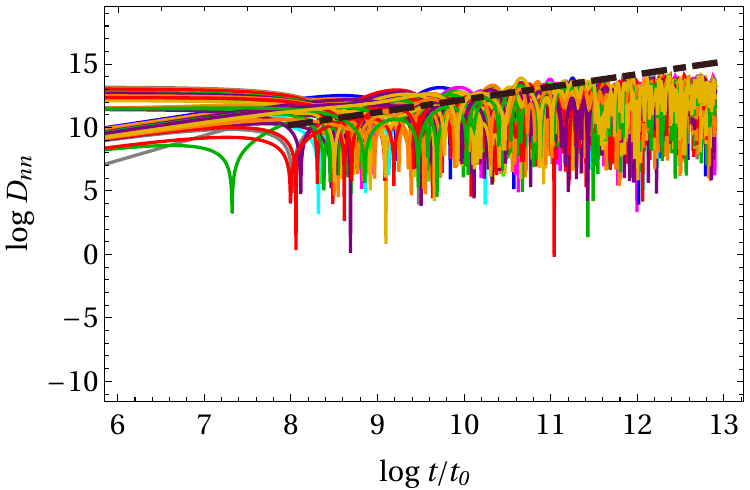}
(b)\includegraphics[width=0.45\textwidth]{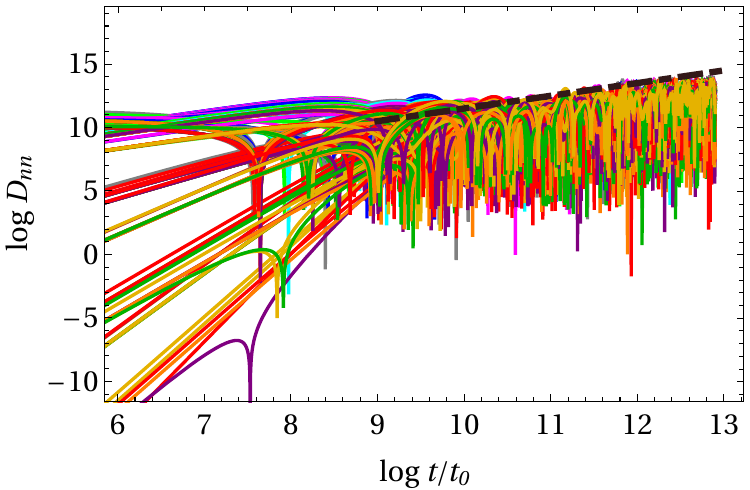}\\
(c)\includegraphics[width=0.45\textwidth]{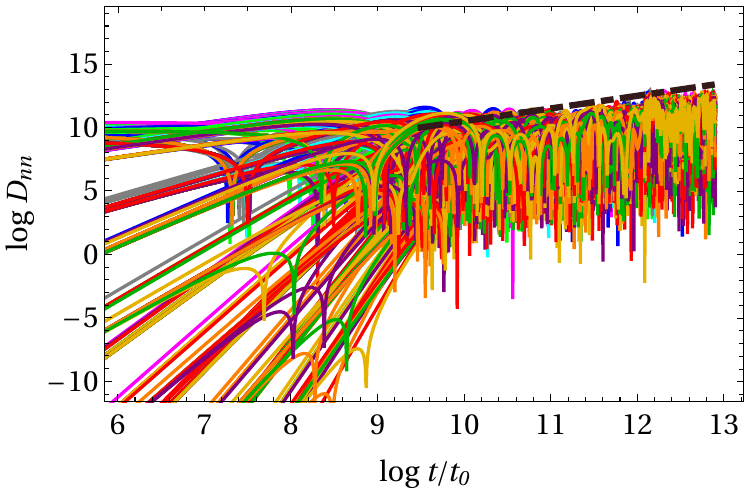}
(d)\includegraphics[width=0.45\textwidth]{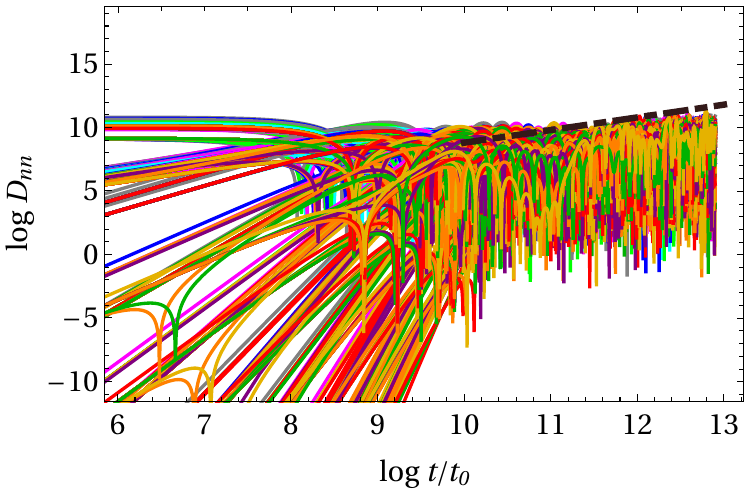}
\caption{\label{normdifftime} Zoom-in on the normal diffusion regime for $L=10,20,30,40$ (a-d) and $U/J=1$, with $n=3$ as the initially filled site. In all cases $\mu=0$. Thick gray dash-dotted lines give the normal diffusion law $D_{nn}\sim t$. The onset of normal diffusion is roughly $\log (t/t_0)\approx 8.0,9.0,9.5,10.0$ for the four increasing lengths. We do not give the color legend as the number of sites is huge (up to $40$) and individual curves are not of much importance anyway as they all collapse to the single linear law.}
\end{figure}

Finally, the time it takes for the system to start thermalizing grows with the size of the system, again in accordance with the intuition that a large system requires more time to homogenize. This is seen from Fig.~\ref{normdifftime}. It is hard to pinpoint the onset of normal diffusion as it is not a sharp transition. We may roughly say that for $L=10,20,30,40$ the diffusive regime starts at $\log t/t_0\approx 8.0,9.3,10.0,10.5$ -- later and later. It is also clear that for $L$ large saturation is reached, i.e. for an almost infinite system the exact number of particles ceases to matter.

The bottom line of this subsection is that, unlike superdiffusion, normal diffusion is sensitive to dynamics and nonintegrability, and also to the localization at high Coulomb repulsion. It takes very long time to kick in, the larger the system the longer the time. Superdiffusion thus does not by itself lead to thermalization. The one thing that superdiffusion and normal diffusion have in common is that neither will happen if the chain is already more or less homogeneous from the beginning. To understand this, we will study the dependence of transport on the choice of the observable -- so far we have looked solely at the actions/occupation numbers.

\section{\label{secother}Transport of other observables}

The obvious generalization of the results obtained so far is to look at different operators. It is clear that any local function $f(P,Q)$ of $P$ and $Q$ variables will show some variant of the general superdiffusion laws (\ref{newbiemix}) or (\ref{newbie}), with a different power determined by the powers of $P$ and $Q$ in $f$ (or its expansion). Most natural observables like local energy will be of this kind. For example, the local energy per site
\be
\label{localen}h_j=-J\left(P_jP_{j+1}+Q_jQ_{j+1}\right)-\frac{U}{8}\left(P_j^2+Q_j^2\right)^2-\frac{\mu}{2}\left(P_j^2+Q_j^2\right)
\ee
behaves the same way as the number density, as it includes the same term $\sim (P^2+Q^2)^2$. This is also easily checked numerically. Since the energy flow is quite important and is often used as a benchmark of anomalous diffusion \cite{Ljubotina:2023}, it implies that the superdiffusive regime that we find must be seen in such benchmarks. This is useful but does not introduce new physics: the mechanism must be the same.

\begin{figure}[H]
(a)\includegraphics[width=0.27\textwidth]{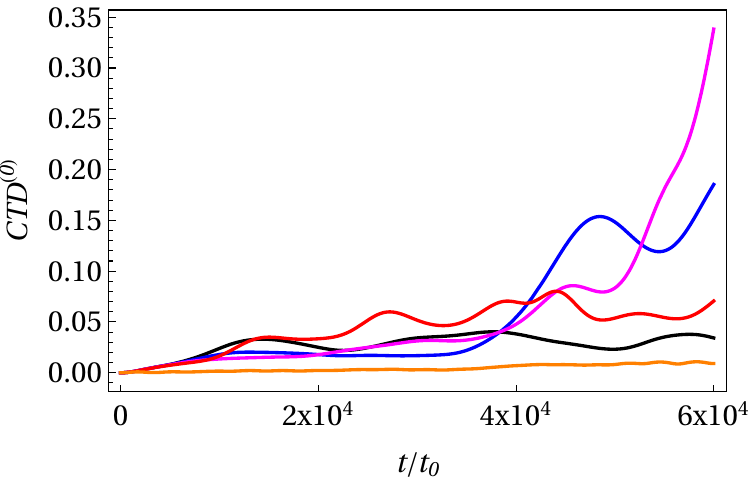}
(b)\includegraphics[width=0.27\textwidth]{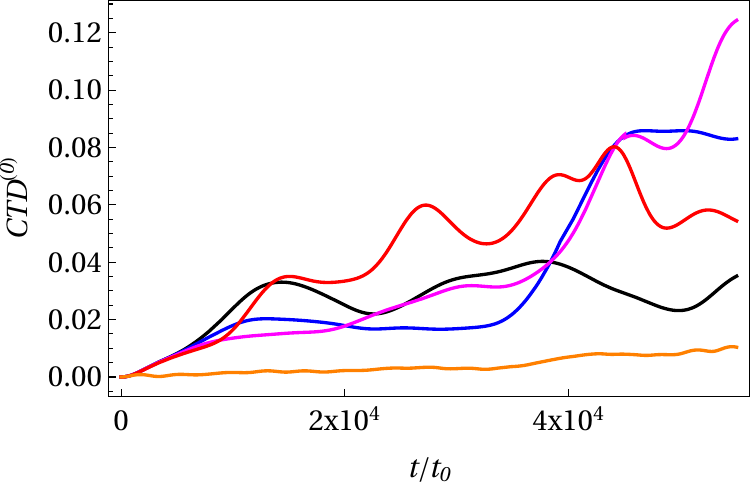}
(c)\includegraphics[width=0.38\textwidth]{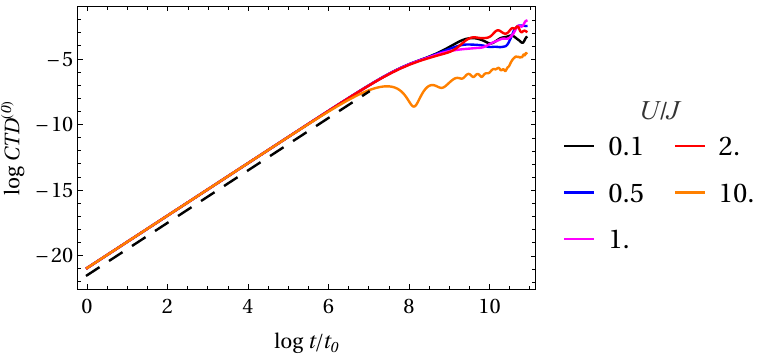}
\caption{\label{ctd1} Correlation transport distance for $U/J=0.1,1,2,5,10$ (dark to light lines), with the site $n=3$ initially filled, in the leading (TWA) approximation (a) and with first-order corrections to TWA (b). In (c) we zoom in at early times, observing a universal quadratic early growth (until roughly $t/t_0\approx\exp(8)\approx 3000$) for all $U/J$ values.}
\end{figure}

In order to see different behavior, one must look at \emph{strongly nonlocal} operators. We have seen that the localization of the initial wavepacket on a subset of sites is crucial for transport, both anomalous and normal; therefore, quantities which depend on all sites will likely yield different behavior. A convenient quantity to characterize the (super)diffusivity over the whole chain is the correlation transport distance (CTD), defined and measured in \cite{2019Natur.573..385R}, and used in \cite{Duenas:2024xxy} precisely for the Bose-Hubbard model. It sums the two-point correlators over all pairs of points, over the whole chain:
\be
\ell(t)=\sum_{n=1}^L\sum_ll\left(\langle I_n(t)I_{n+l}(t)\rangle-\langle I_n(t)\rangle\langle I_{n+l}(t)\rangle\right)\label{ctd}
\ee
Being a sum of two-point functions over the whole chain, it is a very nonlocal observable, and yet it can be measured which is an additional advantage. In Fig.~\ref{ctd1} we study its behavior for a number of $U/J$ values, both with and without quantum jumps (as we have not studied it before). On precisely the same timescale on which superdiffusion rules, CTD shows pristine quadratic scaling:
\be
\ell(t)\sim t^2,
\ee
which in fact could be expected -- this was found in \cite{Duenas:2024xxy} with a full quantum computation (although there is a priori no guarantee that the TWA regime keeps the same scaling). But one important lesson is that this early-time quadratic regime precisely corresponds to the superdiffusive epoch -- it is the consequence of the laws (\ref{newbie}-\ref{newbiemix}). The second lesson is that, unlike in \cite{Duenas:2024xxy} where this early-time behavior is followed by a nonuniversal and in general anomalous diffusion and finally by linear growth, here in the semiclassical regime the linear growth is followed by erratic and nonuniversal behavior and finally by saturation. Also, we do not see any systematic dependence of CTD on $U/J$. Therefore, the early time/high energy regime is not very sensitive to quantum effects (which is also seen from the fact that the behavior of CTD is very similar with and without quantum jumps). One might speculate what will the evolution of $D_{mn}$ look like in an exact quantum computation, which we leave for further work.

\subsection{\label{secothertrans}Unitary transformations from superdiffusion to nondiffusion}

We will now consider a new set of variables and discuss their diffusion properties. Start by defining $a_n\equiv\sqrt{I_n}$. Then the equations of motion for the actions (Eq.~(\ref{bhiphieoms1}) become linear in $\Vec{a}$ and thus can be written in matrix form:
\be
\imath\frac{d\Vec{a}}{dt}=\hat{A}\Vec{a}\label{EOM0}
\ee
with
\be
\label{amat}\hat{A}=-\imath J\begin{pmatrix}
0 & \sin\phi_{1,2} &0 &\ldots& \ldots& \ldots \\
 \sin\phi_{2,1} & 0 &  \sin\phi_{2,3} & 0& \ldots&\ldots \\
 0 &  \sin\phi_ {3,2} & 0 &  \sin\phi_{3,4} & 0 &\ldots \\
 \dots&\ldots&\ldots&\ldots&\ldots&\ldots\\
\ldots & \ldots & \ldots & \sin\phi_{L-1,L-2} &0 &\sin\phi_{L-1,L}  \\
\ldots&\ldots &\ldots &\ldots &  \sin\phi_{L,L-1} & 0
\end{pmatrix},
\ee
where $\phi_{i,j}=\phi_i-\phi_j$. The matrix $\hat{A}$ is Hermitian and one can diagonalize it via a unitary transformation: $\hat{A}=\hat{T}^\dagger\hat{D}\hat{T}$. In the new basis, Eq.~(\ref{EOM0}) becomes
\begin{equation}
\imath\frac{d\Vec{a}}{dt}=\hat{T}^\dagger\hat{D}\hat{T}\Vec{a}. \label{EOM1}
\end{equation}
The matrices $\hat{A}$, $\hat{T}$, $\hat{T}^\dagger$ and $\hat{D}$ are all time-dependent, since their elements are time-dependent functions. In the new basis $\Vec{\alpha}\equiv\hat{T}\Vec{a}$, Eq.~(\ref{EOM1}) is transformed as:
\begin{equation}
    \frac{d\Vec{\alpha}}{dt}=-\imath\hat{D}\Vec{\alpha}+\frac{d\hat{T}}{dt}\Vec{a}.\label{alphaEOM1}
\end{equation}
Since the matrix $\hat{A}$ is a function of angles only, the elements of the matrix $\hat{T}$ are likewise functions of angles only, which suggests that their dynamics is "fast" (compared to the dynamics of actions) because their derivatives can take large values. On the other hand, the vector $\Vec{\alpha}$ dynamically behaves as a vector of action variables which evolve slowly, and the matrix $\hat{D}$ is a function of sines of angles, without derivatives. In conclusion, the second term can be treated as a fast and thus effectively stochastic contribution, i.e. as a noise term. If we indeed treat the $\frac{d\hat{T}}{dt}\Vec{a}$ term in Eq.~(\ref{alphaEOM1}) as an uncorrelated white-in-time noise term, we can solve the equations in the Ito formalism, finding:
\be
\label{alphasol}\alpha_n(t)=\alpha_{n}(0)e^{-\imath\int_0^t\Delta_n dt}+e^{-\imath\int_0^t\Delta_n dt}\int_0^t\eta_n(t')e^{\imath\int_0^{t'}\Delta_n dt^{''}}dt',
\ee
where $\Delta_n$ are the eigenvalues of $\hat{A}$ and the noise term $\eta_n(t)$ models the behavior of the fast term in Eq.~(\ref{alphaEOM1}): $\frac{dT_{nm}}{dt}a_m\mapsto\eta_n(t)$.

We can also assign a physical meaning to the $\alpha$ variables. Let us calculate their total Euclidean norm:
\begin{equation}
    \sum_{n=1}^L|\alpha_n|^2=\sum_{n=1}^L \alpha_n\alpha^*_n=\sum_{n,m,k=1}^LT_{nm}a_mT^*_{nk}a_k=\sum_{m,k=1}^L\delta_{mk}a_ma_k=\sum_{k=1}^La_k^2=1.
\end{equation}
From the last equality, the $\alpha$-variables satisfy the number conservation constraint of our system. This comes as no surprise as they are obtained by a rotation in the complex "$a$-space" starting from the original actions. Since the equations of motion for $\alpha$'s are diagonal in $\alpha$'s, they acquire the meaning of "heavy" degrees of freedom which only interact through the "gas" of light and fast angle excitations.

\begin{figure}[H] 
\centering
\includegraphics[width=.9\linewidth]{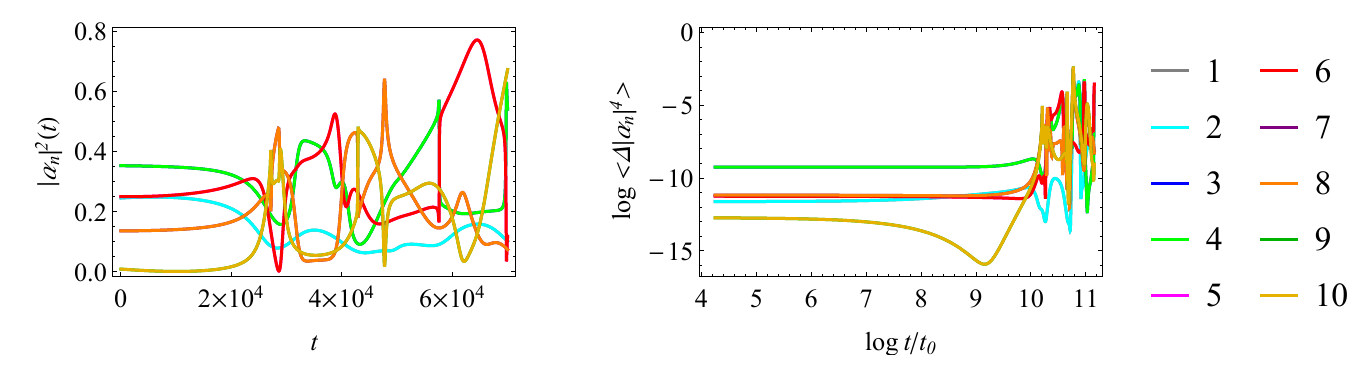}
\caption{Time evolution of $|\alpha_n|^2$ in time (left) and the log-log plot of their second moment (right) with $U/J=0.125$ and $L=10$ (superfluid regime). In the superfluid regime the new variables retain similar fast dynamics as the original actions $I_n$, however the superdiffusive regime is now absent and the late-time diffusion is replaced by erratic behavior still without a clear trend. Very early times ($t/t_0<\exp(4)$) are left out from the left figure for practical reasons only (for these early times there is no transport either). The color code (lighter and darker shades) denotes the site number.}
\label{superfluidalpha}
\end{figure}

\begin{figure}[H] 
\centering
\includegraphics[width=.9\linewidth]{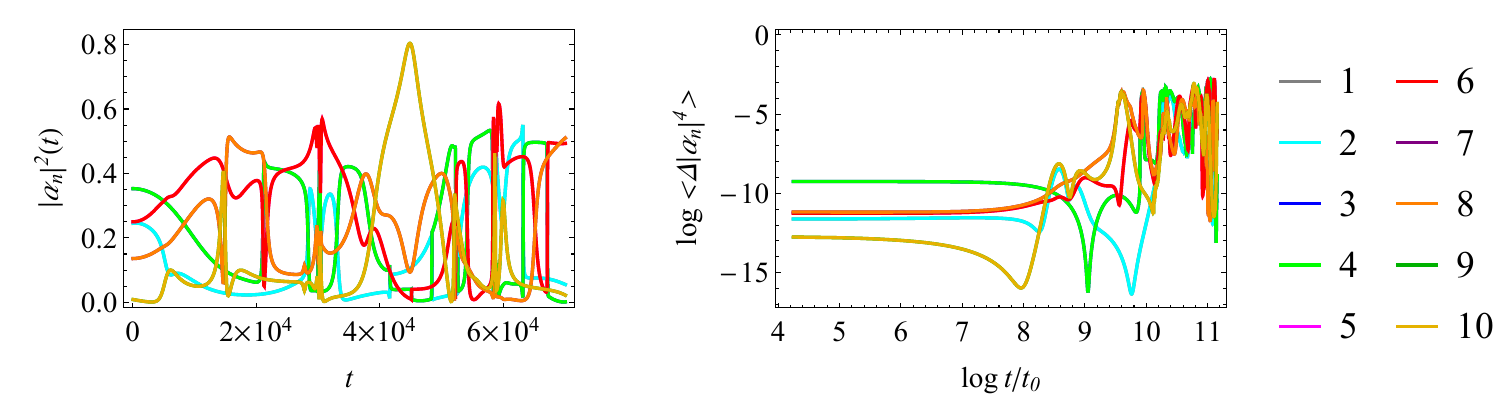}
\caption{Time evolution of $|\alpha|^2$ in time (left) and the log-log plot of their second moment (right) with $U/J=10$ and $L=10$ (Mott regime). In the Mott regime the new variables behave erratically just like in the superfluid regime; they are not nearly static like the occupation numbers $I_n$. The transport remains identical, i.e. non-existent in both regimes. The color code (lighter and darker shades) denotes the site number.}
\label{mottalpha}
\end{figure}

Now we move to our main motivation for introducing the new variables. We will discuss the dispersion of the squared module of the $\alpha$-variables:
\begin{equation}
\sigma^2(|\alpha_n|^2)\equiv\langle |\alpha_n|^4(t)\rangle-\langle |\alpha_n|^2(t)\rangle^2.
\end{equation}
As we have previously shown, these quantities satisfy the particle number conservation and thus can be interpreted as number density of some effective excitations (which are highly nonlocal functions of the physical number variables $I_n$). Analytic diagonalization of the matrix $\hat{A}$ is impractical but we can easily do it numerically: we solve the Hamiltonian equations, diagonalizing the matrix $\hat{A}$ at each time step, for the whole integration time. The results are given in Figs.~\ref{superfluidalpha} and \ref{mottalpha}.

We observe that in these variables neither the early superdiffusion nor the late diffusion survives. This is not surprising given the presence of oscillating terms $\exp(\imath\Delta_n)$ in the solution (\ref{alphasol}) -- the $\alpha$ density oscillates forever, without ever equilibrating. One more thing of interest is the degeneracy of solutions in Figs.~\ref{superfluidalpha}-\ref{mottalpha}: we only see $L/2$ different solutions for the $L$ sites. This is because the sites whose eigenvalues are equal in absolute value have the same dynamics, and the eigenvalues of the matrix $\hat{A}$ come in pairs $\pm\Delta_n$, thus indeed we have at most $L/2$ different evolutions. In the future we plan to explore further consequences of this hidden symmetry.

We can also understand analytically why is there no transport for the $\alpha$ variables. Let us take a look at the derivative of $|\alpha_n|^2$ with respect to time:
\begin{equation}
    \frac{d|\alpha_n|^2}{dt}=\alpha_n\frac{d\alpha_n^*}{dt}+\alpha_n^*\frac{d\alpha_n}{dt}=\alpha_n(\imath\Delta_n\alpha_n^*)+\alpha_n\frac{dT_{nm}^*}{dt}a_m+\mathrm{c.c.}=\alpha_n\frac{dT_{nm}^*}{dt}a_m+\mathrm{c.c.}=\sum_{m,k} a_ma_k\frac{d}{dt}\left( T_{mn}^\dagger T_{nk}\right).\end{equation}
The sum over $m$ and $k$ destroys any simple scaling, thus there will be no scaling behavior, either normal or anomalous: in the $\alpha$-picture we do not see any diffusion. This is a very illustrative result: only via a unitary transformation of variables we were able to kill the discrete and robust superdiffusion exponents (and indeed to kill all transport completely). This happens even though the $\alpha$-variables have the same interpretation of numbers/fillings (in a formal sense, for some effective excitations).

We have already seen that sufficiently localized initial conditions are necessary for transport, and the same holds for the observables: only a certain class of operators can exhibit transport (operators which are "simple" functions of the actions). A generic operator will behave differently. These two factors (the "geometry" of the initial state and of the observable) are in fact related: a complicated function of the occupation numbers that depends on many (mutually distant) sites does not see the locality of the initial state that we found to be essential to the existence of anomalous exponents, precisely because of its complicated dependence on many sites. The distinction between the superdiffusive and anomalous regime is irrelevant here.

\section{\label{secchaos}Transport versus statistics and dynamics of the chain}

\subsection{Mixing entropies}

It remains to explore how the transport properties we have found react to the statistics and dynamics of the system -- we have found evidence that any transport requires sufficiently localized initial state and observable and hints that normal diffusion requires strong chaos but we have not made this quantitative yet. Let us now explore both theses -- dependence on the initial distribution and dependence on chaos -- in a more quantitative way.

\begin{figure}[H]
\centering
(a)\includegraphics[width=0.45\textwidth]{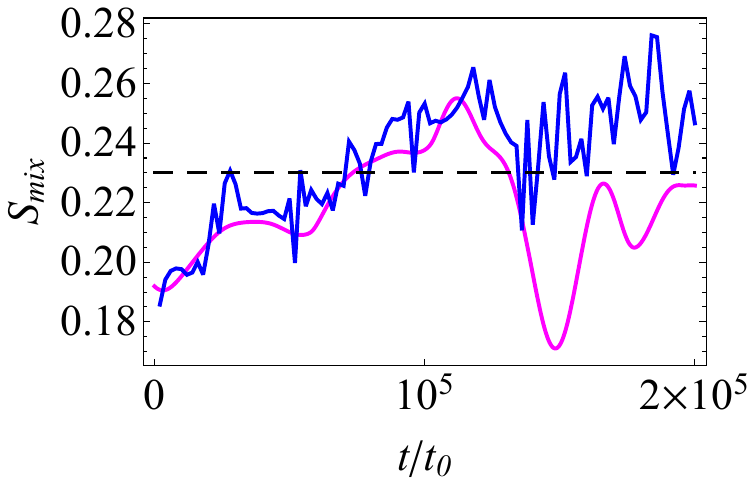}
(b)\includegraphics[width=0.45\textwidth]{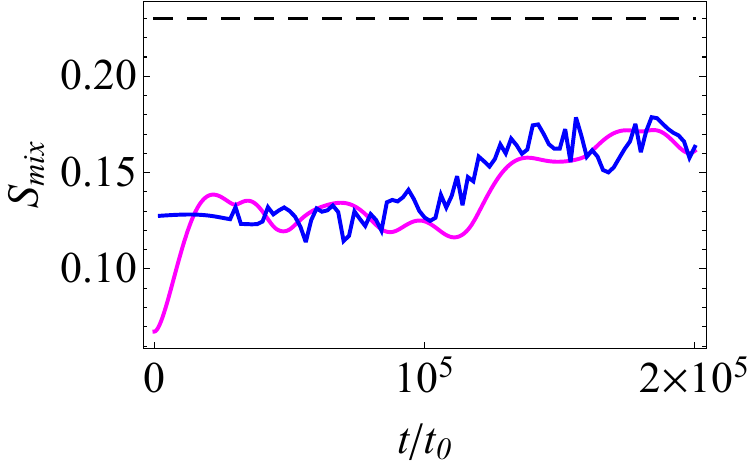}
\caption{Mixing entropies for the chain of length $L=10$ in the superfluid regime ($U/J=0.125$) with $\mu/J=0.10$, with random initial conditions (left) and for initially filled sites $I_3(t=0)=I_5(t=0)=1/2$ (right). The blue line (darker) represents mixing entropy for the $\alpha$ variables and the magenta line (lighter) represents the mixing entropy for the actions. The black dashed line indicates the theoretical maximum for the mixing entropy from Eq.~(\ref{mixentmax}). In (a), the near-uniform initial distribution makes $S_\mathrm{mix}$ nearly maximal from the beginning (see the scale on the $y$-axis) whereas for an initially localized distribution (b) the growth starts far away from thew maximum. In addition, the action variables (magenta) in (b) undergo a period of fast mixing at early times, which precisely accounts for the epoch of superdiffusive transport. The fact that computed mixing entropies can oscillate above $S_{\mathrm{mix}}^{\mathrm{max}}$ is due to numerical fluctuations, i.e. the roundoff error in the number conservation constraint.}
\label{superfluidentropy}
\end{figure}

To that end we employ the mixing entropy, a well-known concept in chemical physics, used in \cite{Richaud_2018} to study one-dimensional cold-atom systems:
\begin{equation}
S^I_{\mathrm{mix}}=-\frac{1}{L}\sum_{n=1}^L\frac{I_n}{I_1+I_2+\cdots+I_L}\log\left(\frac{I_n}{I_1+I_2+\cdots+I_L} \right)=-\frac{1}{L}\sum_nI_n\log I_n,\label{mixent}
\end{equation}
where the denominator is replaced by unity in the second equality since we work with the constraint $\sum_{n=1}^LI_n=1$. The interpretation is clear -- a large mixing entropy implies that most or all actions are substantially nonzero and the dynamics mixes all of them. It can be shown that this quantity indeed satisfies the necessary axiomatic properties for an entropy. We can also define the mixing for the $\alpha$ variables as follows:
\begin{equation}
S^{\alpha}_{\mathrm{mix}}=-\frac{1}{L}\sum_{n=1}^L|\alpha_n|^2\log|\alpha_n|^2,
\end{equation}
as they also satisfy the identical constraint as the actions. In this specific case this entropy is nothing but the Shannon entropy for our system as the action $I_n(t)$ is the percentage of bosons at the site $n$ at a particular moment of time $t$. The mixing entropy has the upper bound which is attained for a uniform distribution, i.e. equal occupation numbers for all sites:
\begin{equation}
S_{\mathrm{mix}}^{\mathrm{max}}=\frac{\log L}{L}.\label{mixentmax}
\end{equation}
The mixing entropy shows the dependence of transport on initial conditions in a very transparent way. From Fig.~\ref{superfluidentropy}, showing the time evolution of $S_{\mathrm{mix}}$ for small and large $U/J$, for the occupation numbers $I_n$ (magenta) and for $\vert\alpha_n\vert^2$ (blue), with uniform initial conditions in the left panels and for two initially occupied sites in the right panels, we draw two main conclusions:
\begin{enumerate}
\item Superdiffusion corresponds to the regime of fast and monotonous growth of the mixing entropy. At about the same time when superdiffision stops, this growth also stops. For the uniform initial state, and for the $\alpha$ variables which are nearly always more or less uniformaly distributed, this growth is never present.
\item Uniformly distributed occupation yields entropy close to the maximum of Eq.~(\ref{mixentmax}); this maximum is quickly reached and the rest is just oscillatory behavior. On the other hand, initially localized states start far away from the saturation value and exhibit a period of sharp $S_\mathrm{mix}$ growth.
\item None of the above phenomena significantly depends on $U/J$.
\end{enumerate}
Fig.~\ref{mottentropy} in Appendix \ref{secappfigs} also shows that none of the above depends significantly on $U/J$. All of this is exactly reflected in the criteria for the existence of the superdiffusive regime (the same early time epoch, necessity of localized initial states and localized operators, independence on $U/J$); same criteria apply for the normal diffusion regime and thermalization, \emph{except} the third thesis -- as we have seen, the normal regime is suppressed when $U/J$ is too small or too large to allow strong chaos. 

In summary, the mixing entropy serves as a perfect benchmark of anomalous diffusion. It is a bit surprising that it does not detect thermalization and equilibration due to normal diffusion: the latter starts when $S_\mathrm{mix}$ almost stops growing. We will discuss this in more detail in the last section.

\subsection{Chaos and spectrum}

Now we wish to understand to what extent transport relates to chaos: so far we have the hints that anomalous diffusion is completely insensitive to it whereas the normal diffusion does become more pronounced and kicks in more quickly when nonintegrability is high, and the homogenization of the system definitely strongly depends on the nonintegrability parameter $U/J$. 

In our previous work \cite{Markovic:2023gtx} we have found that classical Lyapunov exponents are the largest for intermediate $U/J$ and for large $\mu/J$ \footnote{The former is logical as both extreme limits ($U/J=0$ and $U/J\to\infty$) are integrable, and the latter is explained by the fact that for growing $\mu$ the system is closer and closer to degenerate (second derivatives of $H$ with respect to actions are by a factor $U/\mu$ smaller than the first derivatives for $\mu/U$ large).}, and for initially full sites, but the dependence on all three is weak. This is unlike quantum chaos, which is found in \cite{Pausch:2025zgd,PauschOptimalRoute,Kolovsky:2016} to depend strongly on $U/J$; the reason is likely that quantum chaos is nearly always weaker than classical.

\begin{figure}[h]
\includegraphics[width=0.5\textwidth]{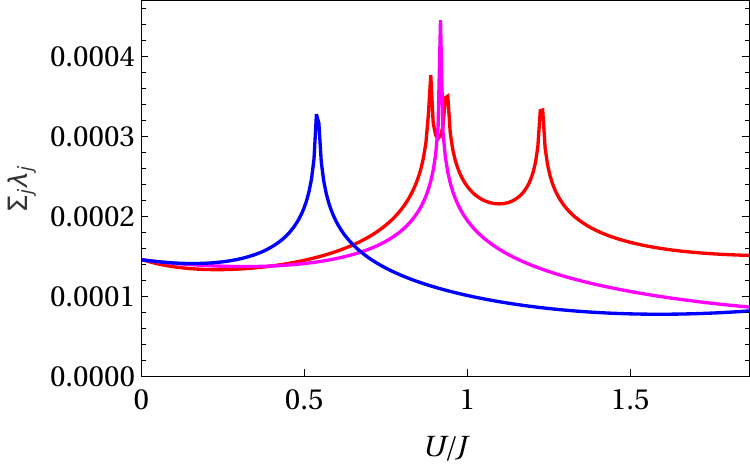}
\caption{\label{ftle} The sum of positive FTLE in a chain of length $L$ and chemical potential $\mu=0$ for the integration time $t/t_0=20000$ as a function of $U/J$, for three different initial conditions: $n=3$ initially filled (blue), $n=3,6,9$ initially filled (magenta) and homogeneous chain (all sites with equal filling, red). FTLE clearly peak around some $U/J$ of order unity, when the competition of hopping and Coulomb blockade is strong and the system is far from integrable. The reason that the exponents are nonzero even for the integrable case $U/J=0$ is the finite-time calculation. In the limit $t\to\infty$ the exponent for this point would converge to zero.} 
\end{figure}

Quantum jumps at leading order do not give a correction to Lyapunov exponents because the evolution equation for the actions (\ref{i1exp}) does not receive a contribution from the quantum jumps. Hence our earlier findings remain valid, but now we want to focus on the most chaotic regime around $U/J\approx 1$ and to show that Lyapunov exponents show a peak (although a small one) for the same value of $U/J\sim 1$ for which normal diffusion kicks in the fastest and the time variation of the occupation number is smallest at long times. To that end, we plot the sum of all positive finite-time Lyapunov exponents (FTLE) in Fig.~\ref{ftle}: they show a small but clear maximum for some intermediate value of $U/J$, for any initial condition. Fig.~\ref{ftle} is computed at $\mu=0$; at finite chemical potential the picture is similar but the peaks are less and less prominent. The reason we compute FTLE instead of true Lyapunov exponents is computational efficiency: impractically long calculation times are often needed for all the exponents to converge to their asymptotic values. The location of the peak for one initially filled site ($U/J=0.5$) precisely coincides with the minimum of the time fluctuation (Fig.~\ref{vartplot}) and the most homogeneous long-time distribution (Fig.~\ref{densityplot}) with the same initial state.

\begin{figure}[H]
(a)\includegraphics[width=0.27\textwidth]{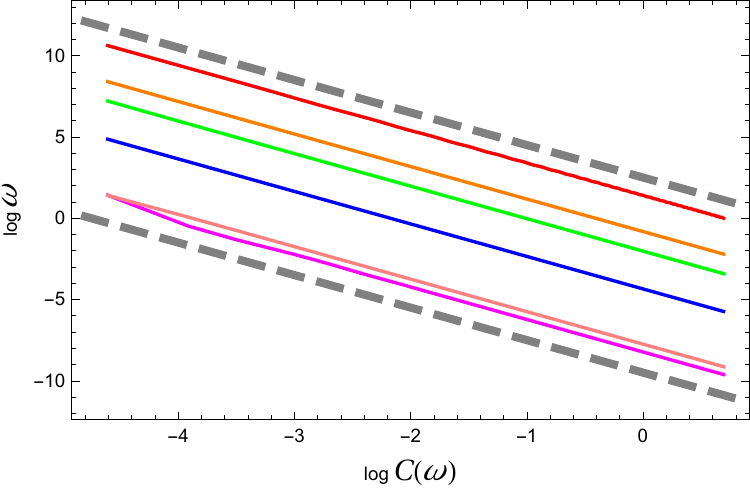}
(b)\includegraphics[width=0.27\textwidth]{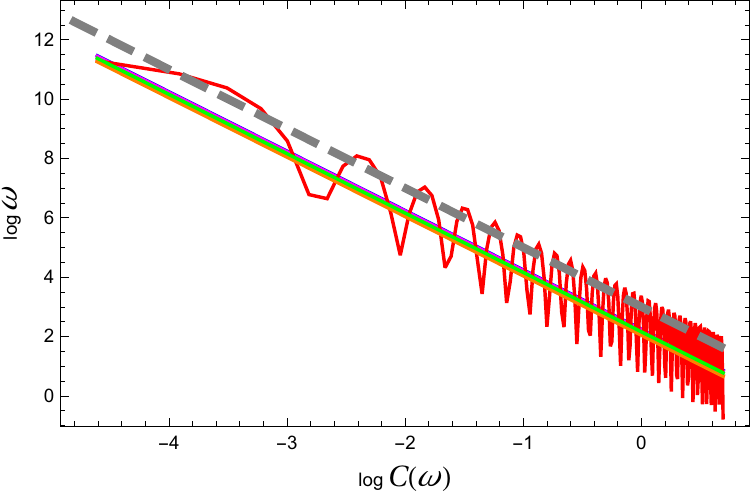}
(c)\includegraphics[width=0.40\textwidth]{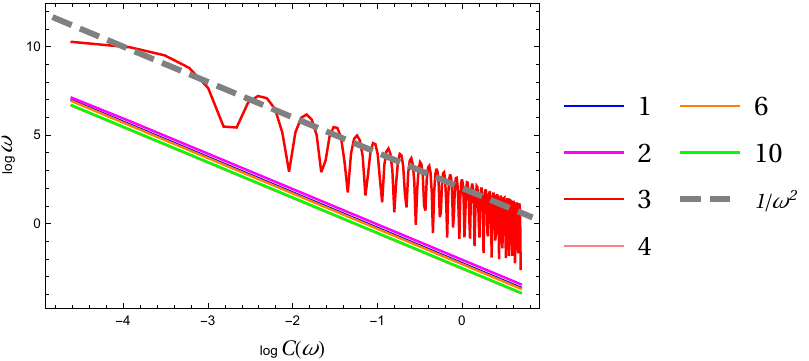}\\
(d)\includegraphics[width=0.27\textwidth]{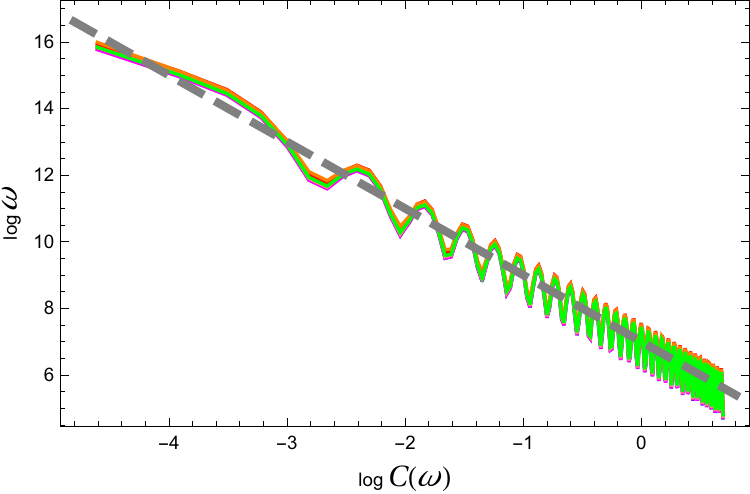}
(e)\includegraphics[width=0.27\textwidth]{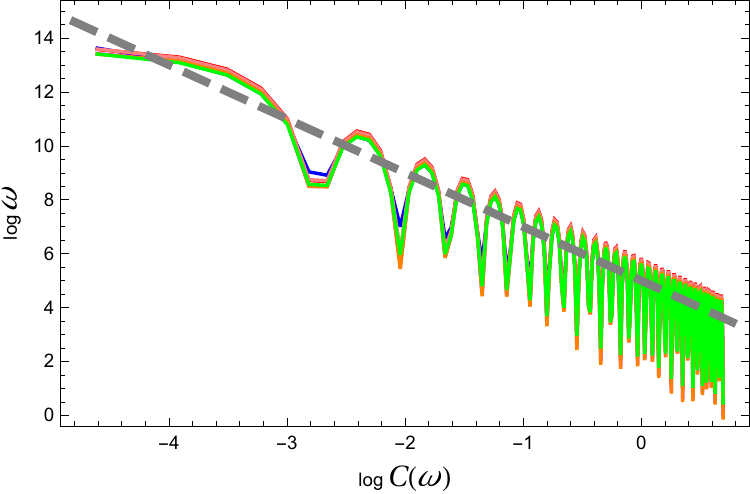}
(f)\includegraphics[width=0.40\textwidth]{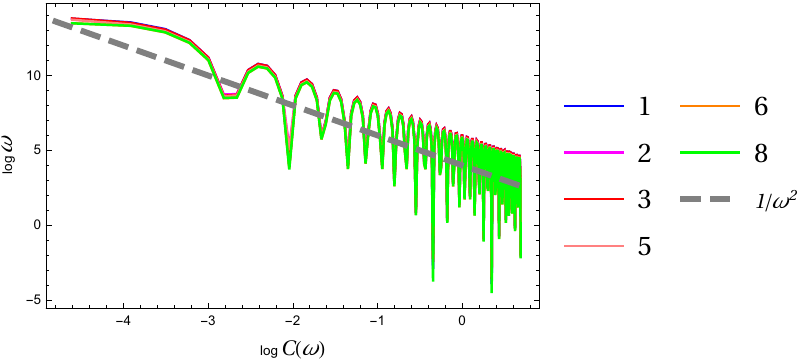}
\caption{\label{powerspec} Log-log plot of the power spectrum of the occupation number $I_n$ as a function of Fourier frequency $C(\omega)$, for $U/J$ equal to $0.1$ (a, d), $1.0$ (b,e) and $10.0$ (c, f) and for a number of different sites $n=1,2,3,6,8,9$ (dark to bright). The top row (a-c) is for initially filled sites $n=3,6,9$ and the bottom row (d-f) for a uniform initial distribution. The chain length is $L=10$. We also plot reference lines with the slope $-2$ (gray thick dashed lines), to show that the numerical data are highly consistent with the Brownian spectrum $C(\omega)\sim 1/\omega^2$, expected for developed chaos.}
\end{figure}

But the definite proof that semiclassical chaos is always strong, at most weakly dependent on $U/J$, is the power spectrum given in Fig.~\ref{powerspec}. The spectrum always shows perfect $1/\omega^2$ behavior of Brownian motion, for all three $U/J$ values and, interestingly, both for localized and uniform initial state. This is in partial conflict with the previous results on chaos -- previously we have found at least a weak peak in chaoticity for $U/J\approx 1$, but the spectrum is completely insensitive to $U/J$; the spectrum is more complicated for initially filled sites but the scaling is unchanged. What is most surprising is that the Brownian spectrum is usually a sufficient condition for normal diffusion, described by the Fokker-Planck equation. The only resolution we can think of is that normal diffusion will always be present, but in cases such as uniform initial state or a nonlocal observable the diffusion coefficient will be extremely small. Anomalous diffusion is then clearly not due to strong correlations (which would violate the $1/\omega^2$ law) or integrability but a finite-time effect, seen only for very large $\omega$ in the spectrum.

Finally, one might wonder why we do not compute OTOC and the spectral form factor, the foremost indicators of chaos for a quantum system. The reason is that in the TWA approach that we take the spectral form factor only has the dip regime instead of the dip-ramp-plateau structure which encapsulates full information on quantum chaos (see Appendix \ref{secappsff}), while OTOC is trivial (equal to TOC up to a sign) in our case, as we show in detail in the Appendix \ref{secappotoc}.

\section{\label{secconc}Discussion and conclusions}

Our sharpest finding is the existence of superdiffusion with universal integer exponents, determined solely by the boundary conditions. The exponents are sensitive to quantum jumps -- they reduce from $4l$ to $l$ (where $l$ is the distance from the nearest initially occupied site). Superdiffusion is the consequence of a scaling symmetry of the Bose-Hubbard Hamiltonian expressed in action-angle variables, in particular of the fact that filling an empty site is a nonperturbative process (i.e. the density current is formally infinite at $t=0$). This nonperturbativeness is also the reason of the discontinuous change of the superdiffusion exponents due to quantum jumps.

The superdiffusive regime is completely insensitive to chaos and thermalization. While we now fully understand where it comes from, it is unclear if it has deeper implications. It would certainly be interesting to check if it can be reproduced in a full quantum calculation (even if with further changes to the scaling exponents), which we plan to do in further work. Another interesting implication is that precisely this early-time regime is most approachable in experiments: it should thus be possible to confirm it experimentally.

On the other hand, the homogeneity of the final state, i.e. the ergodicity of late-time dynamics and the suppression of recurrences and oscillations, is strongly correlated with the chaos strength. This agrees with the usual expectations that diffusion and thermalization are a consequence of chaos. The existence of normal diffusion as such is much less dependent on chaos -- this is somewhat odd. One would expect that normal diffusion necessarily ends with a completely mixed state, while for weaker chaos ($U/J\ll 1$ or $U/J\gg 1$) we sometimes find pristine late-time normal diffusion but with significant spatial and temporal fluctuations of the late-time state. Another thing to understand better is the fact that the inclusion of quantum jumps makes normal diffusion more prominent: the naive expectation would be that the hydrodynamic regime is always very close to classical.

Finally, chaos is maximal when $U/J$, as the nonintegrability parameter, is of order unity, far away from both integrable limits. That the dependence on $U/J$ is weak and the chaos nearly always strong is likely due to the large size of the system (we have considered the chains with $10\leq L\leq 50$ sites), so it is not much of a surprise. Our results are in rough agreement with similar work in \cite{PauschOptimalRoute,Pausch:2025zgd} where the authors compare the purely classical approach (integration of classical equations of motion) with the full quantum approach. These earlier works find stronger dependence of chaos on $U/J$ and energy, but they consider shorter chains. Still, one cannot discard the possibility that TWA with quantum jumps has different chaotic properties from both the classical limit and the full quantum evolution.

A particularly important question is the timescale at which the superdiffusion ceases. In the numerics, it is always roughly between $\log t/t_0\approx 8$ and $\log t/t_0\approx 10$, as can be appreciated from Figs. \ref{superfluidTWA}-\ref{normdiff}, \ref{superfluidalpha}-\ref{mottalpha} and in particular from Fig.~\ref{normdifftime}. In the discussion below Eq.~(\ref{scallawfin}), we have reproduced this timescale: the derivation of the scaling law relies on the Taylor expansion of the occupation of initially empty sites, so it requires simply $t<1$ in appropriate units. Importantly, neither the derivation leading to Eq.~(\ref{scallawfin}) nor the numerical results in Fig.~\ref{normdifftime} predict any dependence of the crossover time on the degree of nonintegrability, but the numerics does suggest that it becomes longer for longer chains. On the other hand, recent state-of-the-art results on slow thermalization and metastable states \cite{Yin:2024hjm} predict the onset of thermalization to scale exponentially with the nonintegrability parameter $\epsilon$. We thus conclude that our superdiffusive regime is strongly related to the special initial conditions and cannot be understood as the decay of a metastable state. It would be interesting to check what happens in the limit $L\to\infty$ and if this case some of the sites may remain superdiffusive forever (although $L\to\infty$ is admittedly not realistic in experiments).

One interesting question is if the superdiffusive regime can be understood as prethermalization \footnote{We thank Fabrizio Minganti for discussions on this point.}. From the results of this paper, it does not look so: the prethermal state is a long-living but ultimately unstable state into which the system settles before reaching true thermal equilibrium. On the other hand, the superidiffusive regime does not result in any specific universal state, and its timescale is not broadly separated from the normal diffusion timescale, i.e. the intermediate regime between superdiffusion and diffusion is quite short, whereas the key to prethermalization is the wide separation of timescales, so the prethermalized state lives for some significant time before thermalizing (see, e.g., \cite{Mori:2018}). On the other hand, the fact that superdiffusion lasts longer in larger systems does agree with the prethermalization concept. Still, we believe that it is mainly the consequence of a hidden scaling symmetry.

Finally, it is very telling that the existence of transport (normal or anomalous) strongly depends on the choice of the observable. It is not that surprising that the special scale invariance leading to superdiffusion is not preserved upon a change of variables, however the fact that normal diffusion is also lost is unexpected: it goes against the logic of eigenstate thermalization. This is a potentially very deep issue which we also plan to address in the future.

\acknowledgments

We are grateful to Fabrizio Minganti, Filippo Ferrari, Andrea Richaud, Ana Hudomal and Ivana Vasi\'c for stimulating discussions. The work on Sections \ref{secint}, \ref{secchaos} and \ref{secconc} (chaos analysis, supervision and interpretation) was supported by Russian Science Foundation Grant No. 24-72-10061 [https://rscf.ru/project/24-72-10061/]  and performed at Steklov Mathematical Institute of Russian Academy of Sciences (Mihailo \v{C}ubrovi\'c). The rest of the paper was funded by the Institute of Physics Belgrade, through the grant by the Ministry of Science, Technological Development, and Innovation of the Republic of Serbia.

\appendix

\section{Numerical equivalence between the Langevin-equation approach and the momentum-integrated approach to computing the operator expectation values}\label{secappequiv}

\begin{figure}[H]
(a)\includegraphics[width=0.45\textwidth]{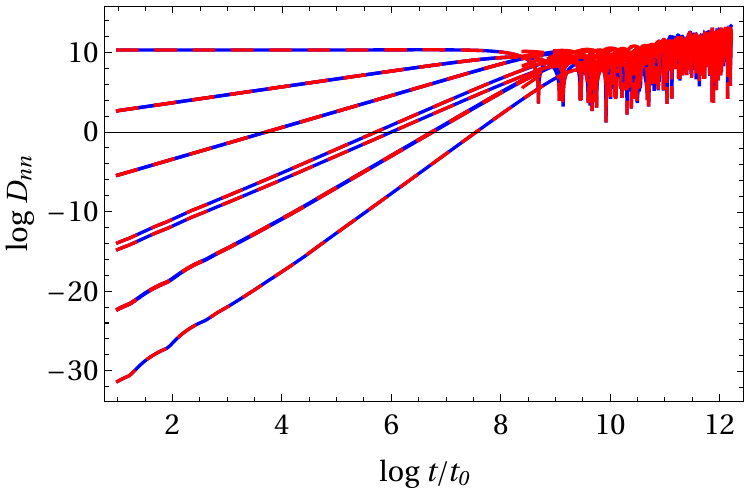}
(b)\includegraphics[width=0.45\textwidth]{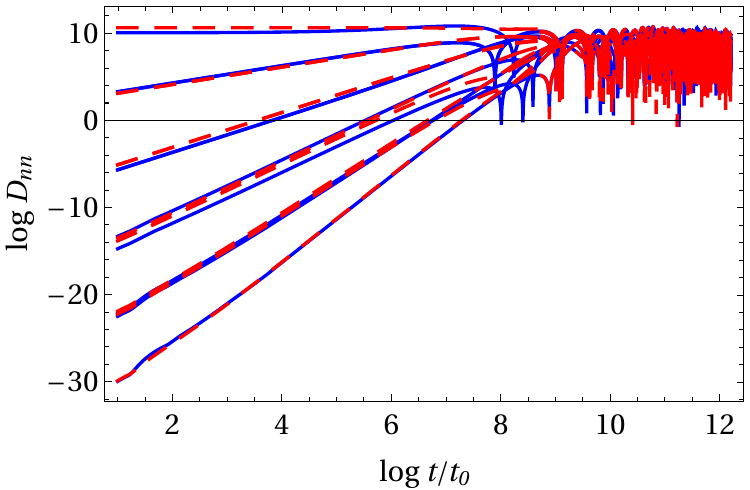}\\
(c)\includegraphics[width=0.45\textwidth]{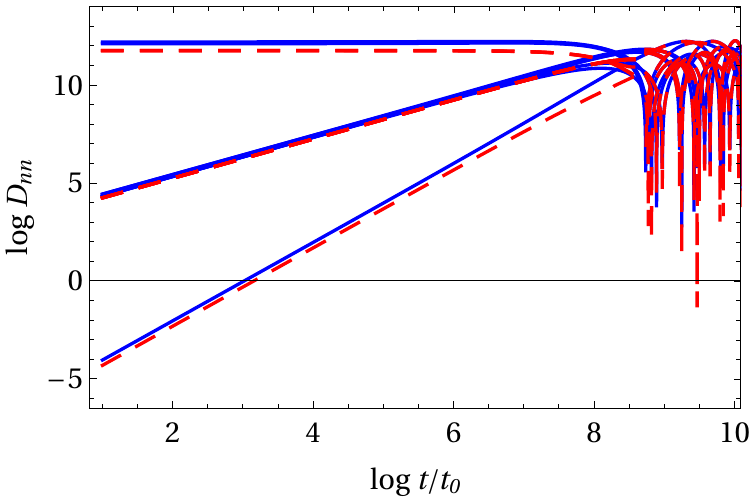}
(d)\includegraphics[width=0.45\textwidth]{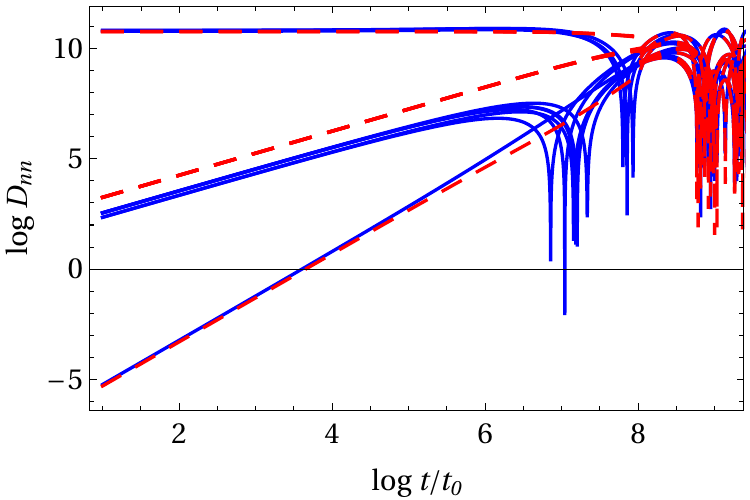}
\caption{\label{twacomp} Evolution of the squared deviation $D_{nn}$ for $U/J=0.5$ (a, b) and $U/J=8$ (c, d), for initially filled site $n=3$ (a, c) and for initially filled sites $n=3,6,9$ (b, d). In both cases $\mu/J=0.1$ and $L=10$. Blue (darker) lines are computed by integral (diffusive) approach, i.e. from Eqs.~(\ref{i1exp}\ref{i2expmixed}) as in the rest of the paper, whereas the red (lighter) lines are computed by the Langevin approach, i.e. by directly implementing quantum jumps. We do not distinguish different sites by color as our goal here is just to compare the two calculations. The agreement is very good indeed.}
\end{figure}

Here we demonstrate that the direct quantum trajectory calculation, with random jumps inserted at random time instants (one may call it the "Langevin approach" to beyond-TWA methods), is numerically equivalent to the time-integrated formalism that we use throughout the paper (we may call this the "diffusion approach"). Analytically, we know from start that the two methods have to be equivalent but here we see that their numerical characteristics and quantitative results are also very similar. In Fig.~\ref{twacomp} we plot the evolution of the correlation functions $D_{nn}$ for four different setups. The agreement is nearly perfect, which justifies the diffusive approach as an equivalent formulation of the Langevin approach. Only in Fig.~\ref{twacomp}(d) the difference is non-negligible in the sense that some of the fluctuations are suppressed in the Langevin approach. Nevertheless, the superdiffusive scaling is very robust.

\section{\label{secappfigs}Additional numerical results}

In this Appendix we supply a few figures providing additional numerical evidence of our findings. They contain no new concepts compared to the results in the main text, just additional examples to better corroborate our conclusions.

\subsection{Superdiffusive scaling}

\begin{figure}[H]
(a)\includegraphics[width=0.385\textwidth]{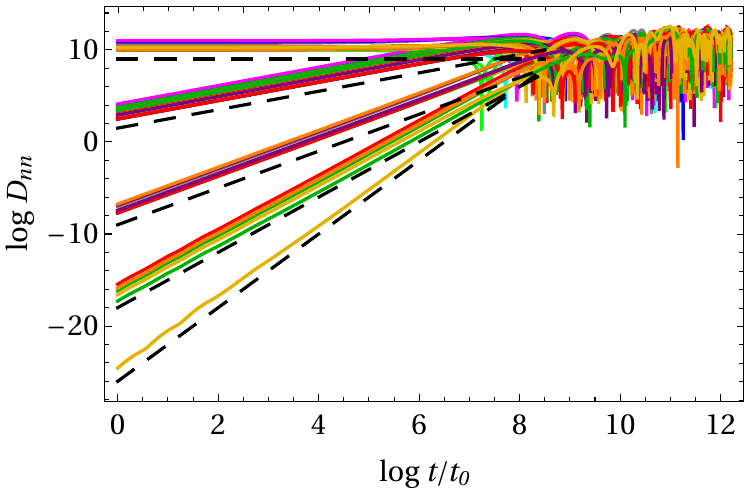}
(b)\includegraphics[width=0.515\textwidth]{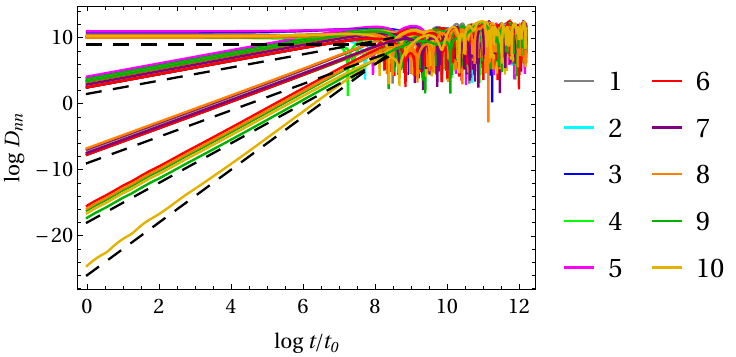}
\caption{\label{restransTWA} Log-log plot of the dispersion $D_{nn}=\langle I_n^2\rangle-\langle I_n\rangle^2$, with $L=10$, $\mu/J=0.05$ and $U/J=0.1$. In (a) the initial fillings are resonant
i.e. commensurable $\langle I_3(t=0)\rangle=\langle I_5(t=0)\rangle=1/2$, and in (b) they are nonresonant $\langle I_3(t=0)\rangle:\langle I_5(t=0)\rangle=\sqrt{2}$. The law of dispersion is the same as before and insensitive to the resonances.}
\end{figure}

In Fig.~\ref{restransTWA} we show the transport for two initially filled sites ($n=3$ and $n=5$) instead of one. In (a) the state is resonant, i.e. the occupation numbers have the ratio $\langle I_3(t=0)\rangle:\langle I_5(t=0)\rangle=1:1$; in the second case the ratio is $\langle I_3(t=0)\rangle:\langle I_5(t=0)\rangle=\sqrt{2}:1$ and therefore nonresonant (the ratio is irrational) \footnote{The terms resonant and non-resonant are justified in the classical limit: it turns out, as we have shown in \cite{Markovic:2023gtx}, that the ratio of frequencies is proportional to the ratio of actions, i.e. occupation numbers.}. Both cases show the same scaling, in accordance with the general law of Eq.~(\ref{newbie}). This is different from the leading-order TWA approximation in \cite{Markovic:2023gtx}, where a resonant initial state has a different set of exponents. The reason is the fact that we integrate over a pseudodistribution of finite support, so the special ratio of the initial expectation values is not that important.

\begin{figure}[H]
(a)\includegraphics[width=0.3\textwidth]{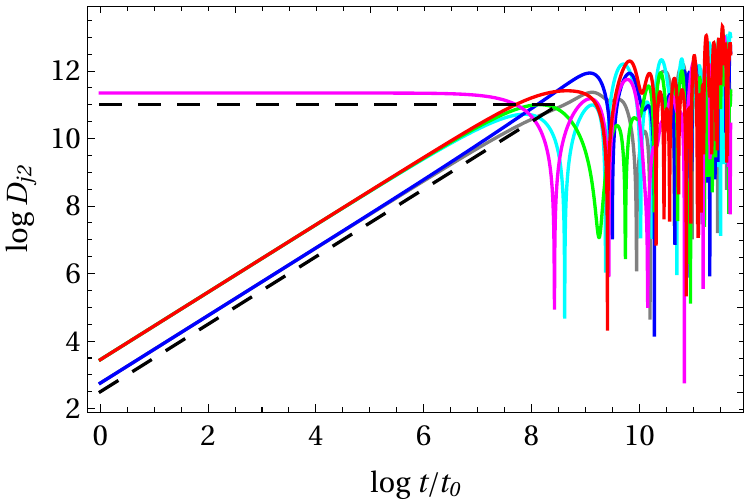}
(b)\includegraphics[width=0.3\textwidth]{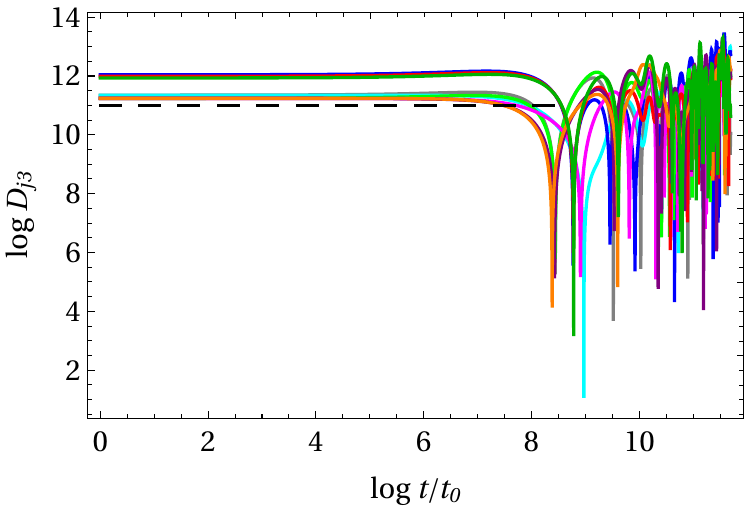}
(c)\includegraphics[width=0.3\textwidth]{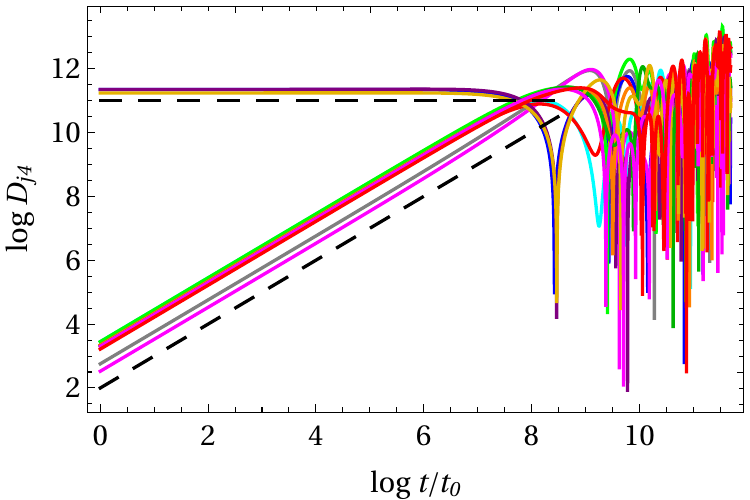}
(d)\includegraphics[width=0.3\textwidth]{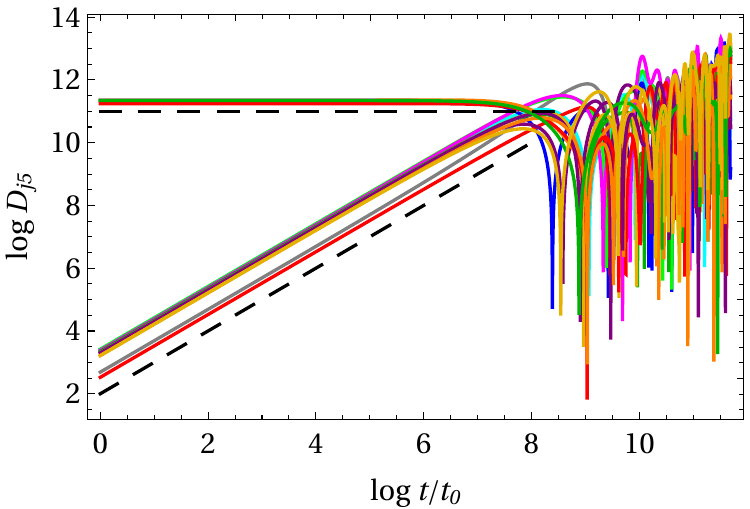}
(e)\includegraphics[width=0.3\textwidth]{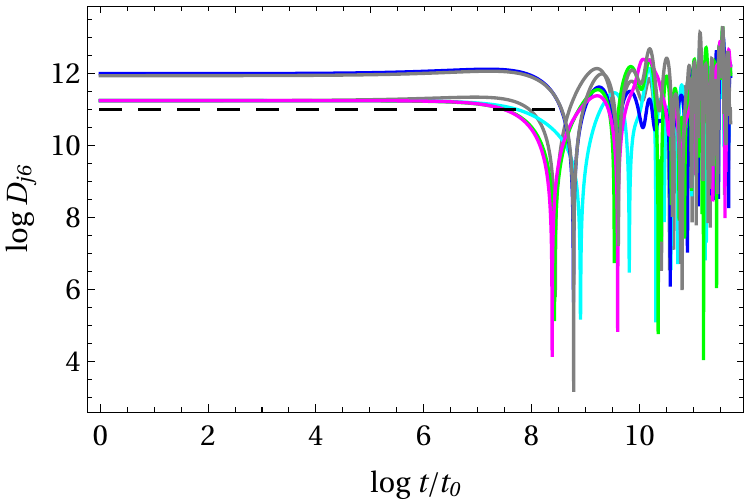}
(f)\includegraphics[width=0.3\textwidth]{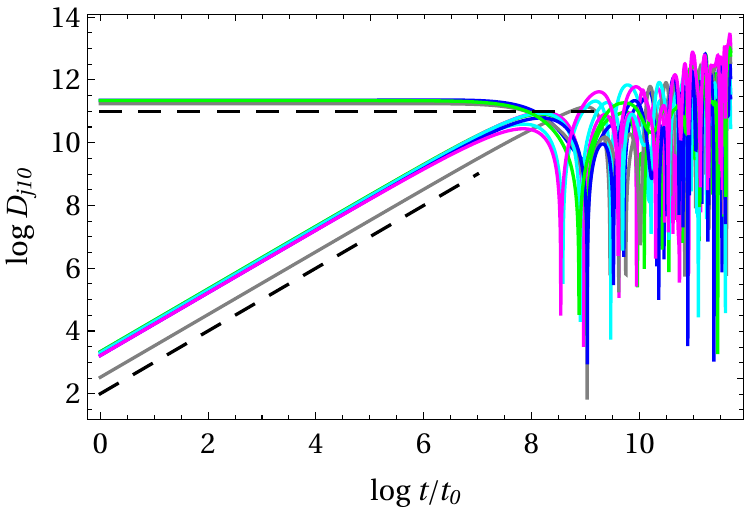}
\caption{\label{mixtransTWA3} Log-log plot of two-site correlation functions $D_{nj}\equiv\langle I_nI_j\rangle-\langle I_n\rangle\langle I_j\rangle$ with $n=2,3,4,5,7,10$ (panels a-e) and $j$ ranging over the whole chain ($1$ to $10$) in a chain of length $L=10$, with 
$U/J=0.1$, $\mu/J=0.05$. The only initially filled sites are $n=3$, $n=6$ and $n=9$, with the filling ratios $2:1:1$. The color code for $j$ is the same as in the previous figures.}
\end{figure}

Fig.~\ref{mixtransTWA3} provides another example of the scaling law (\ref{restransTWA}) for the two-site correlators, for different parameters and initial conditions compared to Fig.~\ref{mixtransTWA1}. The scaling is again very good.

\subsection{Superdiffusive vs. normal diffusive mixing}

\begin{figure}[H]
(a)\includegraphics[width=0.27\textwidth]{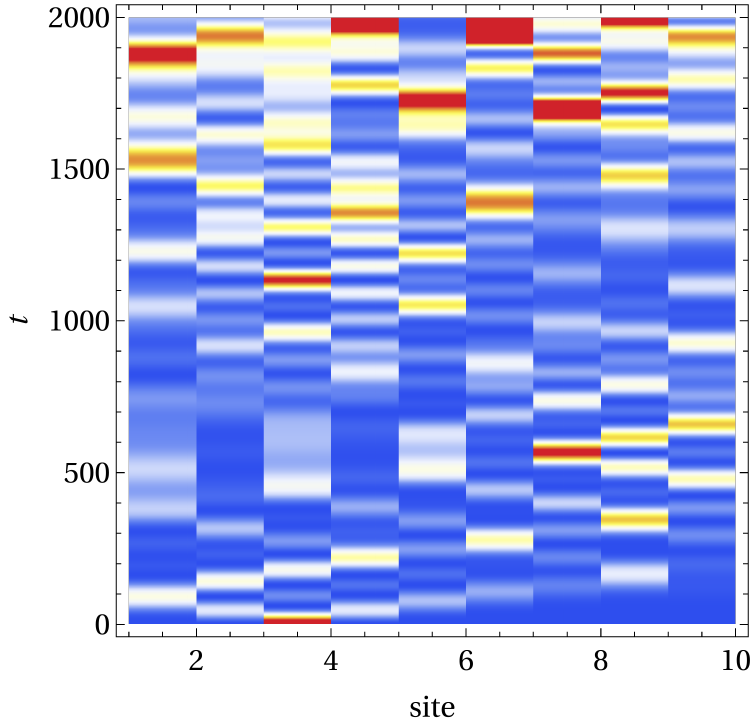}
(b)\includegraphics[width=0.27\textwidth]{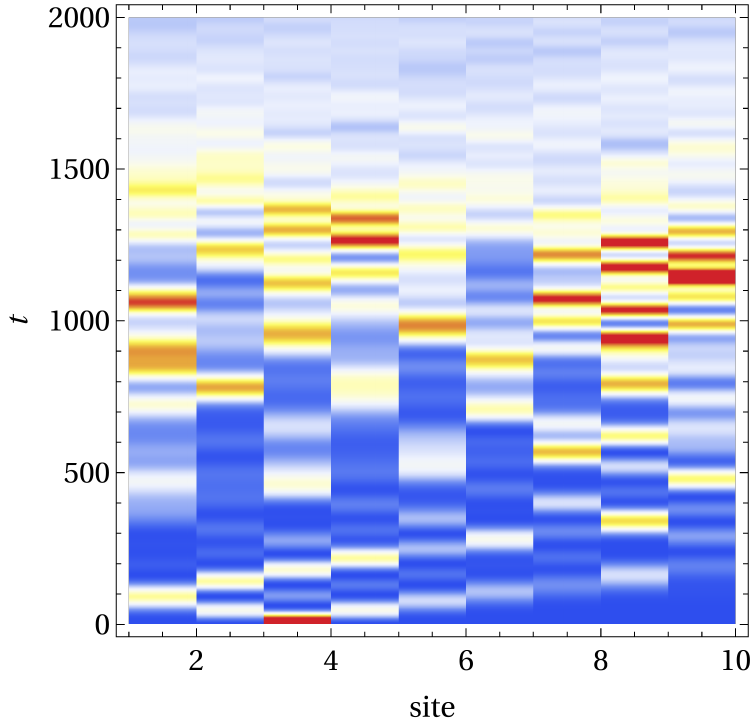}
(c)\includegraphics[width=0.27\textwidth]{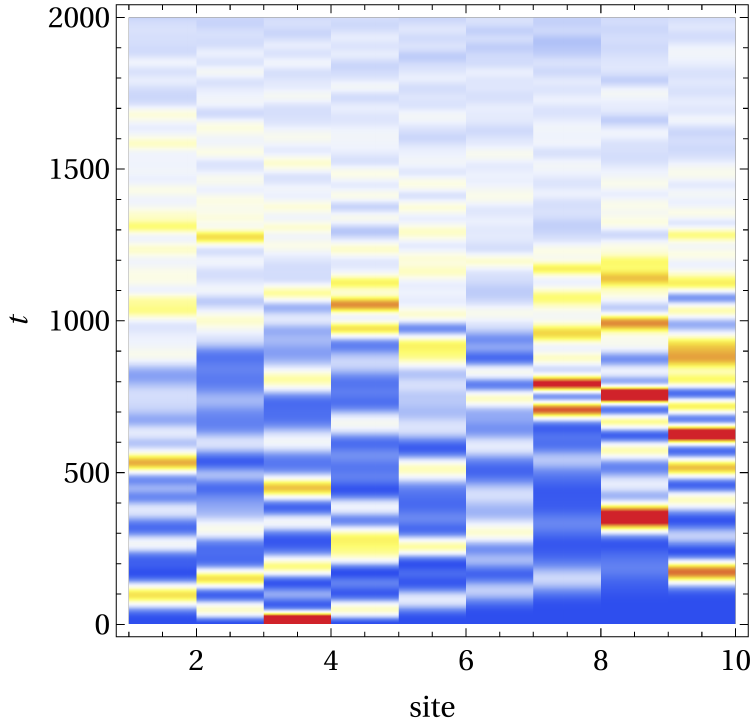}\\
(d)\includegraphics[width=0.27\textwidth]{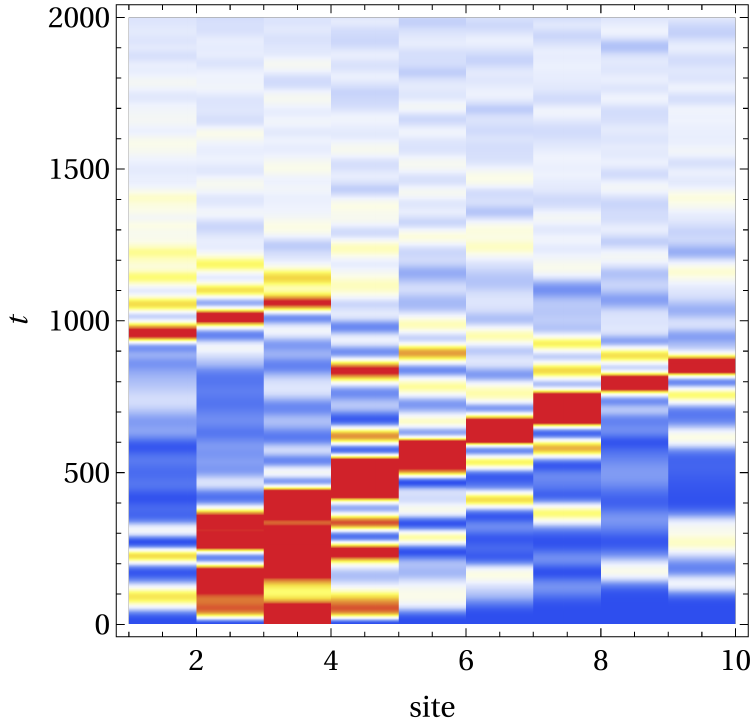}
(e)\includegraphics[width=0.27\textwidth]{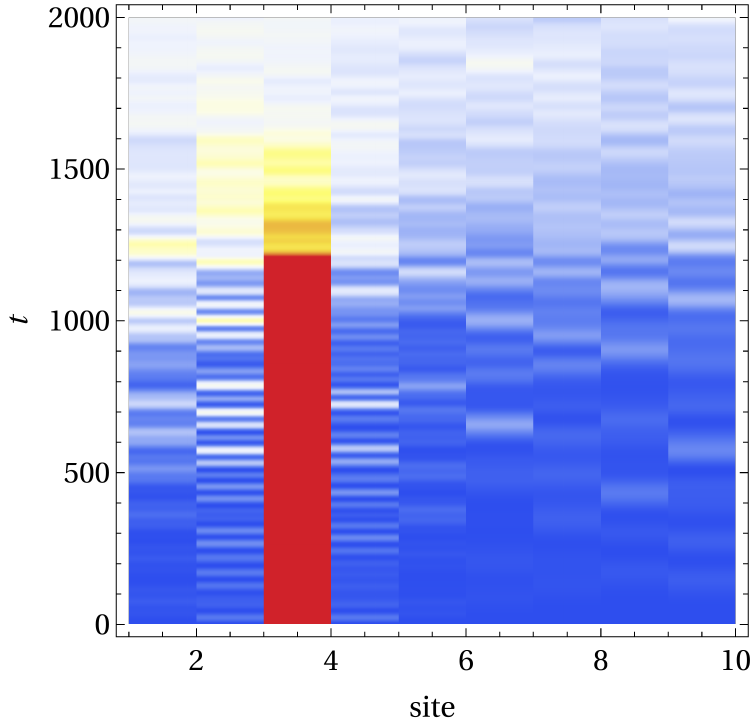}
(f)\includegraphics[width=0.31\textwidth]{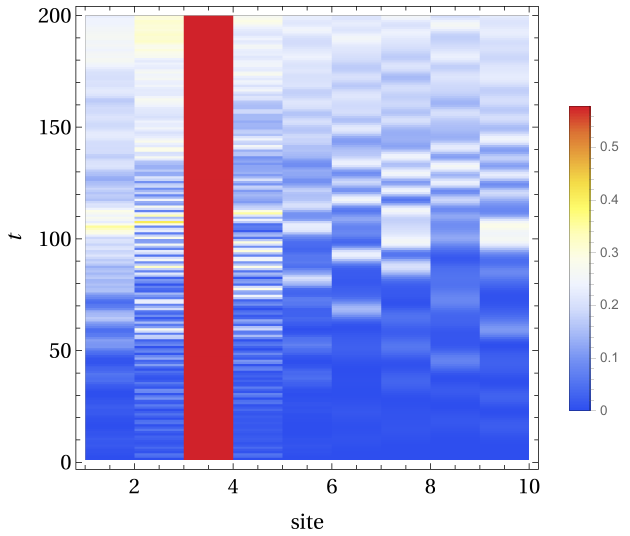}
\caption{\label{densityplot} Number density for the whole chain of length $L=10$, with the initially filled site $n=3$, with $U/J=0.1,0.5,2.0,4.0,8.0,10.0$ and $\mu=0$. At small $U/J$ the density oscillates forever, at large $U/J$ it retains very long (or maybe eternal?) memory of the initial conditions. For the intermediate $U/J$ values, which correspond to most nonintegrable dynamics, the long-time distribution is uniform and the system is ergodic. For practical reasons, the unit of time (y-axis) is not the usual computational unit $t_0$ but $100t_0$. The color scale encodes the occupation number $I_n$ at site $n$, from low (dark) to high (bright).}
\end{figure}

Here we find further support for the conclusion that equilibration is only efficient for strongly chaotic ($U/J\sim 1$) regime, in accordance with the comments in relation to Fig.~\ref{vartplot}. The evolution of density in Fig.~\ref{densityplot} shows that equilibration does not happen for large $U/J$ -- expectedly as it is the insulator regime, but it also fails for small $U/J$, when the superfluid oscillates for a very long time. The occupation number only becomes uniform across the chain for $0.5\lesssim U/J\lesssim 2.0$, the values similar but not quite identical to those that show best linear fits in Fig.~\ref{normdiff}, and the minimum of the variation in Fig.~\ref{vartplot}. The final equilibrium seems more sensitive to chaos than the existence of normal diffusion itself \footnote{Admittedly, from numerics we cannot tell if the uniform distribution is never reached for small and large $U/J$, or it just takes longer time. However, Fig.~\ref{normdiff} suggests that the dispersion itself does not grow monotonically for large $U/J$ but goes up and down. For small $U/J$ the question is indeed open.}.

\subsection{Mixing entropy independence of $U/J$}

\begin{figure}[H]
\includegraphics[width=0.45\textwidth]{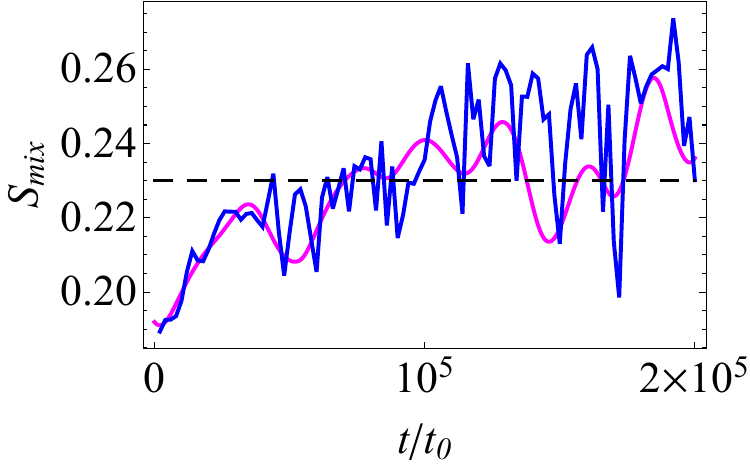}
\includegraphics[width=0.45\textwidth]{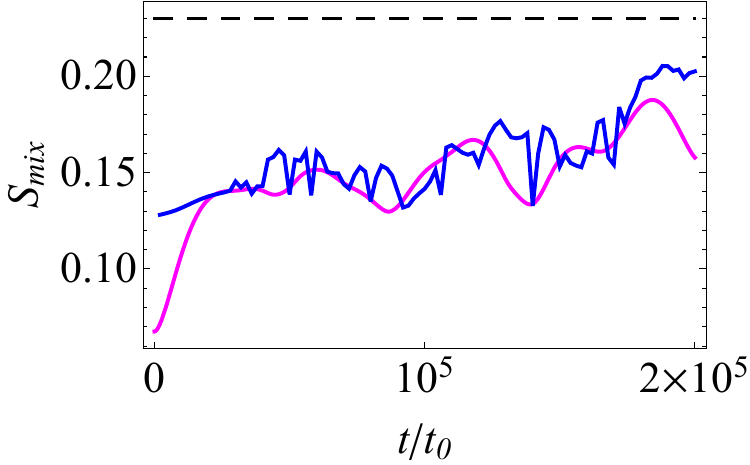}
\caption{\label{mottentropy} Mixing entropies for the chain of length $L=10$ in the Mott regime ($U/J=4$) with $\mu/J=0.10$, with random initial conditions (left) and for initially filled sites $I_3(t=0)=I_5(t=0)=1/2$ (right). The blue line (darker) represents mixing entropy for the $\alpha$ variables and the magenta line (lighter) represents the mixing entropy for the actions. The black dashed line indicates the theoretical maximum for the mixing entropy from Eq.~(\ref{mixentmax}). The overall picture is very similar to the previous figure, i.e. the Mott/superfluid regime does not influence the mixing.} 
\end{figure}

In Fig.~\ref{mottentropy} we find exactly the same story as in Fig.~\ref{superfluidentropy}, even though we now have large $U/J$ and the insulator regime. This is because the mixing entropy describes the superdiffusive regime which is rather insensitive to $U/J$ or chaos.

\section{\label{secappsff}Spectral form factor and chaos}

\begin{figure}[H]
\includegraphics[width=0.385\textwidth]{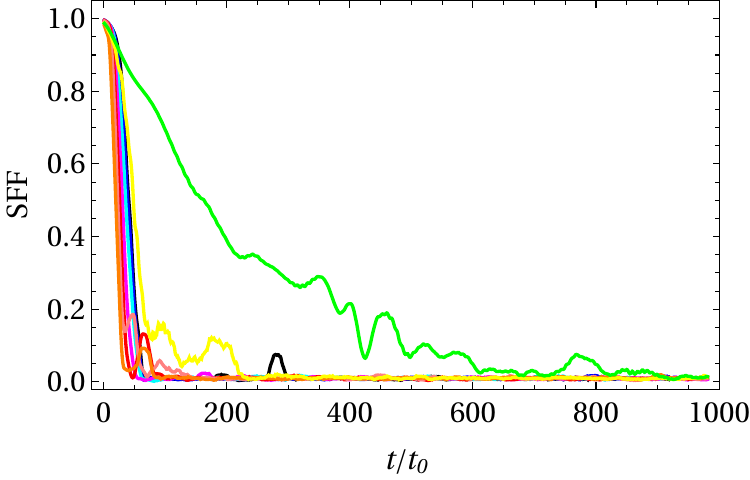}
\includegraphics[width=0.515\textwidth]{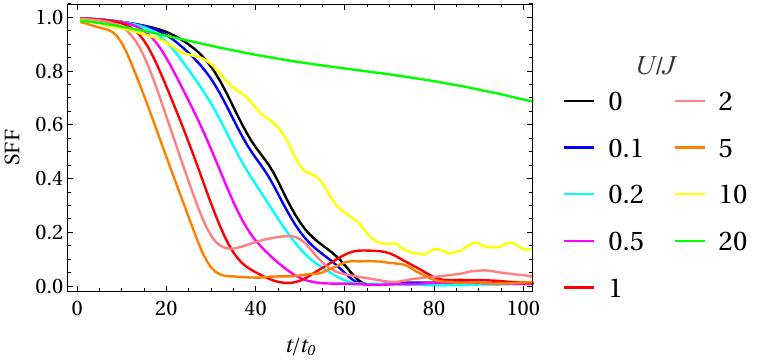}
\caption{\label{sff1} Semiclassical spectral form factor for the ensemble of quantum trajectories drawn from the Wigner pseudodistribution for the Gaussian coherent state of Eq.~(\ref{w0coherent}), for a range of $U/J$ values, encoded by the colors from dark (low $U$) to light (high $U$). In the semiclassical regime the only structure that can be seen is the dip, which is the fastest when chaos is the strongest, which expectedly happens for intermediate values of $U/J$ (here between $2$ and $5$). Initially filled sites are at $n=3,6,9$.} 
\end{figure}

Even though the spectral form factor (SFF) only makes full sense in a full quantum computation, when individual states and energy levels can be discerned, already in the semiclassical regime the dip epoch can indicate the strength of chaos, which offers some additional insight and further corroborates the dependence of chaos on $U/J$. In Fig.~\ref{sff1} we show the evolution of SFF. To remind, SFF is defined as
\be
\mathrm{SFF}(t)=\frac{\langle Z(\beta-\imath t)Z(\beta+\imath t)\rangle}{\langle Z(\beta)\rangle^2},~~Z(\beta\pm\imath t)\equiv\sum_\alpha e^{(-\beta\pm\imath t)E_\alpha},\label{sff}
\ee
where $\beta$ is the inverse temperature and $\alpha$ counts the states, with energies $E_\alpha$. Normally, one sums the partition function over the whole Hilbert space. In our TWA-based calculation, there is no Hilbert space, but the orbits for different initial conditions generated from $W_0$, as in Eq.~(\ref{w0coherent}), give us a proxy of the state space. We thus sum over the states generated from $W_0$ (the same that we use as initial conditions for the integration of the equations of motion) and with the weights drawn from $W_0$. Finally, we assume that the system is locally at infinite temperature ($\beta=0$), which is justified in the semiclassical limit, when all the states are explored in accordance with their Wigner pseudoweights, without an additional Gibbs thermal factor; the interpretation is that semiclassically the temperature is much larger than the microscopic (level spacing) scale. Altogether, this yields the prescription:
\be
\mathrm{SFF}(t)=\frac{\langle Z(-\imath t)Z(\imath t)\rangle}{\mathcal{N}^2}=\frac{\int d\mathbf{P}_0\int d\mathbf{Q}_0W_0\left(\mathbf{P}_0,\mathbf{Q}_0\right)e^{-\imath tH\left(\mathbf{P}_0,\mathbf{Q}_0;\mathbf{P}(t),\mathbf{Q}(t)\right)}\int d\mathbf{P}_0\int d\mathbf{Q}_0W_0\left(\mathbf{P}_0,\mathbf{Q}_0\right)e^{\imath tH\left(\mathbf{P}_0,\mathbf{Q}_0;\mathbf{P}(t),\mathbf{Q}(t)\right)}}{\left(\int d\mathbf{P}_0\int d\mathbf{Q}_0W_0\left(\mathbf{P}_0,\mathbf{Q}_0\right)\right)^2}.\label{sfftwa}
\ee
The textbook dip-ramp-plateau structure of quantum chaotic systems cannot be expected in the semiclassical regime: the ramp comes from level repulsion (Wigner-Dyson level dynamics) -- but there are no discrete levels in our setup. The plateau also comes from the discreteness of the spectrum. Only the dip is expected to survive.

Roughly speaking, the earlier and faster the dip, the stronger the chaos. According to Fig.~\ref{sff1}, the strongest chaos is found for intermediate values of $U/J$ though not quite the same as the peak of Lyapunov exponents in Fig.~\ref{ftle} -- the latter is located at $U/J=1$ while the dip is the fastest between $U/J=2$ and $U/J=5$. Still, a rough agreement exists.

\section{\label{secappotoc}No OTOC in truncated Wigner approximation}

Here we show that the usual notion of out-of-time ordered correlator cannot be calculated in TWA for a system which conserves the particle number, at least not without some modification of its usual definition. To remind, OTOC is motivated by considering the expectation value of the module squared of the commutator of two operators $[A,B]$. A frequent choice is to take two canonically conjugate operators, in this case $A\mapsto I_n$ and $B\mapsto\phi_n$. But the argument is completely general so we will just use the generic canonically conjugate variables $p_m$ and $q_n$. We will show that TOC and OTOC turn out equal in absolute value and opposite in sign, i.e. that the whole 4-point function $\langle\vert[A,B]\vert^2\rangle$ vanishes. Start from the whole 4-point function for $p_m$ and $q_n$:
\begin{equation}
C_{mn}(t)=\langle\vert\left[\hat{p}_m(t),\hat{q}_n(0)\right]\vert^2\rangle=-\langle\left(\hat{q}_n\left(0\right)\hat{p}_m\left(t\right)-\hat{p}_m\left(t\right)\hat{q}_n\left(0\right)\right)^2\rangle.
\end{equation}
Since we have two identical operators we can use the Bopp rules to get:
\begin{equation}
\left(\hat{q}_n(0)\hat{p}_m(t)-\hat{p}_m(t)\hat{q}_n(0)\right)=q_n(0)p_m(t)+\frac{\imath\hbar}{2}\frac{\partial p_m(t)}{\partial p_n(0)}-p_m(t)q_n(0)+\frac{\imath\hbar}{2}\frac{\partial q_n(0)}{\partial p_m(t)}=\frac{\imath\hbar}{2}\left(\frac{\partial p_m(t)}{\partial p_n(0)}+\frac{\partial q_n(0)}{\partial q_m(t)}\right).
\end{equation}
Higher-order corrections to TWA will not change this conclusion as the above result is just a c-number, not an operator. Therefore:
\begin{equation}
    C_{mn}(t)=\frac{\hbar^2}{4}\left\langle\left(\frac{\partial p_m(t)}{\partial p_n(0)}+\frac{\partial q_n(0)}{\partial q_m(t)}\right)^2\right\rangle.
\end{equation}
In principle it looks like this might take some arbitrary nonzero value, but taking into account the number conservation and averaging one finds:
\begin{equation}
    C_{mn}(t)=\frac{\hbar^2}{4}\left\langle\left(\frac{\partial p_m(\mathbf{p}_0,\mathbf{q}_0,t)}{\partial p_n(0)}+\left(\frac{\partial q_m(\mathbf{p}_0,\mathbf{q}_0,t)}{\partial q_n(0)}\right)^{-1}\right)^2\right\rangle.\label{cabapp}
\end{equation}
The dependence on $p_n(0)$ and $q_n(0)$ is purely fictional: there are $2L$ variables but with one constraint -- number conservation (actually two since the energy is conserved as well) and their dependence can be eliminated, so in fact, the whole correlation function $C_{mn}$ vanishes on shell. Since $C_{mn}=2\times\mathrm{TOC}+2\times\mathrm{OTOC}$, it means that OTOC is equal to TOC in absolute value and contains no new information. Notice that the number conservation is crucial for (\ref{cabapp}) to be zero: if the number is not conserved then the value is in general nonzero. This happens, for example, in \cite{Ferrari:2024ogi}, where dissipation breaks the number conservation and OTOC contains valuable information.

\bibliography{bhRefs}

\begin{thebibliography}{49}%
\makeatletter
\providecommand \@ifxundefined [1]{%
 \@ifx{#1\undefined}
}%
\providecommand \@ifnum [1]{%
 \ifnum #1\expandafter \@firstoftwo
 \else \expandafter \@secondoftwo
 \fi
}%
\providecommand \@ifx [1]{%
 \ifx #1\expandafter \@firstoftwo
 \else \expandafter \@secondoftwo
 \fi
}%
\providecommand \natexlab [1]{#1}%
\providecommand \enquote  [1]{``#1''}%
\providecommand \bibnamefont  [1]{#1}%
\providecommand \bibfnamefont [1]{#1}%
\providecommand \citenamefont [1]{#1}%
\providecommand \href@noop [0]{\@secondoftwo}%
\providecommand \href [0]{\begingroup \@sanitize@url \@href}%
\providecommand \@href[1]{\@@startlink{#1}\@@href}%
\providecommand \@@href[1]{\endgroup#1\@@endlink}%
\providecommand \@sanitize@url [0]{\catcode `\\12\catcode `\$12\catcode `\&12\catcode `\#12\catcode `\^12\catcode `\_12\catcode `\%12\relax}%
\providecommand \@@startlink[1]{}%
\providecommand \@@endlink[0]{}%
\providecommand \url  [0]{\begingroup\@sanitize@url \@url }%
\providecommand \@url [1]{\endgroup\@href {#1}{\urlprefix }}%
\providecommand \urlprefix  [0]{URL }%
\providecommand \Eprint [0]{\href }%
\providecommand \doibase [0]{https://doi.org/}%
\providecommand \selectlanguage [0]{\@gobble}%
\providecommand \bibinfo  [0]{\@secondoftwo}%
\providecommand \bibfield  [0]{\@secondoftwo}%
\providecommand \translation [1]{[#1]}%
\providecommand \BibitemOpen [0]{}%
\providecommand \bibitemStop [0]{}%
\providecommand \bibitemNoStop [0]{.\EOS\space}%
\providecommand \EOS [0]{\spacefactor3000\relax}%
\providecommand \BibitemShut  [1]{\csname bibitem#1\endcsname}%
\let\auto@bib@innerbib\@empty
\bibitem [{\citenamefont {Lin}\ and\ \citenamefont {Motrunich}(2017)}]{Lin:2017}%
  \BibitemOpen
  \bibfield  {author} {\bibinfo {author} {\bibfnamefont {C.-J.}\ \bibnamefont {Lin}}\ and\ \bibinfo {author} {\bibfnamefont {O.~I.}\ \bibnamefont {Motrunich}},\ }\bibfield  {title} {\bibinfo {title} {Quasiparticle explanation of the weak-thermalization regime under quench in a nonintegrable quantum spin chain},\ }\href {https://doi.org/10.1103/PhysRevA.95.023621} {\bibfield  {journal} {\bibinfo  {journal} {Phys. Rev. A}\ }\textbf {\bibinfo {volume} {95}},\ \bibinfo {pages} {023621} (\bibinfo {year} {2017})}\BibitemShut {NoStop}%
\bibitem [{\citenamefont {Lin}\ and\ \citenamefont {Motrunich}(2019)}]{Lin:2019}%
  \BibitemOpen
  \bibfield  {author} {\bibinfo {author} {\bibfnamefont {C.-J.}\ \bibnamefont {Lin}}\ and\ \bibinfo {author} {\bibfnamefont {O.~I.}\ \bibnamefont {Motrunich}},\ }\bibfield  {title} {\bibinfo {title} {Exact quantum many-body scar states in the rydberg-blockaded atom chain},\ }\href {https://doi.org/10.1103/PhysRevLett.122.173401} {\bibfield  {journal} {\bibinfo  {journal} {Phys. Rev. Lett.}\ }\textbf {\bibinfo {volume} {122}},\ \bibinfo {pages} {173401} (\bibinfo {year} {2019})}\BibitemShut {NoStop}%
\bibitem [{\citenamefont {{Kolovsky}}\ and\ \citenamefont {{Buchleitner}}(2004)}]{Kolovsky:2004}%
  \BibitemOpen
  \bibfield  {author} {\bibinfo {author} {\bibfnamefont {A.~R.}\ \bibnamefont {{Kolovsky}}}\ and\ \bibinfo {author} {\bibfnamefont {A.}~\bibnamefont {{Buchleitner}}},\ }\bibfield  {title} {\bibinfo {title} {{Quantum chaos in the Bose-Hubbard model}},\ }\href {https://doi.org/10.1209/epl/i2004-10265-7} {\bibfield  {journal} {\bibinfo  {journal} {EPL (Europhysics Letters)}\ }\textbf {\bibinfo {volume} {68}},\ \bibinfo {pages} {632} (\bibinfo {year} {2004})},\ \Eprint {https://arxiv.org/abs/cond-mat/0403213} {arXiv:cond-mat/0403213 [cond-mat.soft]} \BibitemShut {NoStop}%
\bibitem [{\citenamefont {Kolovsky}(2007)}]{Kolovsky:2007}%
  \BibitemOpen
  \bibfield  {author} {\bibinfo {author} {\bibfnamefont {A.~R.}\ \bibnamefont {Kolovsky}},\ }\bibfield  {title} {\bibinfo {title} {Semiclassical quantization of the bogoliubov spectrum},\ }\href {https://doi.org/10.1103/PhysRevLett.99.020401} {\bibfield  {journal} {\bibinfo  {journal} {Phys. Rev. Lett.}\ }\textbf {\bibinfo {volume} {99}},\ \bibinfo {pages} {020401} (\bibinfo {year} {2007})}\BibitemShut {NoStop}%
\bibitem [{\citenamefont {Graefe}\ and\ \citenamefont {Korsch}(2007)}]{Graefe:2007}%
  \BibitemOpen
  \bibfield  {author} {\bibinfo {author} {\bibfnamefont {E.~M.}\ \bibnamefont {Graefe}}\ and\ \bibinfo {author} {\bibfnamefont {H.~J.}\ \bibnamefont {Korsch}},\ }\bibfield  {title} {\bibinfo {title} {Semiclassical quantization of an $n$-particle bose-hubbard model},\ }\href {https://doi.org/10.1103/PhysRevA.76.032116} {\bibfield  {journal} {\bibinfo  {journal} {Phys. Rev. A}\ }\textbf {\bibinfo {volume} {76}},\ \bibinfo {pages} {032116} (\bibinfo {year} {2007})}\BibitemShut {NoStop}%
\bibitem [{\citenamefont {Trimborn}\ \emph {et~al.}(2008)\citenamefont {Trimborn}, \citenamefont {Witthaut},\ and\ \citenamefont {Korsch}}]{Trimborn:2008}%
  \BibitemOpen
  \bibfield  {author} {\bibinfo {author} {\bibfnamefont {F.}~\bibnamefont {Trimborn}}, \bibinfo {author} {\bibfnamefont {D.}~\bibnamefont {Witthaut}},\ and\ \bibinfo {author} {\bibfnamefont {H.~J.}\ \bibnamefont {Korsch}},\ }\bibfield  {title} {\bibinfo {title} {Exact number-conserving phase-space dynamics of the $m$-site bose-hubbard model},\ }\href {https://doi.org/10.1103/PhysRevA.77.043631} {\bibfield  {journal} {\bibinfo  {journal} {Phys. Rev. A}\ }\textbf {\bibinfo {volume} {77}},\ \bibinfo {pages} {043631} (\bibinfo {year} {2008})}\BibitemShut {NoStop}%
\bibitem [{\citenamefont {{Kolovsky}}(2016)}]{Kolovsky:2016}%
  \BibitemOpen
  \bibfield  {author} {\bibinfo {author} {\bibfnamefont {A.~R.}\ \bibnamefont {{Kolovsky}}},\ }\bibfield  {title} {\bibinfo {title} {{Bose-Hubbard Hamiltonian: Quantum chaos approach}},\ }\href {https://doi.org/10.1142/S0217979216300097} {\bibfield  {journal} {\bibinfo  {journal} {International Journal of Modern Physics B}\ }\textbf {\bibinfo {volume} {30}},\ \bibinfo {eid} {1630009} (\bibinfo {year} {2016})},\ \Eprint {https://arxiv.org/abs/1507.03413} {arXiv:1507.03413 [quant-ph]} \BibitemShut {NoStop}%
\bibitem [{\citenamefont {Fischer}\ \emph {et~al.}(2016)\citenamefont {Fischer}, \citenamefont {Hoffmann},\ and\ \citenamefont {Wimberger}}]{Fischer:PhysRevA.93.043620}%
  \BibitemOpen
  \bibfield  {author} {\bibinfo {author} {\bibfnamefont {D.}~\bibnamefont {Fischer}}, \bibinfo {author} {\bibfnamefont {D.}~\bibnamefont {Hoffmann}},\ and\ \bibinfo {author} {\bibfnamefont {S.}~\bibnamefont {Wimberger}},\ }\bibfield  {title} {\bibinfo {title} {Spectral analysis of two-dimensional bose-hubbard models},\ }\href {https://doi.org/10.1103/PhysRevA.93.043620} {\bibfield  {journal} {\bibinfo  {journal} {Phys. Rev. A}\ }\textbf {\bibinfo {volume} {93}},\ \bibinfo {pages} {043620} (\bibinfo {year} {2016})}\BibitemShut {NoStop}%
\bibitem [{\citenamefont {Sorg}\ \emph {et~al.}(2014)\citenamefont {Sorg}, \citenamefont {Vidmar}, \citenamefont {Pollet},\ and\ \citenamefont {Heidrich-Meisner}}]{Sorg:PhysRevA.90.033606}%
  \BibitemOpen
  \bibfield  {author} {\bibinfo {author} {\bibfnamefont {S.}~\bibnamefont {Sorg}}, \bibinfo {author} {\bibfnamefont {L.}~\bibnamefont {Vidmar}}, \bibinfo {author} {\bibfnamefont {L.}~\bibnamefont {Pollet}},\ and\ \bibinfo {author} {\bibfnamefont {F.}~\bibnamefont {Heidrich-Meisner}},\ }\bibfield  {title} {\bibinfo {title} {Relaxation and thermalization in the one-dimensional bose-hubbard model: A case study for the interaction quantum quench from the atomic limit},\ }\href {https://doi.org/10.1103/PhysRevA.90.033606} {\bibfield  {journal} {\bibinfo  {journal} {Phys. Rev. A}\ }\textbf {\bibinfo {volume} {90}},\ \bibinfo {pages} {033606} (\bibinfo {year} {2014})}\BibitemShut {NoStop}%
\bibitem [{\citenamefont {Bychek}\ \emph {et~al.}(2020)\citenamefont {Bychek}, \citenamefont {Muraev}, \citenamefont {Maksimov},\ and\ \citenamefont {Kolovsky}}]{Kolovsky:2020}%
  \BibitemOpen
  \bibfield  {author} {\bibinfo {author} {\bibfnamefont {A.~A.}\ \bibnamefont {Bychek}}, \bibinfo {author} {\bibfnamefont {P.~S.}\ \bibnamefont {Muraev}}, \bibinfo {author} {\bibfnamefont {D.~N.}\ \bibnamefont {Maksimov}},\ and\ \bibinfo {author} {\bibfnamefont {A.~R.}\ \bibnamefont {Kolovsky}},\ }\bibfield  {title} {\bibinfo {title} {Open bose-hubbard chain: Pseudoclassical approach},\ }\href {https://doi.org/10.1103/PhysRevE.101.012208} {\bibfield  {journal} {\bibinfo  {journal} {Phys. Rev. E}\ }\textbf {\bibinfo {volume} {101}},\ \bibinfo {pages} {012208} (\bibinfo {year} {2020})}\BibitemShut {NoStop}%
\bibitem [{\citenamefont {{McCormack}}\ \emph {et~al.}(2021)\citenamefont {{McCormack}}, \citenamefont {{Nath}},\ and\ \citenamefont {{Li}}}]{McCormack:2021Photo...8..554M}%
  \BibitemOpen
  \bibfield  {author} {\bibinfo {author} {\bibfnamefont {G.}~\bibnamefont {{McCormack}}}, \bibinfo {author} {\bibfnamefont {R.}~\bibnamefont {{Nath}}},\ and\ \bibinfo {author} {\bibfnamefont {W.}~\bibnamefont {{Li}}},\ }\bibfield  {title} {\bibinfo {title} {{Hyperchaos in a Bose-Hubbard Chain with Rydberg-Dressed Interactions}},\ }\href {https://doi.org/10.3390/photonics8120554} {\bibfield  {journal} {\bibinfo  {journal} {Photonics}\ }\textbf {\bibinfo {volume} {8}},\ \bibinfo {eid} {554} (\bibinfo {year} {2021})},\ \Eprint {https://arxiv.org/abs/2108.09683} {arXiv:2108.09683 [cond-mat.quant-gas]} \BibitemShut {NoStop}%
\bibitem [{\citenamefont {Kollath}\ \emph {et~al.}(2007)\citenamefont {Kollath}, \citenamefont {L\"auchli},\ and\ \citenamefont {Altman}}]{Altman:PhysRevLett.98.180601}%
  \BibitemOpen
  \bibfield  {author} {\bibinfo {author} {\bibfnamefont {C.}~\bibnamefont {Kollath}}, \bibinfo {author} {\bibfnamefont {A.~M.}\ \bibnamefont {L\"auchli}},\ and\ \bibinfo {author} {\bibfnamefont {E.}~\bibnamefont {Altman}},\ }\bibfield  {title} {\bibinfo {title} {Quench dynamics and nonequilibrium phase diagram of the bose-hubbard model},\ }\href {https://doi.org/10.1103/PhysRevLett.98.180601} {\bibfield  {journal} {\bibinfo  {journal} {Phys. Rev. Lett.}\ }\textbf {\bibinfo {volume} {98}},\ \bibinfo {pages} {180601} (\bibinfo {year} {2007})}\BibitemShut {NoStop}%
\bibitem [{\citenamefont {Pausch}\ \emph {et~al.}(2021{\natexlab{a}})\citenamefont {Pausch}, \citenamefont {Carnio}, \citenamefont {Buchleitner},\ and\ \citenamefont {Rodr\'\i{}guez}}]{Pausch2body}%
  \BibitemOpen
  \bibfield  {author} {\bibinfo {author} {\bibfnamefont {L.}~\bibnamefont {Pausch}}, \bibinfo {author} {\bibfnamefont {E.~G.}\ \bibnamefont {Carnio}}, \bibinfo {author} {\bibfnamefont {A.}~\bibnamefont {Buchleitner}},\ and\ \bibinfo {author} {\bibfnamefont {A.}~\bibnamefont {Rodr\'\i{}guez}},\ }\bibfield  {title} {\bibinfo {title} {{Chaos in the Bose\textendash{}Hubbard model and random two-body Hamiltonians}},\ }\href {https://doi.org/10.1088/1367-2630/ac3c0d} {\bibfield  {journal} {\bibinfo  {journal} {New J. Phys.}\ }\textbf {\bibinfo {volume} {23}},\ \bibinfo {pages} {123036} (\bibinfo {year} {2021}{\natexlab{a}})},\ \Eprint {https://arxiv.org/abs/2109.06236} {arXiv:2109.06236 [quant-ph]} \BibitemShut {NoStop}%
\bibitem [{\citenamefont {Pausch}\ \emph {et~al.}(2021{\natexlab{b}})\citenamefont {Pausch}, \citenamefont {Carnio}, \citenamefont {Rodr\'{\i}guez},\ and\ \citenamefont {Buchleitner}}]{PauschSpectrum}%
  \BibitemOpen
  \bibfield  {author} {\bibinfo {author} {\bibfnamefont {L.}~\bibnamefont {Pausch}}, \bibinfo {author} {\bibfnamefont {E.~G.}\ \bibnamefont {Carnio}}, \bibinfo {author} {\bibfnamefont {A.}~\bibnamefont {Rodr\'{\i}guez}},\ and\ \bibinfo {author} {\bibfnamefont {A.}~\bibnamefont {Buchleitner}},\ }\bibfield  {title} {\bibinfo {title} {Chaos and ergodicity across the energy spectrum of interacting bosons},\ }\href {https://doi.org/10.1103/PhysRevLett.126.150601} {\bibfield  {journal} {\bibinfo  {journal} {Phys. Rev. Lett.}\ }\textbf {\bibinfo {volume} {126}},\ \bibinfo {pages} {150601} (\bibinfo {year} {2021}{\natexlab{b}})}\BibitemShut {NoStop}%
\bibitem [{\citenamefont {Dag}\ \emph {et~al.}(2023)\citenamefont {Dag}, \citenamefont {Mistakidis}, \citenamefont {Chan},\ and\ \citenamefont {Sadeghpour}}]{Dag:2022vqb}%
  \BibitemOpen
  \bibfield  {author} {\bibinfo {author} {\bibfnamefont {C.~B.}\ \bibnamefont {Dag}}, \bibinfo {author} {\bibfnamefont {S.~I.}\ \bibnamefont {Mistakidis}}, \bibinfo {author} {\bibfnamefont {A.}~\bibnamefont {Chan}},\ and\ \bibinfo {author} {\bibfnamefont {H.~R.}\ \bibnamefont {Sadeghpour}},\ }\bibfield  {title} {\bibinfo {title} {{Many-body quantum chaos in stroboscopically-driven cold atoms}},\ }\href {https://doi.org/10.1038/s42005-023-01258-1} {\bibfield  {journal} {\bibinfo  {journal} {Commun. Phys.}\ }\textbf {\bibinfo {volume} {6}},\ \bibinfo {pages} {136} (\bibinfo {year} {2023})},\ \Eprint {https://arxiv.org/abs/2210.03840} {arXiv:2210.03840 [cond-mat.quant-gas]} \BibitemShut {NoStop}%
\bibitem [{\citenamefont {Dahan}\ \emph {et~al.}(2022)\citenamefont {Dahan}, \citenamefont {Arwas},\ and\ \citenamefont {Grosfeld}}]{Dahan:2022classical}%
  \BibitemOpen
  \bibfield  {author} {\bibinfo {author} {\bibfnamefont {D.}~\bibnamefont {Dahan}}, \bibinfo {author} {\bibfnamefont {G.}~\bibnamefont {Arwas}},\ and\ \bibinfo {author} {\bibfnamefont {E.}~\bibnamefont {Grosfeld}},\ }\bibfield  {title} {\bibinfo {title} {Classical and quantum chaos in chirally-driven, dissipative bose-hubbard systems},\ }\href@noop {} {\bibfield  {journal} {\bibinfo  {journal} {npj quantum information}\ }\textbf {\bibinfo {volume} {8}},\ \bibinfo {pages} {1} (\bibinfo {year} {2022})}\BibitemShut {NoStop}%
\bibitem [{\citenamefont {Pausch}\ \emph {et~al.}(2022)\citenamefont {Pausch}, \citenamefont {Buchleitner}, \citenamefont {Carnio},\ and\ \citenamefont {Rodr\'\i{}guez}}]{PauschOptimalRoute}%
  \BibitemOpen
  \bibfield  {author} {\bibinfo {author} {\bibfnamefont {L.}~\bibnamefont {Pausch}}, \bibinfo {author} {\bibfnamefont {A.}~\bibnamefont {Buchleitner}}, \bibinfo {author} {\bibfnamefont {E.~G.}\ \bibnamefont {Carnio}},\ and\ \bibinfo {author} {\bibfnamefont {A.}~\bibnamefont {Rodr\'\i{}guez}},\ }\bibfield  {title} {\bibinfo {title} {{Optimal route to quantum chaos in the Bose\textendash{}Hubbard model}},\ }\href {https://doi.org/10.1088/1751-8121/ac7e0b} {\bibfield  {journal} {\bibinfo  {journal} {J. Phys. A}\ }\textbf {\bibinfo {volume} {55}},\ \bibinfo {pages} {324002} (\bibinfo {year} {2022})},\ \Eprint {https://arxiv.org/abs/2205.04209} {arXiv:2205.04209 [quant-ph]} \BibitemShut {NoStop}%
\bibitem [{\citenamefont {Nakerst}\ and\ \citenamefont {Haque}(2023)}]{Nakerst:2022prc}%
  \BibitemOpen
  \bibfield  {author} {\bibinfo {author} {\bibfnamefont {G.}~\bibnamefont {Nakerst}}\ and\ \bibinfo {author} {\bibfnamefont {M.}~\bibnamefont {Haque}},\ }\bibfield  {title} {\bibinfo {title} {{Chaos in the three-site Bose-Hubbard model: Classical versus quantum}},\ }\href {https://doi.org/10.1103/PhysRevE.107.024210} {\bibfield  {journal} {\bibinfo  {journal} {Phys. Rev. E}\ }\textbf {\bibinfo {volume} {107}},\ \bibinfo {pages} {024210} (\bibinfo {year} {2023})},\ \Eprint {https://arxiv.org/abs/2203.09953} {arXiv:2203.09953 [quant-ph]} \BibitemShut {NoStop}%
\bibitem [{\citenamefont {Markovi\'c}\ and\ \citenamefont {\v{C}ubrovi\'c}(2024)}]{Markovic:2023gtx}%
  \BibitemOpen
  \bibfield  {author} {\bibinfo {author} {\bibfnamefont {D.}~\bibnamefont {Markovi\'c}}\ and\ \bibinfo {author} {\bibfnamefont {M.}~\bibnamefont {\v{C}ubrovi\'c}},\ }\bibfield  {title} {\bibinfo {title} {{Chaos and anomalous transport in a semiclassical Bose-Hubbard chain}},\ }\href {https://doi.org/10.1103/PhysRevE.109.034213} {\bibfield  {journal} {\bibinfo  {journal} {Phys. Rev. E}\ }\textbf {\bibinfo {volume} {109}},\ \bibinfo {pages} {034213} (\bibinfo {year} {2024})},\ \Eprint {https://arxiv.org/abs/2308.14720} {arXiv:2308.14720 [quant-ph]} \BibitemShut {NoStop}%
\bibitem [{\citenamefont {Minganti}\ \emph {et~al.}(2023)\citenamefont {Minganti}, \citenamefont {Savona},\ and\ \citenamefont {Biella}}]{Minganti2023dissipativephase}%
  \BibitemOpen
  \bibfield  {author} {\bibinfo {author} {\bibfnamefont {F.}~\bibnamefont {Minganti}}, \bibinfo {author} {\bibfnamefont {V.}~\bibnamefont {Savona}},\ and\ \bibinfo {author} {\bibfnamefont {A.}~\bibnamefont {Biella}},\ }\bibfield  {title} {\bibinfo {title} {Dissipative phase transitions in {$n$}-photon driven quantum nonlinear resonators},\ }\href {https://doi.org/10.22331/q-2023-11-07-1170} {\bibfield  {journal} {\bibinfo  {journal} {{Quantum}}\ }\textbf {\bibinfo {volume} {7}},\ \bibinfo {pages} {1170} (\bibinfo {year} {2023})}\BibitemShut {NoStop}%
\bibitem [{\citenamefont {Ferrari}\ \emph {et~al.}(2023)\citenamefont {Ferrari}, \citenamefont {Gravina}, \citenamefont {Eeltink}, \citenamefont {Scarlino}, \citenamefont {Savona},\ and\ \citenamefont {Minganti}}]{Ferrari:2023zfe}%
  \BibitemOpen
  \bibfield  {author} {\bibinfo {author} {\bibfnamefont {F.}~\bibnamefont {Ferrari}}, \bibinfo {author} {\bibfnamefont {L.}~\bibnamefont {Gravina}}, \bibinfo {author} {\bibfnamefont {D.}~\bibnamefont {Eeltink}}, \bibinfo {author} {\bibfnamefont {P.}~\bibnamefont {Scarlino}}, \bibinfo {author} {\bibfnamefont {V.}~\bibnamefont {Savona}},\ and\ \bibinfo {author} {\bibfnamefont {F.}~\bibnamefont {Minganti}},\ }\bibfield  {title} {\bibinfo {title} {{Steady-state quantum chaos in open quantum systems}},\ }\href@noop {} {\  (\bibinfo {year} {2023})},\ \Eprint {https://arxiv.org/abs/2305.15479} {arXiv:2305.15479 [quant-ph]} \BibitemShut {NoStop}%
\bibitem [{\citenamefont {Ferrari}\ \emph {et~al.}(2024)\citenamefont {Ferrari}, \citenamefont {Minganti}, \citenamefont {Aron},\ and\ \citenamefont {Savona}}]{Ferrari:2024ogi}%
  \BibitemOpen
  \bibfield  {author} {\bibinfo {author} {\bibfnamefont {F.}~\bibnamefont {Ferrari}}, \bibinfo {author} {\bibfnamefont {F.}~\bibnamefont {Minganti}}, \bibinfo {author} {\bibfnamefont {C.}~\bibnamefont {Aron}},\ and\ \bibinfo {author} {\bibfnamefont {V.}~\bibnamefont {Savona}},\ }\bibfield  {title} {\bibinfo {title} {{Chaos and spatial prethermalization in driven-dissipative bosonic chains}},\ }\href@noop {} {\  (\bibinfo {year} {2024})},\ \Eprint {https://arxiv.org/abs/2409.12225} {arXiv:2409.12225 [quant-ph]} \BibitemShut {NoStop}%
\bibitem [{\citenamefont {Lu}\ \emph {et~al.}(2025)\citenamefont {Lu}, \citenamefont {Graf}, \citenamefont {Heller}, \citenamefont {Keski-Rahkonen},\ and\ \citenamefont {Dag}}]{Lu:2025bqn}%
  \BibitemOpen
  \bibfield  {author} {\bibinfo {author} {\bibfnamefont {Z.}~\bibnamefont {Lu}}, \bibinfo {author} {\bibfnamefont {A.~M.}\ \bibnamefont {Graf}}, \bibinfo {author} {\bibfnamefont {E.~J.}\ \bibnamefont {Heller}}, \bibinfo {author} {\bibfnamefont {J.}~\bibnamefont {Keski-Rahkonen}},\ and\ \bibinfo {author} {\bibfnamefont {C.~B.}\ \bibnamefont {Dag}},\ }\bibfield  {title} {\bibinfo {title} {{Anti-scarring from eigenstate stacking in a chaotic spinor condensate}},\ }\href@noop {} {\  (\bibinfo {year} {2025})},\ \Eprint {https://arxiv.org/abs/2501.17856} {arXiv:2501.17856 [quant-ph]} \BibitemShut {NoStop}%
\bibitem [{\citenamefont {Pausch}\ \emph {et~al.}(2025)\citenamefont {Pausch}, \citenamefont {Carnio}, \citenamefont {Buchleitner},\ and\ \citenamefont {Rodriguez}}]{Pausch:2025zgd}%
  \BibitemOpen
  \bibfield  {author} {\bibinfo {author} {\bibfnamefont {L.}~\bibnamefont {Pausch}}, \bibinfo {author} {\bibfnamefont {E.~G.}\ \bibnamefont {Carnio}}, \bibinfo {author} {\bibfnamefont {A.}~\bibnamefont {Buchleitner}},\ and\ \bibinfo {author} {\bibfnamefont {A.}~\bibnamefont {Rodriguez}},\ }\bibfield  {title} {\bibinfo {title} {{How to seed ergodic dynamics of interacting bosons under conditions of many-body quantum chaos}},\ }\href@noop {} {\  (\bibinfo {year} {2025})},\ \Eprint {https://arxiv.org/abs/2501.13556} {arXiv:2501.13556 [quant-ph]} \BibitemShut {NoStop}%
\bibitem [{\citenamefont {Haake}(2006)}]{Haake:book}%
  \BibitemOpen
  \bibfield  {author} {\bibinfo {author} {\bibfnamefont {F.}~\bibnamefont {Haake}},\ }\href@noop {} {\emph {\bibinfo {title} {Quantum Signatures of Chaos}}}\ (\bibinfo  {publisher} {Springer-Verlag},\ \bibinfo {address} {Berlin, Heidelberg},\ \bibinfo {year} {2006})\BibitemShut {NoStop}%
\bibitem [{\citenamefont {Swingle}(2018)}]{Swingle:2018ekw}%
  \BibitemOpen
  \bibfield  {author} {\bibinfo {author} {\bibfnamefont {B.}~\bibnamefont {Swingle}},\ }\bibfield  {title} {\bibinfo {title} {{Unscrambling the physics of out-of-time-order correlators}},\ }\href {https://doi.org/10.1038/s41567-018-0295-5} {\bibfield  {journal} {\bibinfo  {journal} {Nature Phys.}\ }\textbf {\bibinfo {volume} {14}},\ \bibinfo {pages} {988} (\bibinfo {year} {2018})}\BibitemShut {NoStop}%
\bibitem [{\citenamefont {Polkovnikov}\ \emph {et~al.}(2002)\citenamefont {Polkovnikov}, \citenamefont {Sachdev},\ and\ \citenamefont {Girvin}}]{PolkovnikSachdev2002}%
  \BibitemOpen
  \bibfield  {author} {\bibinfo {author} {\bibfnamefont {A.}~\bibnamefont {Polkovnikov}}, \bibinfo {author} {\bibfnamefont {S.}~\bibnamefont {Sachdev}},\ and\ \bibinfo {author} {\bibfnamefont {S.~M.}\ \bibnamefont {Girvin}},\ }\bibfield  {title} {\bibinfo {title} {Nonequilibrium gross-pitaevskii dynamics of boson lattice models},\ }\href {https://doi.org/10.1103/PhysRevA.66.053607} {\bibfield  {journal} {\bibinfo  {journal} {Phys. Rev. A}\ }\textbf {\bibinfo {volume} {66}},\ \bibinfo {pages} {053607} (\bibinfo {year} {2002})}\BibitemShut {NoStop}%
\bibitem [{\citenamefont {Polkovnikov}(2003{\natexlab{a}})}]{Polkovnik2003}%
  \BibitemOpen
  \bibfield  {author} {\bibinfo {author} {\bibfnamefont {A.}~\bibnamefont {Polkovnikov}},\ }\bibfield  {title} {\bibinfo {title} {Quantum corrections to the dynamics of interacting bosons: Beyond the truncated wigner approximation},\ }\href {https://doi.org/10.1103/PhysRevA.68.053604} {\bibfield  {journal} {\bibinfo  {journal} {Phys. Rev. A}\ }\textbf {\bibinfo {volume} {68}},\ \bibinfo {pages} {053604} (\bibinfo {year} {2003}{\natexlab{a}})}\BibitemShut {NoStop}%
\bibitem [{\citenamefont {Polkovnikov}(2003{\natexlab{b}})}]{Polkovnik2003Main}%
  \BibitemOpen
  \bibfield  {author} {\bibinfo {author} {\bibfnamefont {A.}~\bibnamefont {Polkovnikov}},\ }\bibfield  {title} {\bibinfo {title} {Evolution of the macroscopically entangled states in optical lattices},\ }\href {https://doi.org/10.1103/PhysRevA.68.033609} {\bibfield  {journal} {\bibinfo  {journal} {Phys. Rev. A}\ }\textbf {\bibinfo {volume} {68}},\ \bibinfo {pages} {033609} (\bibinfo {year} {2003}{\natexlab{b}})}\BibitemShut {NoStop}%
\bibitem [{\citenamefont {Polkovnikov}(2010)}]{PolkovnikAnnals}%
  \BibitemOpen
  \bibfield  {author} {\bibinfo {author} {\bibfnamefont {A.}~\bibnamefont {Polkovnikov}},\ }\bibfield  {title} {\bibinfo {title} {Phase space representation of quantum dynamics},\ }\href {https://doi.org/https://doi.org/10.1016/j.aop.2010.02.006} {\bibfield  {journal} {\bibinfo  {journal} {Annals of Physics}\ }\textbf {\bibinfo {volume} {325}},\ \bibinfo {pages} {1790} (\bibinfo {year} {2010})}\BibitemShut {NoStop}%
\bibitem [{\citenamefont {Claeys}\ and\ \citenamefont {Polkovnikov}(2021)}]{PolkovnikSciPost}%
  \BibitemOpen
  \bibfield  {author} {\bibinfo {author} {\bibfnamefont {P.~W.}\ \bibnamefont {Claeys}}\ and\ \bibinfo {author} {\bibfnamefont {A.}~\bibnamefont {Polkovnikov}},\ }\bibfield  {title} {\bibinfo {title} {{Quantum eigenstates from classical Gibbs distributions}},\ }\href {https://doi.org/10.21468/SciPostPhys.10.1.014} {\bibfield  {journal} {\bibinfo  {journal} {SciPost Phys.}\ }\textbf {\bibinfo {volume} {10}},\ \bibinfo {pages} {014} (\bibinfo {year} {2021})}\BibitemShut {NoStop}%
\bibitem [{\citenamefont {Wurtz}\ and\ \citenamefont {Polkovnikov}(2020)}]{Polkovnik2018}%
  \BibitemOpen
  \bibfield  {author} {\bibinfo {author} {\bibfnamefont {J.}~\bibnamefont {Wurtz}}\ and\ \bibinfo {author} {\bibfnamefont {A.}~\bibnamefont {Polkovnikov}},\ }\bibfield  {title} {\bibinfo {title} {Quantum diffusion in spin chains with phase space methods},\ }\href {https://doi.org/10.1103/PhysRevE.101.052120} {\bibfield  {journal} {\bibinfo  {journal} {Phys. Rev. E}\ }\textbf {\bibinfo {volume} {101}},\ \bibinfo {pages} {052120} (\bibinfo {year} {2020})}\BibitemShut {NoStop}%
\bibitem [{Note1()}]{Note1}%
  \BibitemOpen
  \bibinfo {note} {With slight abuse of terminology, the name action-angle variables is also used in the literature (including our previous work \cite {Markovic:2023gtx}) in nonintegrable cases, the logic being that if the system is sufficiently close to integrable the actions are still adiabatic invariants.}\BibitemShut {Stop}%
\bibitem [{\citenamefont {{Sinatra}}\ \emph {et~al.}(2002)\citenamefont {{Sinatra}}, \citenamefont {{Lobo}},\ and\ \citenamefont {{Castin}}}]{Sinatra:2002JPhB...35.3599S}%
  \BibitemOpen
  \bibfield  {author} {\bibinfo {author} {\bibfnamefont {A.}~\bibnamefont {{Sinatra}}}, \bibinfo {author} {\bibfnamefont {C.}~\bibnamefont {{Lobo}}},\ and\ \bibinfo {author} {\bibfnamefont {Y.}~\bibnamefont {{Castin}}},\ }\bibfield  {title} {\bibinfo {title} {{The truncated Wigner method for Bose-condensed gases: limits of validity and applications}},\ }\href {https://doi.org/10.1088/0953-4075/35/17/301} {\bibfield  {journal} {\bibinfo  {journal} {Journal of Physics B Atomic Molecular Physics}\ }\textbf {\bibinfo {volume} {35}},\ \bibinfo {pages} {3599} (\bibinfo {year} {2002})},\ \Eprint {https://arxiv.org/abs/cond-mat/0201217} {arXiv:cond-mat/0201217 [cond-mat.stat-mech]} \BibitemShut {NoStop}%
\bibitem [{\citenamefont {{Mink}}\ \emph {et~al.}(2022)\citenamefont {{Mink}}, \citenamefont {{Pelster}}, \citenamefont {{Benary}}, \citenamefont {{Ott}},\ and\ \citenamefont {{Fleischhauer}}}]{Pelster:2022ScPP...12...51M}%
  \BibitemOpen
  \bibfield  {author} {\bibinfo {author} {\bibfnamefont {C.}~\bibnamefont {{Mink}}}, \bibinfo {author} {\bibfnamefont {A.}~\bibnamefont {{Pelster}}}, \bibinfo {author} {\bibfnamefont {J.}~\bibnamefont {{Benary}}}, \bibinfo {author} {\bibfnamefont {H.}~\bibnamefont {{Ott}}},\ and\ \bibinfo {author} {\bibfnamefont {M.}~\bibnamefont {{Fleischhauer}}},\ }\bibfield  {title} {\bibinfo {title} {{Variational truncated Wigner approximation for weakly interacting Bose fields: Dynamics of coupled condensates}},\ }\href {https://doi.org/10.21468/SciPostPhys.12.2.051} {\bibfield  {journal} {\bibinfo  {journal} {SciPost Physics}\ }\textbf {\bibinfo {volume} {12}},\ \bibinfo {eid} {051} (\bibinfo {year} {2022})},\ \Eprint {https://arxiv.org/abs/2106.05354} {arXiv:2106.05354 [physics.atom-ph]} \BibitemShut {NoStop}%
\bibitem [{Note2()}]{Note2}%
  \BibitemOpen
  \bibinfo {note} {Explicitly: $\DOTSI \intop \ilimits@ \DOTSI \intop \ilimits@ dS_jdR_j R_j^4 (R_j^2+S_j^2-4)\exp \left (-\protect \frac {R_j^2+S_j^2}{2}\right )=\DOTSI \intop \ilimits@ r dr\DOTSI \intop \ilimits@ d\varphi r^4 \cos ^4\varphi (r^4-4)e^{-\protect \frac {r^2}{2}}=\protect \frac {3\pi }{4}\times 16=12\pi $.}\BibitemShut {Stop}%
\bibitem [{Note3()}]{Note3}%
  \BibitemOpen
  \bibinfo {note} {On the other hand, the simplified derivation has some disadvantages. It is less rigorous and more handwaving, and it does not capture the classically resonant case; but since the resonant case loses its special scaling in the presence of quantum corrections which are of prime importance in this work, we do not regard this as important.}\BibitemShut {Stop}%
\bibitem [{Note4()}]{Note4}%
  \BibitemOpen
  \bibinfo {note} {Of course, since we average over $W_0$, it is in general not true that $\langle I_{n\pm l}^2\rangle =\langle I_{n\pm l}\rangle ^2$. But in the simple approximation that we use the averaging does not change the naive result.}\BibitemShut {Stop}%
\bibitem [{Note5()}]{Note5}%
  \BibitemOpen
  \bibinfo {note} {This is somewhat unexpected, one would rather expect a power law. But in order to be sure one would also need to study more than four $L$ values.}\BibitemShut {Stop}%
\bibitem [{\citenamefont {Ljubotina}\ \emph {et~al.}(2023)\citenamefont {Ljubotina}, \citenamefont {Desaules}, \citenamefont {Serbyn},\ and\ \citenamefont {Papi\ifmmode~\acute{c}\else \'{c}\fi{}}}]{Ljubotina:2023}%
  \BibitemOpen
  \bibfield  {author} {\bibinfo {author} {\bibfnamefont {M.}~\bibnamefont {Ljubotina}}, \bibinfo {author} {\bibfnamefont {J.-Y.}\ \bibnamefont {Desaules}}, \bibinfo {author} {\bibfnamefont {M.}~\bibnamefont {Serbyn}},\ and\ \bibinfo {author} {\bibfnamefont {Z.}~\bibnamefont {Papi\ifmmode~\acute{c}\else \'{c}\fi{}}},\ }\bibfield  {title} {\bibinfo {title} {Superdiffusive energy transport in kinetically constrained models},\ }\href {https://doi.org/10.1103/PhysRevX.13.011033} {\bibfield  {journal} {\bibinfo  {journal} {Phys. Rev. X}\ }\textbf {\bibinfo {volume} {13}},\ \bibinfo {pages} {011033} (\bibinfo {year} {2023})}\BibitemShut {NoStop}%
\bibitem [{\citenamefont {{Rispoli}}\ \emph {et~al.}(2019)\citenamefont {{Rispoli}}, \citenamefont {{Lukin}}, \citenamefont {{Schittko}}, \citenamefont {{Kim}}, \citenamefont {{Tai}}, \citenamefont {{L{\'e}onard}},\ and\ \citenamefont {{Greiner}}}]{2019Natur.573..385R}%
  \BibitemOpen
  \bibfield  {author} {\bibinfo {author} {\bibfnamefont {M.}~\bibnamefont {{Rispoli}}}, \bibinfo {author} {\bibfnamefont {A.}~\bibnamefont {{Lukin}}}, \bibinfo {author} {\bibfnamefont {R.}~\bibnamefont {{Schittko}}}, \bibinfo {author} {\bibfnamefont {S.}~\bibnamefont {{Kim}}}, \bibinfo {author} {\bibfnamefont {M.~E.}\ \bibnamefont {{Tai}}}, \bibinfo {author} {\bibfnamefont {J.}~\bibnamefont {{L{\'e}onard}}},\ and\ \bibinfo {author} {\bibfnamefont {M.}~\bibnamefont {{Greiner}}},\ }\bibfield  {title} {\bibinfo {title} {{Quantum critical behaviour at the many-body localization transition}},\ }\href {https://doi.org/10.1038/s41586-019-1527-2} {\bibfield  {journal} {\bibinfo  {journal} {\nat}\ }\textbf {\bibinfo {volume} {573}},\ \bibinfo {pages} {385} (\bibinfo {year} {2019})},\ \Eprint {https://arxiv.org/abs/1812.06959} {arXiv:1812.06959 [cond-mat.quant-gas]} \BibitemShut {NoStop}%
\bibitem [{\citenamefont {Due\~nas}\ \emph {et~al.}(2025)\citenamefont {Due\~nas}, \citenamefont {Pe\~na},\ and\ \citenamefont {Rodr\'{\i}guez}}]{Duenas:2024xxy}%
  \BibitemOpen
  \bibfield  {author} {\bibinfo {author} {\bibfnamefont {O.}~\bibnamefont {Due\~nas}}, \bibinfo {author} {\bibfnamefont {D.}~\bibnamefont {Pe\~na}},\ and\ \bibinfo {author} {\bibfnamefont {A.}~\bibnamefont {Rodr\'{\i}guez}},\ }\bibfield  {title} {\bibinfo {title} {Propagation of two-particle correlations across the chaotic phase for interacting bosons},\ }\href {https://doi.org/10.1103/PhysRevResearch.7.L012031} {\bibfield  {journal} {\bibinfo  {journal} {Phys. Rev. Res.}\ }\textbf {\bibinfo {volume} {7}},\ \bibinfo {pages} {L012031} (\bibinfo {year} {2025})}\BibitemShut {NoStop}%
\bibitem [{\citenamefont {Richaud}\ and\ \citenamefont {Penna}(2018)}]{Richaud_2018}%
  \BibitemOpen
  \bibfield  {author} {\bibinfo {author} {\bibfnamefont {A.}~\bibnamefont {Richaud}}\ and\ \bibinfo {author} {\bibfnamefont {V.}~\bibnamefont {Penna}},\ }\bibfield  {title} {\bibinfo {title} {Phase separation can be stronger than chaos},\ }\href {https://doi.org/10.1088/1367-2630/aae73e} {\bibfield  {journal} {\bibinfo  {journal} {New Journal of Physics}\ }\textbf {\bibinfo {volume} {20}},\ \bibinfo {pages} {105008} (\bibinfo {year} {2018})}\BibitemShut {NoStop}%
\bibitem [{Note6()}]{Note6}%
  \BibitemOpen
  \bibinfo {note} {The former is logical as both extreme limits ($U/J=0$ and $U/J\to \infty $) are integrable, and the latter is explained by the fact that for growing $\mu $ the system is closer and closer to degenerate (second derivatives of $H$ with respect to actions are by a factor $U/\mu $ smaller than the first derivatives for $\mu /U$ large).}\BibitemShut {Stop}%
\bibitem [{\citenamefont {Yin}\ \emph {et~al.}(2025)\citenamefont {Yin}, \citenamefont {Surace},\ and\ \citenamefont {Lucas}}]{Yin:2024hjm}%
  \BibitemOpen
  \bibfield  {author} {\bibinfo {author} {\bibfnamefont {C.}~\bibnamefont {Yin}}, \bibinfo {author} {\bibfnamefont {F.~M.}\ \bibnamefont {Surace}},\ and\ \bibinfo {author} {\bibfnamefont {A.}~\bibnamefont {Lucas}},\ }\bibfield  {title} {\bibinfo {title} {{Theory of Metastable States in Many-Body Quantum Systems}},\ }\href {https://doi.org/10.1103/PhysRevX.15.011064} {\bibfield  {journal} {\bibinfo  {journal} {Phys. Rev. X}\ }\textbf {\bibinfo {volume} {15}},\ \bibinfo {pages} {011064} (\bibinfo {year} {2025})},\ \Eprint {https://arxiv.org/abs/2408.05261} {arXiv:2408.05261 [math-ph]} \BibitemShut {NoStop}%
\bibitem [{Note7()}]{Note7}%
  \BibitemOpen
  \bibinfo {note} {We thank Fabrizio Minganti for discussions on this point.}\BibitemShut {Stop}%
\bibitem [{\citenamefont {Mori}\ \emph {et~al.}(2018)\citenamefont {Mori}, \citenamefont {Ikeda}, \citenamefont {Kaminishi},\ and\ \citenamefont {Ueda}}]{Mori:2018}%
  \BibitemOpen
  \bibfield  {author} {\bibinfo {author} {\bibfnamefont {T.}~\bibnamefont {Mori}}, \bibinfo {author} {\bibfnamefont {T.~N.}\ \bibnamefont {Ikeda}}, \bibinfo {author} {\bibfnamefont {E.}~\bibnamefont {Kaminishi}},\ and\ \bibinfo {author} {\bibfnamefont {M.}~\bibnamefont {Ueda}},\ }\bibfield  {title} {\bibinfo {title} {Thermalization and prethermalization in isolated quantum systems: a theoretical overview},\ }\href {https://doi.org/10.1088/1361-6455/aabcdf} {\bibfield  {journal} {\bibinfo  {journal} {Journal of Physics B: Atomic, Molecular and Optical Physics}\ }\textbf {\bibinfo {volume} {51}},\ \bibinfo {pages} {112001} (\bibinfo {year} {2018})}\BibitemShut {NoStop}%
\bibitem [{Note8()}]{Note8}%
  \BibitemOpen
  \bibinfo {note} {The terms resonant and non-resonant are justified in the classical limit: it turns out, as we have shown in \cite {Markovic:2023gtx}, that the ratio of frequencies is proportional to the ratio of actions, i.e. occupation numbers.}\BibitemShut {Stop}%
\bibitem [{Note9()}]{Note9}%
  \BibitemOpen
  \bibinfo {note} {Admittedly, from numerics we cannot tell if the uniform distribution is never reached for small and large $U/J$, or it just takes longer time. However, Fig.~\ref {normdiff} suggests that the dispersion itself does not grow monotonically for large $U/J$ but goes up and down. For small $U/J$ the question is indeed open.}\BibitemShut {Stop}%
\end{thebibliography}%
\end{document}